\documentclass[a4paper,12pt]{article}
\usepackage[utf8]{inputenc}
\usepackage[T1]{fontenc}
\usepackage{subfig}
\usepackage{hyperref}
\usepackage{amsmath,amssymb,amsfonts}
\usepackage{cite}
\usepackage[pdftex,dvipsnames]{xcolor}
\usepackage{graphicx}
\usepackage{caption}
\usepackage{titlesec}
\usepackage{color,soul}
\usepackage[normalem]{ulem}
\usepackage[numbers,sort&compress]{natbib}
\usepackage{fancyhdr}

\newcommand{\ep}{\varepsilon}
\def\res{\mathop{\mathrm{Res}}}
\newcommand{\eref}[1]{(\ref{#1})}
\newcommand{\rmi}{\mathrm{i}}
\newcommand{\rme}{\mathrm{e}}
\newcommand{\e}{\mathrm{e}}
\newcommand{\fref}[1]{figure \ref{#1}}
\newcommand{\rmd}{\mathrm{d}}
\renewcommand{\Re}{\mathrm{Re}}
\renewcommand{\Im}{\mathrm{Im}}
\newcommand{\atp}[2]{\genfrac{}{}{0pt}{}{#1}{#2}}
\renewcommand{\binom}[2]{{#1 \choose #2}}

\newcommand*{\doi}[1]{Doi:\ \href{http://dx.doi.org/#1}{#1}}

\renewcommand*{\url}[1]{Link:\ \href{#1}{#1}}

\hypersetup{
    unicode=true,          
    pdftoolbar=true,        
    pdfmenubar=true,        
    pdfstartview={FitH},    
    pdftitle={General asymptotic expansions of the hypergeometric function with two large parameters},    
    pdfauthor={Mislav Cvitković, Ana-Sunčana Smith, Jayant Pande},     
    pdfsubject={General asymptotic expansions of the hypergeometric function with two large parameters},   
    pdfcreator={Mislav Cvitković},   
    pdfproducer={Mislav Cvitković}, 
    pdfkeywords={hypergeometric function, asymptotic expansion, method of steepest descent, large parameters, special functions}, 
    colorlinks=true,       
    linkcolor=blue,          
    citecolor=purple,        
    filecolor=magenta,      
    urlcolor=magenta           
}

\numberwithin{equation}{section}
\titleformat{\section}{\normalfont\sffamily\Large\bfseries}{\thesection}{1em}{}
\titleformat{\subsection}{\normalfont\sffamily\large\bfseries}{\thesubsection}{1em}{}
\titleformat{\subsubsection}{\normalfont\sffamily\bfseries}{\thesubsubsection}{1em}{}
\begin{document}

\thispagestyle{plain}
\begin{center}
{\Large{{\bf{\textsf{{ }\\{\hspace{1cm} }\\ Asymptotic expansions of the hypergeometric \\ \vspace{0.03cm} function with two large parameters --- application to \\ \vspace{0.03cm} the partition function of a lattice gas in a field of \\ \vspace{0.03cm} traps \\ \vspace{0.03cm}}}}}}
\end{center}
\begin{center}
\vskip6pt
Mislav Cvitkovi\'c$^{1,2,\dagger}$, Ana-Sun\v{c}ana\ Smith$^{1,2}$, Jayant Pande$^{1}$

\vskip12pt

$^1${\footnotesize{\emph{PULS Group, Department of Physics \& Cluster of Excellence: EAM, Friedrich-Alexander University Erlangen-N\"urnberg, N\"agelsbachstr.\ 49b, Erlangen, Germany}}}\\
$^2${\footnotesize{\emph{Division of Physical Chemistry, Ruđer Bošković Institute, Bijeni\v{c}ka c.\ 54, Zagreb, Croatia}}}\\
\vskip4pt
$^\dagger${\footnotesize{Author for correspondence. E--mail:\ \href{mailto:mislav.cvitkovic@fau.de}{mislav.cvitkovic@fau.de}. }}
\vskip18pt
\end{center}

\begin{abstract}
The canonical partition function of a two-dimensional lattice gas in a field of randomly placed traps, like many other problems in physics, evaluates to the Gauss hypergeometric function ${}_2F_1(a,b;c;z)$ in the limit when one or more of its parameters become large. This limit is difficult to compute from first principles, and finding the asymptotic expansions of the hypergeometric function is therefore an important task. While some possible cases of the asymptotic expansions of ${}_2F_1(a,b;c;z)$ have been provided in the literature, they are all limited by a narrow domain of validity, either in the complex plane of the variable or in the parameter space. Overcoming this restriction, we provide new asymptotic expansions for the hypergeometric function with two large parameters, which are valid for the entire complex plane of $z$ except for a few specific points. We show that these expansions work well even when we approach the possible singularity of ${}_2F_1(a,b;c;z)$, $|z|=1$, where the current expansions typically fail. Using our results we determine asymptotically the partition function of a lattice gas in a field of traps in the different possible physical limits of few/many particles and few/many traps, illustrating the applicability of the derived asymptotic expansions of the HGF in physics. 
\end{abstract}

\section{Introduction and outline of the problem}

Consider the following physical problem. There is a gas of $p$ identical hard particles performing a random walk in a two-dimensional lattice with $N$ nodes containing $t$ randomly distributed traps. The particles interact with each other via a hard wall potential, and their interaction with the traps is modelled by probabilities to bind ($P_\mathrm{on}$) and unbind ($P_\mathrm{off}$) when a particle visits a trap. The basic quantity that determines the behaviour of the system and any of its physical properties (such as the mean number of bound particles) is the canonical partition function $Z$. How do we calculate it succinctly?

\pagestyle{fancy}
\lhead{{\footnotesize{\textsf{Asymptotic expansions of the HGF with two large parameters}}}}
\chead{}
\rhead{{\footnotesize{\textsf{M.\ Cvitkovi\'{c} \emph{et al.}}}}}
\lfoot{}
\cfoot{\thepage}
\rfoot{}
\renewcommand{\headrulewidth}{0.4pt}
\renewcommand{\footrulewidth}{0pt}
\addtolength{\headheight}{10pt}
\addtolength{\headsep}{-10pt}

For the system, $Z$ depends on the concentration of particles and of traps, and on the binding energy $E_{\mathrm{b}} = -k_{\mathrm{B}} T \ln (P_{\mathrm{on}}/P_{\mathrm{off}})$, in the following way:
\begin{equation}\label{eq:partfun}
Z = \sum_{n=0}^{\mathrm{min}(p,t)} {N-t \choose p-n} {t \choose n} \sum_{k=0}^{n} {n \choose k} \exp \left( -\frac{k E_{\mathrm{b}}}{k_{\mathrm{B}} T} \right). 
\end{equation}
The first binomial coefficient specifies the number of ways to distribute the particles that are out of traps, the second one gives the number of ways to distribute the particles that are in the traps (denoted by $n$), and the third one the number of ways to distribute the particles that get bound to the traps upon entering the traps (given by $k$). By some straightforward manipulation (see \ref{sec:appE}), the partition function \eref{eq:partfun} evaluates to 
\begin{equation}\label{eq:partitionfunction}
Z = {N-t \choose p} \cdot F \! \left( \atp{-p,-t}{N-p-t+1} ; 1-\delta_{1,P_{\mathrm{on}}} + \frac{P_\mathrm{on}}{P_\mathrm{off}} \right), 
\end{equation}
\sloppy where $\delta_{i,j}$ is the Kronecker delta function, and $F \big( \atp{a,b}{c} ; z \big)$ (alternatively denoted as ${}_2F_1(a,b;c;z)$ in the literature) is the Gauss hypergeometric function, defined by the series 
\begin{equation}\label{eq:hgfdefsum}
F \! \left( \atp{a,b}{c} ; z \right) = 1 + \frac{ab}{c} z + \frac{a (a + 1) b (b + 1) }{ c (c + 1) \, 2! } z^2 + \dots 
\end{equation}
on the disk $|z| < 1$ in the complex plane, and by analytic continuation outside this disk. 

The above discussion provides just one of the many instances in which the Gauss hypergeometric function --- hereafter abbreviated as the HGF --- appears in various fields of physics 
\cite{Seifert:1947:302,
Lighthill:1947:196,
Cherry:1950:190,
Gustafsson:2006:226,
Rawlins:1999:236,
Jones2000:237,
Thorsley:2001:244,
Eckardt:1930:230,
Nordsieck:1954:245,
Gavrila:1967:233,
Dekar:1999:232,
Mace:1995,
Jha:2010:246,
Haubold:1995:260,
Atkinson:1988:235,
Kalmykov:2008:259,
Tai:2010:258,
Middleton:2011:239,
Bekenstein:2015:241,
Domazet:2014:214,
Witte:1995304,
Leclair:1996:240,
Goncharenko:2004:243,
Burkhardt:1991:234,
Fuks:1999:238,
Viswanathan:2015:237} 
and mathematics \cite{Thorne:1597:303,
Denzler:1997:301,
Cohl:2010:259,
Guo:2014:251,
Egorychev:1977:227,
Anderson:1997:228,
Koornwinder:1984:228,
Adolphson:1999:252,
Kazalicki:2013:253,
Olver:2010:219}. 
In many of the physical applications of the HGF $F \big( \atp{a,b}{c} ; z \big)$, some of its parameters ($a$, $b$ and/or $c$) adopt very large values \cite{Seifert:1947:302,Lighthill:1947:196,Cherry:1950:190,Rawlins:1999:236,Jones2000:237,Thorsley:2001:244,Eckardt:1930:230,Gavrila:1967:233,Dekar:1999:232,Mace:1995,Haubold:1995:260,Bekenstein:2015:241,Witte:1995304,Leclair:1996:240,Goncharenko:2004:243,Fuks:1999:238,Thorne:1597:303,Denzler:1997:301}.  For instance, in the above-described problem studying a lattice gas in a field of traps, either the number of particles $p$ or the number of traps $t$ or both can become enormous (up to $\sim 10^{23}$) in systems of realistic size, meaning that at least two parameters of the HGF in \eref{eq:partitionfunction} may be large. Unfortunately, determining the large-parameter limit of the HGF by first principles, either analytically or computationally, using \eref{eq:hgfdefsum} is very difficult, due to both a slow convergence of the series and the hefty coefficients involved when the parameters of the HGF are large. For these reasons it is a crucial task to find the asymptotic expansions (hereafter referred to as the AEs) of the HGF when some of its parameters are large, which can then be used to solve the underlying physical problems. 

The first attempt to find the AE of an HGF with large parameters was made by Laplace \cite{Laplace:1825:225} who gave the AE of the Legendre polynomial for large $ n $:
\begin{equation}
P_n(\cos\vartheta) = F \! \left( \atp{ -n, n+1}{1} ; \frac{1-\cos\vartheta}{2} \right) \sim 
\sqrt{\frac{2}{n\pi\sin\vartheta}} \cos \left[\! \left( n+\frac{1}{2} \right)\vartheta -\frac{\pi}{4} \right]
\end{equation} 
as $n\to\infty$, which was a special case of an HGF with two large parameters. The first extensive study of the AEs of the HGF in general was performed by Watson \cite{Watson:1918:172}, who used the method of steepest descent (hereafter denoted as the MSD) to evaluate the general asymptotic form of the HGF,
\begin{equation}\label{eq:asydef}
 F \! \left( \atp{a + \ep_1 \lambda, b + \ep_2 \lambda}{c + \ep_3 \lambda} ; z \right) \quad \textrm{as } |\lambda| \to \infty \quad  \mathrm{for} \quad \ep_i \in \{ 0, 1, -1 \},
\end{equation}
for the cases $(\ep_1,\ep_2,\ep_3) = (0,0,1)$, $(1,-1,0)$ and $(0,-1,1)$. He then used the transformation formulae of the HGF to express the remaining cases of $\ep_i \in \{ 0, 1, -1 \}$ in terms of the evaluated ones. 

The problem with Watson's expansions is their relatively narrow region of validity in the $ z $--complex plane--for instance, they are invalid in the neighbourhood of the critical points of the HGF. Furthermore, the conditions on the parameters when using the trans\-for\-ma\-tion formulae of the HGF strongly restrict the expansions obtained after applying the transformations. After some advances in \cite{Lighthill:1947:196} and \cite{Cherry:1950:190} for a particular HGF occurring in fluid flow theory, the most exhaustive treatment of the cases $\ep_i \in \{ 0, 1, -1 \}$ in \eref{eq:asydef} was provided recently by Olde Daalhuis \cite{Daalhuis:2003:220,Daalhuis:2003:221,Daalhuis:2010:222,FaridKhwaja:2013:286,FaridKhwaja:2014:223,FaridKhwaja:2016:287} and Jones \cite{Jones:2001:224}. The resulting AEs took the form of series of Airy \cite{Daalhuis:2003:221,Daalhuis:2010:222,FaridKhwaja:2014:223}, parabolic cylinder \cite{Daalhuis:2003:220,Daalhuis:2010:222,FaridKhwaja:2014:223}, Bessel \cite{FaridKhwaja:2014:223,Jones:2001:224}, Hankel \cite{FaridKhwaja:2014:223}, and/or Kummer \cite{FaridKhwaja:2013:286,FaridKhwaja:2014:223,FaridKhwaja:2016:287} functions, depending on the value of $\ep_i$.

The AEs of the HGF with more general values of $\ep_i$ have been studied only for particular HGFs appearing in different problems in physics. Some examples are  the solution of the compressible gas flow dynamics problem \cite{Seifert:1947:302},  the persistence problem of the solutions of the sine-Gordon equation in \cite{Denzler:1997:301}, the Legendre functions $ P_{n}^{-m}(z)$ and $ Q_{n}^{-m} (z)$ for $n,m\to \infty$ in \cite{Thorne:1597:303}, the plaquette expansion for lattice--Hamiltonian systems \cite{Witte:1995304}, and  the quantal description of the scattering of charged particles in atomic systems in \cite{Thorsley:2001:244,Nagel:2001:247}. 

Complementary to this, the AEs for a number of problems in physics where the HGF with large parameters appears have not been studied before. A few such examples are the scattering of electromagnetic waves on dielectric obstacles \cite{Jones2000:237} and the hydrogen atom \cite{Gavrila:1967:233}, the Schr\"odinger equation for smooth potentials and the mass step in one dimension \cite{Dekar:1999:232}, the dispersion of plasma with high--energy particles \cite{Mace:1995}, the distributions of mass, pressure and temperature and the total outflow of energy at some distance from the center of the Sun \cite{Haubold:1995:260}, the envelope of the Friedel oscillations caused by a simple impurity in a 1--D Luttinger liquid ($g=1/2$) at finite temperature \cite{Leclair:1996:240}, the exact flow in the deterministic cellular automaton model of traffic \cite{Fuks:1999:238}, and Bekenstein's description of the statistical response of a Kerr black hole with a quantum structure to an incoming quantum radiation \cite{Bekenstein:2015:241}.

An attempt to find the AEs for a general HGF was made recently by Paris \cite{Paris:2013:173,Paris:2013:174}, who reverted to the MSD to obtain the expansion of $ F \big( \atp{a + \ep_1 \lambda, b + \ep_2 \lambda}{c + \ep_3 \lambda}; z \big) $ for $ | \lambda | \to \infty $, with $ \ep_i $ taking any finite value. These expansions, however, have their own restrictions on the regions of validity in the $ z $--plane (e.g.\ $|z|<1/\ep $ for the case $(1,0,0)$ in \cite{Paris:2013:173}), and, as we show later, these restrictions increase when one uses the stated transformations to go from the primary parameter set $ ( \ep_1, \ep_2, \ep_3 ) =  ( \ep, 0, 1) $ to other cases in the class $ \ep_2 = 0 $, and analogously in the classes $ \ep_3 = 0 $ and $ \ep_3 = \pm 1 $. Therefore, in spite of the many efforts that have gone into it, finding the AEs of the general HGF with large parameters, valid for all parameter and variable values, has remained a challenging problem.

In this article, we  overcome this problem by providing closed-form expressions for the AEs of the HGF, with any two of the parameters $a,\ b$ and $c$ large, that are valid over the entire $ z $--plane except for a few points. Our expansions do not suffer from a restriction ubiquitous in the known expansions in the literature, that $1/z$ must be outside the integration loop of the integral representation of $ F $, and work well even in the vicinity of $ | z | = 1 $ where the HGF typically diverges (except for some cases, specified later). This allows us to explicitly calculate the partition function \eref{eq:partitionfunction} for different combinations of large parameters $p$, $t$ and $N$, which represent different physical limits of the lattice gas dynamics.

\section{Calculation of the asymptotic expansions of the HGF}
Our approach to calculate the AEs of the HGF with any two parameters having large values is to use the MSD to calculate $ F \big( \atp{a + \ep \lambda, b}{c + \lambda}; z \big) $ and  $ F \big( \atp{a + \ep \lambda, b + \lambda}{c} ; z \big) $ when $ | \lambda | \to \infty $, for both $ |\ep| < 1 $ and $ |\ep| > 1 $, and then use the transformation rules to reduce any other combination of $( \ep_1, \ep_2, \ep_3 )$, with one $\ep_i$ equal to $0$, to these two cases. When the point $1/z$ appears to lie within the integration loop of the integral representation of $F$, we will either explicitly calculate its contribution or show that it can be neglected.

\subsection{Representations of the HGF and the MSD scheme}

The HGF is the solution of the hypergeometric differential equation [\citenum{Olver:2010:219}, p.\ 394; \citenum{Abramowitz:1972:218}, p.\ 562]
\begin{equation}\label{eq:HGDE}
z(1-z) \frac{\rmd^2 F(z)}{\rmd z^2} +\big[ c- (a+b+1) z \big] \frac{\rmd F (z)}{\rmd z} -ab F(z) = 0. 
\end{equation}
For $ c \neq 0, -1, -2, \dots $, the HGF is defined on the disk $ |z| < 1 $ by the series \eref{eq:hgfdef}, which by introduction of  the Pochhammer symbol
$ (x)_n = \Gamma (x + n) / \Gamma (x) $ can be written concisely as 
\begin{equation}\label{eq:hgfdef}
F \! \left( \atp{a,b}{c} ; z \right) =  \sum_{n=0}^\infty \frac{ (a)_n (b)_n }{ (c)_n n! } z^n. 
\end{equation}
Outside the disk $ |z| < 1 $, the HGF is defined by analytic continuation of \eref{eq:hgfdef}. $ F $ is a multivalued function of $ z $ and is analytic everywhere except possibly at $ z = 0 $, $ z = 1 $ and $ z = \infty $, which may be branch points \cite[p.\ 384]{Olver:2010:219}. The principal branch is the one obtained by introducing a cut from 1 to $ + \infty $ along the real axis, i.e.\ the one in the sector $ | \! \arg (1 - z) | \leq \pi $. In this paper we assume that $z$ belongs to the principal branch if $|z| \geq 1$ and $z$ lies on the branch cut of the HGF, i.e. we assume in that case that $\arg (z) = 0$.

In all the cases we consider, we will assume that $\Re (z) \ge 0$. The case of $\Re (z) < 0$ can then be easily handled by using a transformation formula of the HGF \cite[\S 15.3.3--15.3.5]{Abramowitz:1972:218}
\begin{equation}\label{eq:trans1} 
F \! \left( \atp{a,b}{c} ; z \right) = \frac{ 1 }{ ( 1 - z )^b } F \! \left( \atp{ c - a, b}{c} ; \frac{ z }{ z - 1 } \right) = (1 - z)^{ c - a - b } F \! \left( \atp{ c - a, c - b }{ c } ; z \right)
\end{equation}
to convert the variable $z$ to $z(z-1)^{-1}$. Additionally, a straightforward manipulation of \cite[\S 15.3.6]{Abramowitz:1972:218} using \eref{eq:trans1} and \cite[\S 15.3.9]{Abramowitz:1972:218}, given $c-a-b \notin \mathbb{Z}$ and $ c  \notin \mathbb{N}_0 $, yields a formula
\begin{equation}\label{eq:trans2}
 \frac{ \Gamma (1 - c) \, F  \big( \atp{ a,      b }{ c } ; z\big)} { \Gamma ( a - c + 1 ) \Gamma ( b - c + 1 ) }  =
  \frac{  F  \big( \atp{ a,      b }{ a + b - c + 1 } ; 1 - z \big) }{ \Gamma ( a + b - c + 1 ) } 
 - \frac{ \Gamma (c - 1) }{ \Gamma (a) \Gamma (b) } 
 \frac{ F   \big( \atp{ 1 - a,  1-  b }{2-c } ; z \big) }{  z^{ c-1 } ( 1 - z )^{ a + b - c } }  
\end{equation}
that will be used later. (Here the convention is adopted that $0 \notin \mathbb{N}$ and $\mathbb{N}_0 = \mathbb{N} \cup \{ 0 \}$.)

To apply the MSD, we find it useful to express the HGF in terms of its integral rep\-re\-sen\-ta\-tions \cite[p.\ 388, \S  15.6]{Olver:2010:219}. For different cases different representations are appropriate, due to distinct restrictions on the parameters and the variable $z$. The representations that we use here are: 
\begin{eqnarray}
\! F \! \left( \atp{a,b}{c} ; z \right) &  \!\!\!\!=\!\!\! &  \frac{ \Gamma (c) }{ \Gamma (a) \, \Gamma (c - a) }  \int_0^1    \frac{ t^{a - 1} (1 - t)^{c - a - 1} }{ (1 - z t)^b } \: \mathrm{d} t, 
\quad\!\!\! \textrm{if } \Re (c) > \Re (a) > 0;   \label{eq:intrepA}   \\
\! F \! \left( \atp{a,b}{c} ; z \right) &  \!\!\!\!=\!\!\! &  \frac{ \Gamma (a - c + 1) \, \Gamma (c) }{ 2 \pi \rmi \, \Gamma (a)}  \int_0^{(1+)} \! \frac{ t^{a - 1} (t - 1)^{c - a - 1} }{ (1 - z t)^b } \: \mathrm{d} t , 
\quad\!\!\! \textrm{if } c \! - \! a \notin \mathbb{N}\textrm{, } \Re (a) > 0 ;    \label{eq:intrepB}   \\
\! F \! \left( \atp{a,b}{c} ; z \right) &  \!\!\!\!=\!\!\! &  \rme^{ -a \pi \rmi }   \frac{ \Gamma (1 - a)  \, \Gamma (c) }{ 2 \pi \rmi  \, \Gamma (c - a) }  \int_1^{(0+)}   \frac{ t^{a - 1} (1 - t)^{c - a - 1} }{ (1 - z t)^b } \: \mathrm{d} t,
\quad\!\!\! \textrm{if } a \notin \mathbb N\textrm{, }\! \Re (c \! - \! a) > 0. \qquad    \label{eq:intrepC} 
\end{eqnarray}
The region of validity for each of these equations is $ | \! \arg  (1 - z) | < \pi $.  In \eref{eq:intrepB}, the integration path is the loop that starts at $ t = 0 $, encircles the point $ t = 1 $ in the anti-clockwise direction, terminates at $ t = 0 $, and excludes the point $t = 1/z$. (This loop is denoted by ``$(1+)$''; see dashed black path on figures \ref{fig:3}(b), \ref{fig:4} and \ref{fig:6}(a).) In \eref{eq:intrepC} the integration path is similar, but with the points $ t = 0 $ and $ t = 1 $ swapped (dashed black path on \fref{fig:6}(b)).

In each case that we examine, we express the contour integral from an appropriate representation of the HGF above as $\int_C f(t) \, \e^{ \lambda g(t)} \: \rmd t $, and deform the contour of integration $C$ so that it passes through the saddle point $t_0$ of $\lambda g(t)$ along the steepest descent path. Then we expand $g(t)$ in a Taylor series up to the second order. 
With this approximation, $ \rme^{ \lambda g(t) } $ becomes a Gaussian, which becomes narrower as $ |\lambda| $ gets larger. In the limit $|\lambda| \to  \infty$, $f(t)$ varies very slowly in comparison to $\e^{ \lambda g(t) }$, i.e.\ $ f(t) \approx f(t_0) $, and the vicinity of the saddle point of $\lambda g(t)$ provides the dominant contribution to the integral, which by evaluation of the Gaussian integral approximates to \cite[p.\ 108]{Miller:2006:225}
\begin{equation}\label{eq:MSDdef}
\int_C f(t) \, \rme^{\lambda g(t)} \, \rmd t \sim  \frac{\rme^{\lambda g(t_0) +\rmi \vartheta }}{\sqrt{ | \lambda | }} \left( f(t_0) \sqrt{ \frac{2\pi}{\left| g''(t_0) \right|} } + \mathcal{O}\left(\lambda^{-1}\right) \right) 
 \textrm{ as } |\lambda| \to \infty 
\end{equation}
for the second--order saddle point. Here $\vartheta$ is the angle that the path of steepest descent makes with the real axis, given by the equation $ \Im (\lambda g(t)) = \Im (\lambda g(t_0)) $, which for a general--order saddle point yields the formula \cite[p.\ 105]{Miller:2006:225}
\begin{equation}\label{eq:theta}
\vartheta = \frac{ ( 2 k + 1 ) \pi - \alpha }{ N + 1 },
\end{equation}
where $\alpha  = \arg (\lambda g^{(N)} (t_0) )$, $ N $ is the order of the saddle point, and $ k = 0, 1, \dots, N $. For more details on the MSD consult \cite{Miller:2006:225,Bleistein:1986:265,Bender:1999:266,Murray:1984:267,Wong:2001:291} and the references therein. 

Accuracy of the AEs throughout this paper will be explored by the relative error $R$, defined as
\begin{equation}\label{eq:relerr}
R = 100\cdot \left|1-\frac{\mathrm{AE}}{\mathrm{HGF}} \right|.
\end{equation}

\subsection{Expansions of the HGF for large \texorpdfstring{$a$}{a} and \texorpdfstring{$c$}{c}} \label{sec:ac} 

\subsubsection{Case \texorpdfstring{$ | \ep | \leq 1$}{ε≤1}} \label{ssec:ep<1}

To begin with we will assume that $\ep \ge 0$, and connect this case to that of negative $\ep$ later. This shall be our approach for much of this paper. The representation suitable for $ F \! \big( \atp{ a + \ep \lambda, b}{ c + \lambda } ; z \big) $ when $ \ep \le 1 $ is \eref{eq:intrepA}, in which the HGF reads
\begin{equation}\label{eq:int:ep<1}
F \! \left( \atp{ a + \ep \lambda, b }{ c + \lambda } ; z \right) = \frac{ \Gamma ( c + \lambda ) }{ \Gamma ( a + \ep \lambda ) \, \Gamma ( c - a + ( 1 - \ep ) \lambda ) } \int_0^1 f(t) \, \e^{ \lambda g(t) } \mathrm{d} t,
\end{equation}
where
\begin{align}
f(t) &= \frac{ t^{a - 1} (1 - t)^{c - a - 1} }{ (1 - z t)^b }, \\
g(t) &= \ln \left[ t^{\ep} (1 - t)^{ 1 - \ep } \right], \label{eq:fg:ep<1}
\end{align}
and $ \lim_{|\lambda| \to \infty }| \arg( \lambda ) | < \pi/2$ in order to satisfy the condition of \eref{eq:intrepA}. The function $ g(t) $ has two branch cuts on the real axis given by $ ( - \infty, 0 ] $ and $ [ 1, + \infty ) $. For $ b \neq 0 $ the function $ f(t) $ has the critical point $ t_\mathrm{c} = 1/z $ whose nature depends on $ b $: for $ b \in \mathbb{Z} \backslash \mathbb{N}_0 $ it is a zero of $ f(t) $, for $ b \in \mathbb{N} $ it is a pole of order $ b $, and for $ b \in \mathbb{R} \backslash \mathbb{Z} $ it is a branch point giving a branch cut from $ t_\mathrm{c} $ to infinity in a suitable direction. 

The condition to find the saddle point, $ \lambda g' (t_0) = 0 $, has the solution $ t_0 = \ep $. Consequently, 
\begin{align} 
f(t_0) &= \frac{\ep^{(a-1)} (1-\ep)^{c-a-1}}{(1-\ep z)^b }, \label{eq:ft0def} \\
\rme^{\lambda g(t_0)} &= \ep^{\lambda\ep} (1-\ep)^{(1-\ep)\lambda}, \textrm{ and} \label{eq:egt0def} \\
g'' (t_0) &= \frac{1}{\ep (\ep -1)}, \label{eq:gt0derivative}
\end{align}
for $ \ep \neq 0,\, 1 $, which shows that $t_0$ is a simple saddle point, since $\lambda g'' (t_0) \neq 0 $. Here we assume that $ \ep \neq 0,\, 1 $. This is justified since firstly, for $ \ep = 0 $, the only large parameter in $ F \big( \atp{ a + \ep \lambda, b }{c + \lambda } ; z \big) $ is $c + \lambda$, and then \eref{eq:hgfdef} yields trivially that the asymptotic expansion of the HGF is
\begin{equation}\label{eq:largec}
F \left( \atp{a,b}{c} ; z \right) \sim 1+\frac{ab}{c}z 
\quad \textrm{as } |c| \to \infty 
\end{equation}
to the first order in $c$. Secondly, the case $ \ep = 1 $ has been treated in the literature in a satisfactory manner \cite{Watson:1918:172,Wagner:1947:290,Temme:2003:180,Daalhuis:2003:220,Daalhuis:2003:221,Daalhuis:2010:222,FaridKhwaja:2013:286,FaridKhwaja:2014:223,FaridKhwaja:2016:287,Ferreira:2006:289}. 

For $\ep \notin \{ 0, 1 \} $, $g'' (t_0) = [ \ep ( \ep - 1 ) ]^{-1}$ is purely real and negative and $\alpha = \pi  + \arg(\lambda) $, meaning that the two steepest descent curves emanate from the saddle point $ t_0 = \ep $ with the angles $\vartheta = -\arg(\lambda)/2 $ 
 \begin{figure}[!ht]
\centering
\subfloat[]{\includegraphics[width=0.5\textwidth]{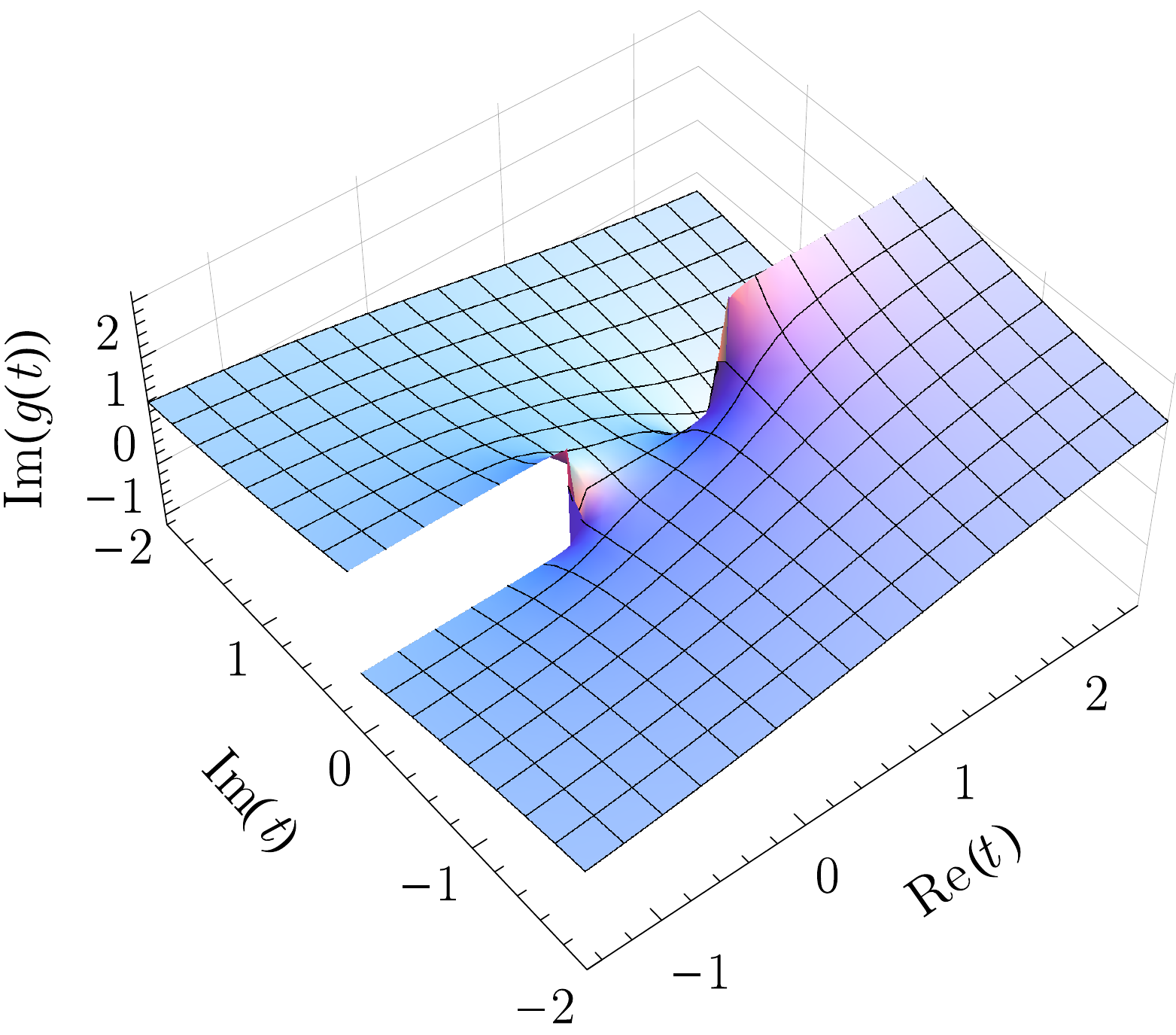}}
\hfill
\subfloat[]{\includegraphics[width=0.44\textwidth]{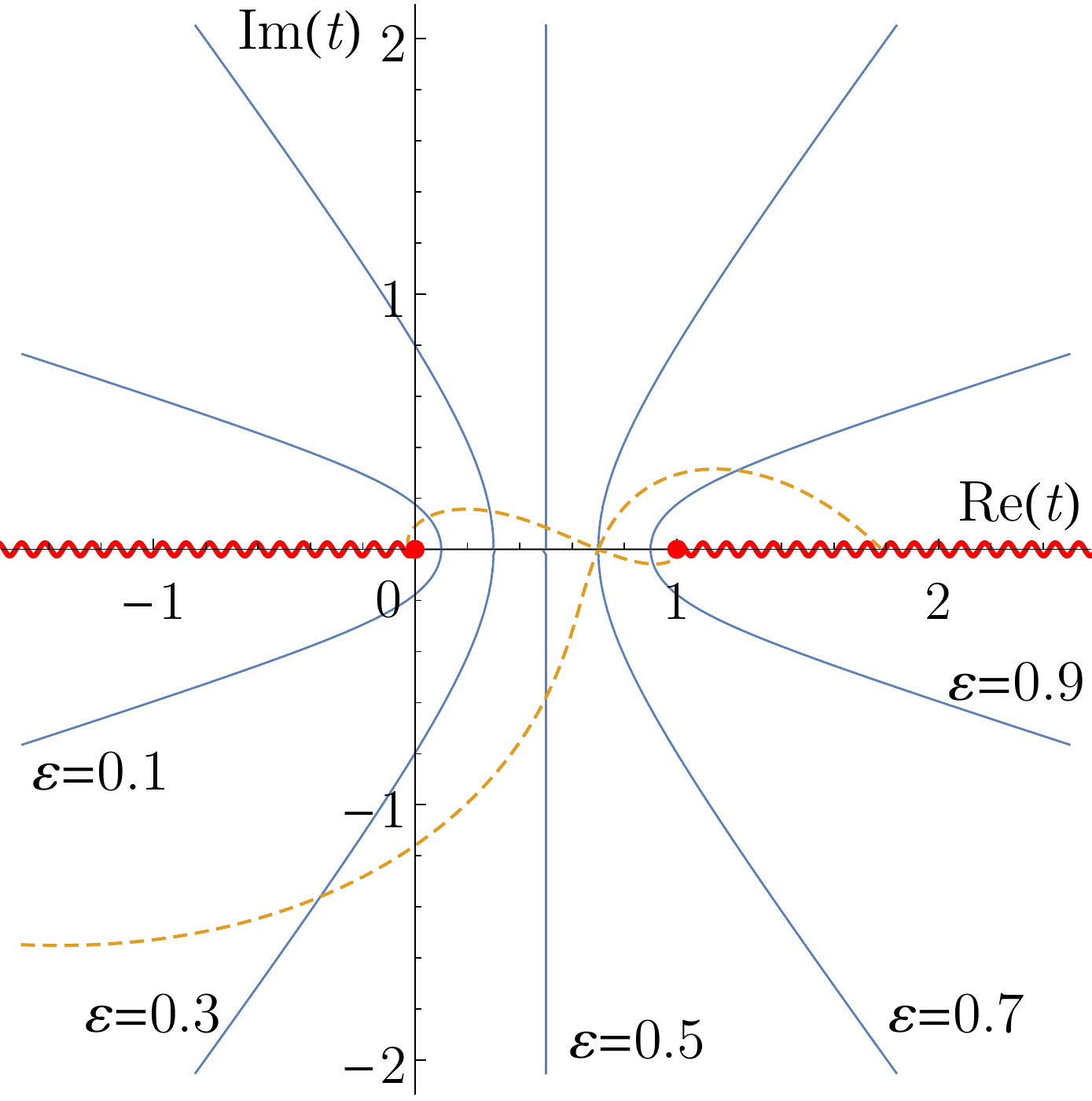}}
\caption{(a) The plot of $ \Im (g(t)) $ over the $ t $--complex plane for $ \ep = 0.3 $. Two branch cuts, $ ( - \infty, 0 ] $ and $ [ 1, \infty ) $, lie on the real axis. (b) The steepest descent (lines on the real axis from 0 to 1) and ascent (hyperbolic) curves of \eref{eq:int:ep<1} for $\lambda = 1$ and different values of $ \ep $ (solid blue) and for $\lambda = 1+\rmi $ and $\ep = 0.7$ (dashed orange). Curly red: the branch cuts. \label{fig:1}}
\end{figure}
and $ \vartheta = \pi -\arg(\lambda)/2  $ to the real axis (see \eref{eq:theta} and \fref{fig:1}), while the steepest ascent curves are perpendicular to the steepest descent curves. We therefore deform the original integration path of \eref{eq:int:ep<1} so that it makes an angle of $\vartheta=-\arg(\lambda)/2 $ to the real axis at the saddle point $\ep$. The integral in \eref{eq:int:ep<1} then, using \eref{eq:ft0def}--\eref{eq:gt0derivative} and \eref{eq:MSDdef}, reads
\begin{equation}
\int_C f(t) \, \rme^{\lambda g(t)} \, \rmd t \sim  \sqrt{\frac{2\pi}{|\lambda|}} \, 
\frac{ (1-\ep)^{c-a+(1-\ep)\lambda -\frac{1}{2}} }{ \ep^{\frac{1}{2}-a-\ep\lambda} \, (1-\ep z)^b} \,  \rme^{-\frac{\rmi}{2}\arg(\lambda)}
\quad \textrm{as }   |\lambda| \to \infty .
\end{equation}
Finally, using Stirling's approximation for $ \Gamma $-functions in the prefactor in \eref{eq:int:ep<1}, the expansion reduces to 
\begin{equation}\label{eq:AE:ep<1}
F \! \left( \atp{ a + \ep \lambda, b }{  c + \lambda } ; z \right) \sim \frac{ 1 }{ ( 1 - \ep z )^b } 
\quad \textrm{as } |\lambda| \to \infty 
\end{equation}
to the leading order.

The AE \eref{eq:AE:ep<1} was previously developed for $ | z | < 1 / \ep $ \cite{Paris:2013:173}. It also follows directly upon using \eref{eq:hgfdef} and approximating the relevant Pochhammer symbols for large $\lambda$, which gives $|\ep z| < 1$ as the condition for the convergence of the series. Here we show that the AE \eref{eq:AE:ep<1} is valid not only for $|z|< 1/\ep$, but for all $z$ except in the vicinity of $z = 1 / \ep $. The validity of the AE is only restricted by the requirement that the critical point $ t_\mathrm{c} = 1/z $, if it is a pole or a branch point (i.e.\ if $b \notin \mathbb{Z} \backslash \mathbb{N} $), should lie outside the integration loop. As the integration path here is not closed, a problem with $ t_\mathrm{c}$ emerges only if it lies right \textit{on} the deformed integration path described above. In this case, we need to deform the path further to avoid $ t_\mathrm{c} $, and include the contribution to the integral arising from the half-turn around $t_{\mathrm{c}}$. For instance, if $ t_{ \mathrm{c} } $ is the pole, then this polar contribution equals $ \pi \rmi \res_{ 1/z } f(t) \rme^{ \lambda g(t) } $. 

It is, however, possible to eliminate the contribution due to $t_{ \mathrm{c}}$ altogether if it is smaller than that due to the saddle point $ t_0 $, for in the limit $|\lambda| \to  \infty$ the smaller of the terms $\rme^{ \lambda g(t_0) }$ and $\rme^{ \lambda g(t_{ \mathrm{c} }) }$ vanishes. The contributions from $t_0$ and $t_{\mathrm{c}}$ can be compared by checking the difference between the real parts of $ g(t) $ at these two points \cite[p.\ 132]{Miller:2006:225} 
which evaluates to 
\begin{equation}\label{eq:rediff1}
\Re ( g( t_{ \mathrm c} ) ) - \Re ( g(t_0) ) = - \frac{\Im (g(t_{\mathrm{c}}))}{\tan (\arg(\lambda))}
\end{equation}
directly from the definition of the steepest descent curve, $ \Im (\lambda g(t)) = \Im (\lambda g (t_0)) $, on which $t_\mathrm{c}$ is assumed to lie. By analysing $\Im (g(t_{\mathrm{c}}))$, we have shown in \ref{sec:appD} that for $ 0 < |t_{\mathrm{c}}| = 1/|z| < \ep  $, i.e.\ for $|z| > 1/\ep$, such that $t_{\mathrm{c}}$ lies on the deformed integration path, $ \Im (g(t_{\mathrm{c}})) > 0$ if $\arg(\lambda) >0$ and $ \Im (g(t_{\mathrm{c}})) < 0$ if $\arg(\lambda) <0$, that is, 
$\Re ( g( t_{ \mathrm c} ) ) < \Re ( g(t_0) ) $ for any $\lambda$. If $\Im (\lambda) = 0$, the steepest descent path is real and \eref{eq:rediff1} is inconclusive. In this case $t_{\mathrm{c}}$, which lies on the path, is real and satisfies the relation $t_{\mathrm{c}} < 1$ ($z > 1$), and the difference is 
\begin{equation}\label{eq:recomp:ep<1}
\Re ( g( t_{ \mathrm c} ) ) - \Re ( g(t_0) ) = \ln \left( \frac{ 1 }{ \ep^{\ep} x } \left| \frac{ x - 1 }{ 1 - \ep } \right|^{ 1 - \ep } \right) = \ln h_{\ep}(x),
\end{equation}
where $x=\Re (z)$.

By an analysis of the argument of the logarithm in \eref{eq:recomp:ep<1}, here denoted as $h_\ep(x)$, it can be shown that $ \Re ( g( t_{ \mathrm c } ) ) < \Re ( g(t_0) )  $ (see \eref{eq:hep3} in \ref{sec:appA}) for this case too.   
Therefore, the contribution to the integral in \eref{eq:int:ep<1} from the critical point $t_\mathrm{c}$, when it lies on the integration path, is negligible for any $\lambda$ and $z$ except in the vicinity of $ z  = 1 / \ep $, when the saddle--point approximation does not work, as the critical point $t_\mathrm{c}= 1/z \to \ep $ approaches the saddle point $t_0 = \ep$.

To conclude, the expansion \eref{eq:AE:ep<1} is valid for $0 < \ep < 1 $, for any $ a, b, c, \Re(\lambda) > 0 $, and for any $ z $ except in the vicinity of $1/\ep$. One notes that the imaginary part of the HGF for real $z>1$, as well as of the AE  \eref{eq:AE:ep<1} for real $z>1/\ep$, becomes non--zero, so from \eref{eq:AE:ep<1} one can directly determine the limiting value of the imaginary part to be $ - \sin( \pi b ) \:  | 1 - \ep z |^{ -b }$ for $z > 1/\ep$. 

\begin{figure}[!ht]
\centering
\hspace{-0.3cm}
\subfloat[]{\includegraphics[width=0.49\textwidth]{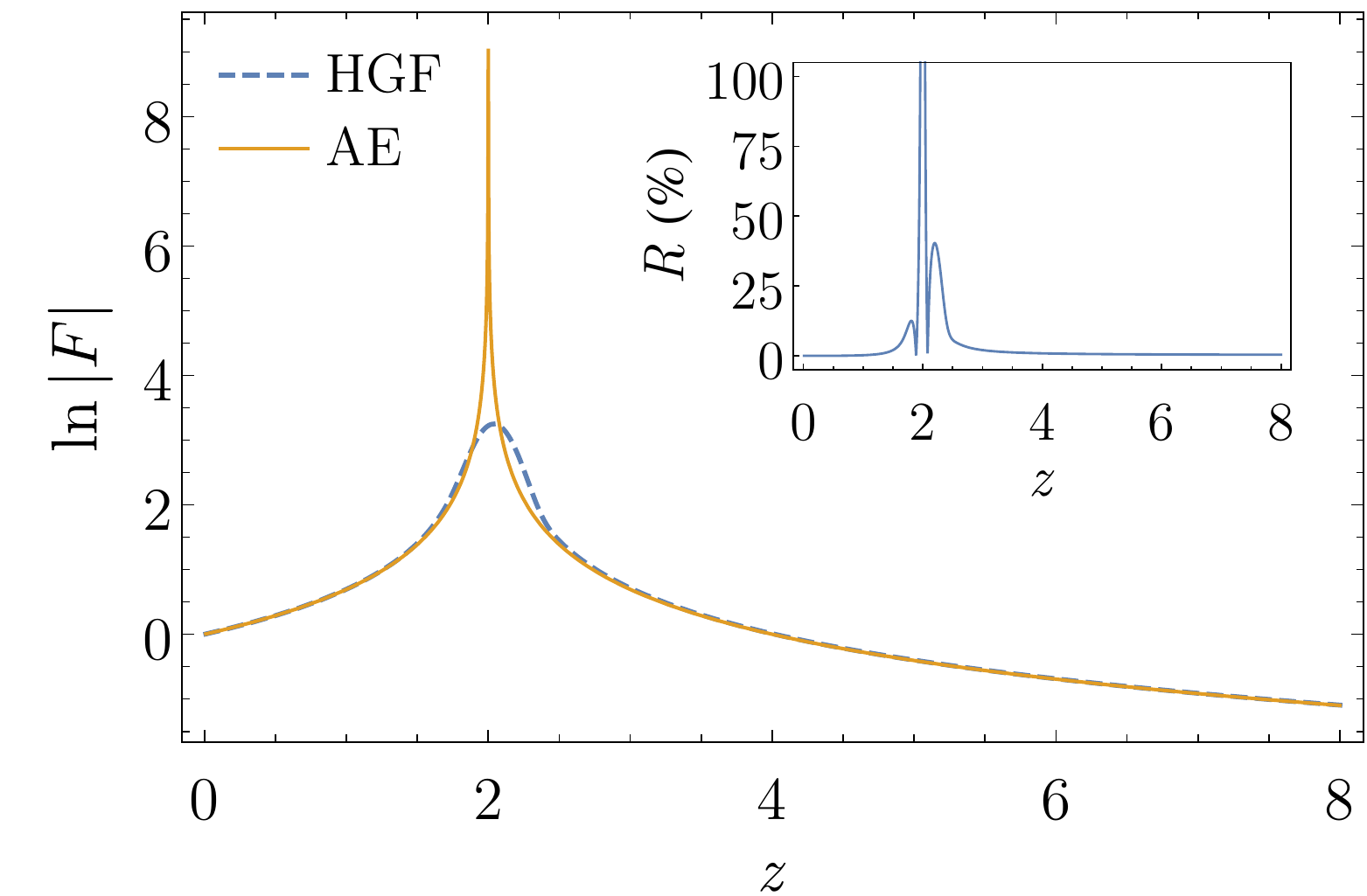}}
\hfill
\subfloat[]{\includegraphics[width=0.49\textwidth]{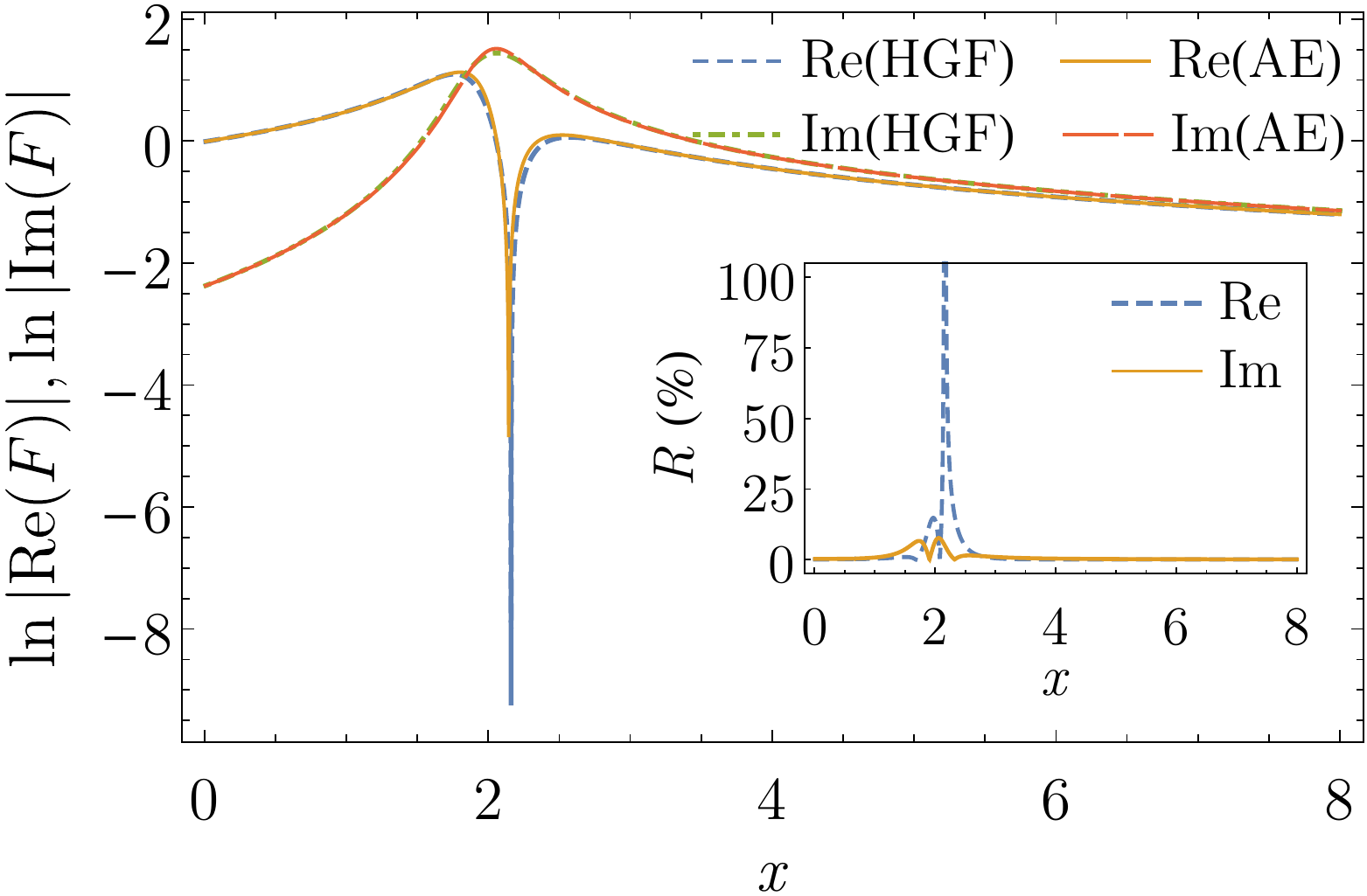}}
\caption{Graphs of the HGF for $ a = 1 $, $ c = 2 $, $ \ep = 0.5 $, $ \lambda = 400 (1+\rmi / 2) $. (a)  $ b = 1 $, $ z \in \mathbb{R} $. (b) $ b = 3/4$, $ z = x + \rmi/4 $, $ x \in \mathbb{R} $. The insets in the plots show the relative error $R$ between the HGF and its AE. Note that in part (a) the HGF looks bounded at $z=1/\ep = 2$ because of the finite value of $\lambda$ used; in the limit $\lambda \to \infty$ the HGF diverges.\label{fig:2} }
\end{figure}

The comparison of the HGF and its AE \eref{eq:AE:ep<1} is shown in \fref{fig:2}. The insets in the plots show the relative error $R$, defined in \eref{eq:relerr}. It is clear that the AE works perfectly for any $ z $ except in the close vicinity of $ z = 1 / \ep $, both for real (\fref{fig:2}(a)) and complex (\fref{fig:2}(b)) $ z $, and for any $ b $. 

To allow $\ep$ to assume negative values, we employ the transformation formulae \eref{eq:trans1} and \eref{eq:trans2}. In particular, the case $ ( - \ep, 0, 1 ) $ can be transformed to $ ( \ep, 0, 1 ) $ by application of \eref{eq:trans1}, while the cases $ ( \pm \ep, 0, -1 ) $ can be transformed to $ ( \ep, 0, 1 ) $ by successive application of \eref{eq:trans1} and \eref{eq:trans2}. With this we have covered all the possibilities of large $ a $ and $ c $ parameters of the HGF for $-1 \le \ep \le 1 $.

 \subsubsection{Case \texorpdfstring{$|\ep| >1$}{ε>1}}\label{ssec:ep>1} 

As before, we will first assume $\ep > 1$ and then connect the case of $\ep < -1$ to this case. The suitable representation of the HGF in the case $(\ep,0,1)$ and $\ep >1$ is \eref{eq:intrepB}, giving
\begin{equation}\label{eq:int:ep>1}
F \! \left( \atp{ a + \ep \lambda, b }{ c + \lambda } ; z \right) = 
\frac{ \Gamma ( c + \lambda ) \, \Gamma ( a - c + ( \ep - 1 ) \lambda + 1 ) }{ 2 \pi \rmi \, \Gamma ( a + \ep \lambda ) } \!
\int_0^{(1+)} \!\!\!  f(t) \, \e^{ \lambda g(t) } \mathrm{d} t
\end{equation}
where we have redefined $ f(t) $ and $ g(t) $ to be 
\begin{align} 
f(t) &= \frac{ t^{ a - 1 } ( t - 1 )^{ c - a - 1 } }{ ( 1 - z t )^b } ,\label{eq:f:ep>1} \\
g(t) &= \ln \left[ t^{\ep} ( t - 1 )^{ 1 - \ep } \right]. \label{eq:g:ep>1}
\end{align}
In addition we must have  $ \lim_{|\lambda|\to\infty} | \arg( \lambda ) | < \pi/2 $ in order to satisfy the conditions under which \eref{eq:intrepB} is valid. The branch cut of $ g(t) $ is now $ ( -\infty, 1 ]$ (equalling $ ( -\infty, 0 ] \cup ( - \infty, 1 ] $), and for $ b\neq 0 $, $ f(t) $ has the critical point $ t_{ \mathrm c } = 1/z $ whose nature is defined by $b$ in the same manner as in section \ref{ssec:ep<1}. 

\begin{figure}[!ht]
\centering
\subfloat[]{\includegraphics[width=0.5\textwidth]{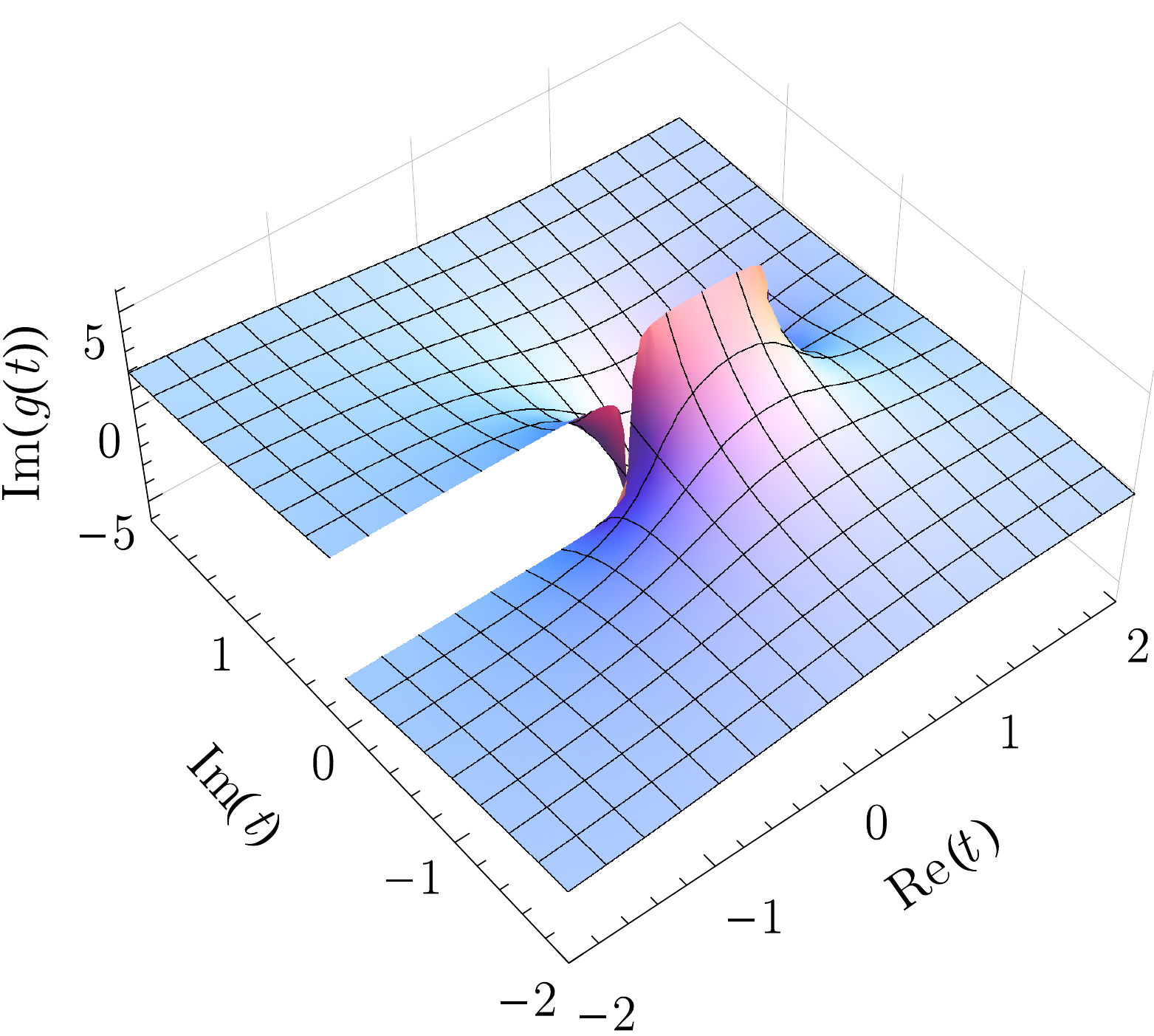}}
\hfill
\subfloat[]{\includegraphics[width=0.44\textwidth]{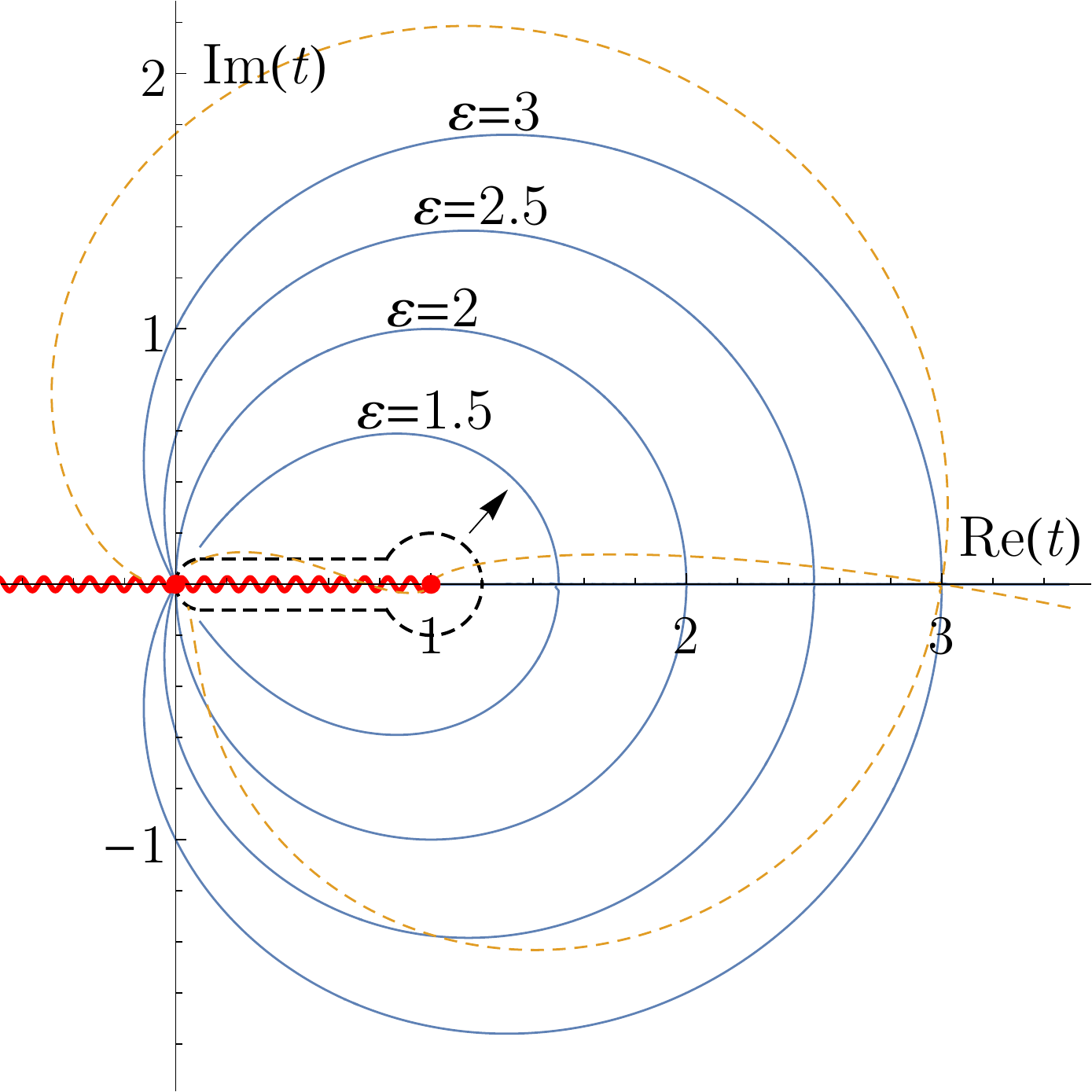}}
\caption{(a) The plot of $ \Im (g(t)) $ over the $ t $--complex plane for $ \ep = 3 $. Two branch cuts, $ ( - \infty, 0 ] $ and $ ( -\infty, 1 ] $, overlap on the real axis. (b) 
The steepest descent (perpendicular to the the real axis) and ascent (lying on the real axis) curves for $\lambda = 1$ and different values of $\ep$ (solid blue) and for $\lambda = 1+\rmi / 3$ and $\ep = 3$ (dashed orange), to which the original integration path of \eref{eq:int:ep>1} (dashed black) is deformed by making use of the Cauchy theorem.   
Curly red: the branch cut.  \label{fig:3}}
\end{figure}

By applying an MSD procedure as in the previous section, it has been shown in the literature that as long as the point $ t_{ \mathrm{c} } $ lies outside the loop of the deformed integration path, the expansion \eref{eq:AE:ep<1} is valid \cite{Paris:2013:173}. This integration path, as before, is that of the steepest descent through the saddle point, which is again $ t_0 = \ep $,  but now $ g'' (t_0) = [ \ep ( \ep - 1 ) ]^{-1}$ is real and positive, so that the steepest descent path makes the angle $ \vartheta = \pi/2  -\arg(\lambda)/2$  (and $ \vartheta = 3\pi/2 -\arg(\lambda)/2$) with the real axis (see \fref{fig:3}). 

The restriction of $t_{\mathrm c}$ to the region outside the loop of integration however strongly limits the validity of the resulting AE. For instance, for real $z$ this condition implies $z< 1/\ep < 1$. Here we show that it is possible to overcome this limitation, by allowing $t_{\mathrm{c}}$ to lie within the loop and calculating the resulting contribution from $t_{\mathrm{c}}$ to the integral in \eref{eq:int:ep>1}.

To begin with, we can discard the case $|z| < 1/\ep$, for then $t_\mathrm{c} = 1/z$ always lies outside the integration loop employed in the application of the MSD (shown as solid paths in \fref{fig:3}(b), which depend on the value of $\ep$) and therefore does not contribute to integral \eref{eq:int:ep>1}. If $|z| > 1/\ep$, then $t_\mathrm{c}$ may lie within the loop depending on the exact value of $z$. If it does then its contribution must be considered. 

The contribution to the AE due to the critical point $t_\mathrm{c}$, in the cases where it lies within the integration loop, depends on the nature of $t_\mathrm{c}$ which is defined by $b$. If $b = 0$, $t_{\mathrm{c}}$ is not a critical point. When $ b \in \mathbb{Z} \backslash \mathbb{N}_0 $, $ t_{ \mathrm{c} } $ is a zero of the integrand and makes no difference to the integral \eref{eq:int:ep>1}. The other possibilities are (i) $ b \in \mathbb{N} $, meaning $ t_{ \mathrm{c} } $ is a pole of order $ b $, and (ii) $ b \in \mathbb{R} \backslash \mathbb{Z} $, meaning $ t_{ \mathrm{c} } $ is a branch point giving a branch cut to infinity in a suitable direction. We will consider these last two nontrivial cases separately.

Let us first assume that $ b \in \mathbb{N} $. Because $ t_{ \mathrm{c} } = 1/z $ is outside the original integration path from the definition of the HGF \eref{eq:int:ep>1} (denoted by $ C $, dashed black path in \fref{fig:3}(b)) and inside the deformed integration path through the saddle point (made for the application of the MSD, denoted by $ C' $, solid paths in \fref{fig:3}(b)), the residue theorem gives
\begin{equation}\label{eq:reseq}
\int_{C'} f(t) \, \rme^{ \lambda g(t) } \rmd t = \int_C f(t) \, \rme^{ \lambda g(t) } \rmd t + 2 \pi \rmi \res_{ 1/z } f(t) \,  \rme^{ \lambda g(t) }.
\end{equation}
If we multiply \eref{eq:reseq} by the prefactor of the integral on the right hand side of \eref{eq:int:ep>1}, then the integral over $C$ becomes the HGF. Futhermore, as $|\lambda| \to \infty$, the integral over $C'$ attains the limiting value of $ ( 1 - \ep z )^{-b} $ (by the application of the MSD), while the residue on the right equals (\ref{sec:appB})
\begin{equation}\label{eq:res}
\res_{ 1/z } f(t) \,  \rme^{ \lambda g(t) } \sim - \frac{ \lambda^{ b - 1 } \, z^{ 1 - c - \lambda } }{ \Gamma (b) \, (1 - z)^{ a + b - c + ( \ep - 1 ) \lambda } } ( \ep z - 1 )^{ b - 1 } 
\quad \textrm{as }  \lambda\to\infty .
\end{equation}
The prefactor of the integral in \eref{eq:int:ep>1} is, in the same limit, by the Stirling approximation
\begin{equation}\label{eq:pf}
\frac{ \Gamma ( c + \lambda ) \, \Gamma ( a - c + ( \ep - 1 ) \lambda + 1 ) }{ 2 \pi \rmi \, \Gamma ( a + \ep \lambda ) } 
\sim - \rmi \sqrt{ \frac{ \lambda }{ 2 \pi } } \frac{ ( \ep - 1 )^{ a - c + ( \ep - 1 ) \lambda + \frac{ 1 }{ 2 } } }{ \ep^{ a + \ep \lambda - \frac{ 1 }{ 2 } } }
\quad \textrm{as }  \lambda\to\infty.
\end{equation}
From \eref{eq:int:ep>1} and \eref{eq:reseq}--\eref{eq:pf} we finally find the asymptotic expansion of the HGF to be 
\begin{multline}\label{eq:AE:ep>1}
F \! \left( \atp{ a + \ep \lambda, b }{ c + \lambda } ; z \right) 
\sim \frac{ 1 }{ ( 1 - \ep z )^b } + 
\frac{ \sqrt{ 2\pi } }{ \Gamma (b) } \frac{ ( \ep - 1 )^{ a - c + \frac{ 1 }{ 2 } } }{ \ep^{ a - \frac{ 1 }{ 2 } } }  \frac{ z^{ 1 - c } \, ( \ep z - 1 )^{ b - 1 } }{ ( 1 - z )^{ a + b - c } }  \lambda^{ b - \frac{ 1 }{ 2 } } \cdot  \\ \cdot \left[ \frac{ 1 }{ \ep^{\ep} z } \left( \frac{ \ep - 1 }{ 1 - z } \right)^{ \ep - 1 } \right]^{ \lambda } 
\quad \textrm{as }  |\lambda | \to\infty.
\end{multline}
The last factor on the right hand side above can be written as $ h_{ \ep }^\lambda (z) $ with $ h_{ \ep } (z) $ as defined before (see discussion following \eref{eq:recomp:ep<1}). 

\begin{figure}[b]
\centering 
   \begin{minipage}[t]{0.46\textwidth}
    \includegraphics[width=\textwidth]{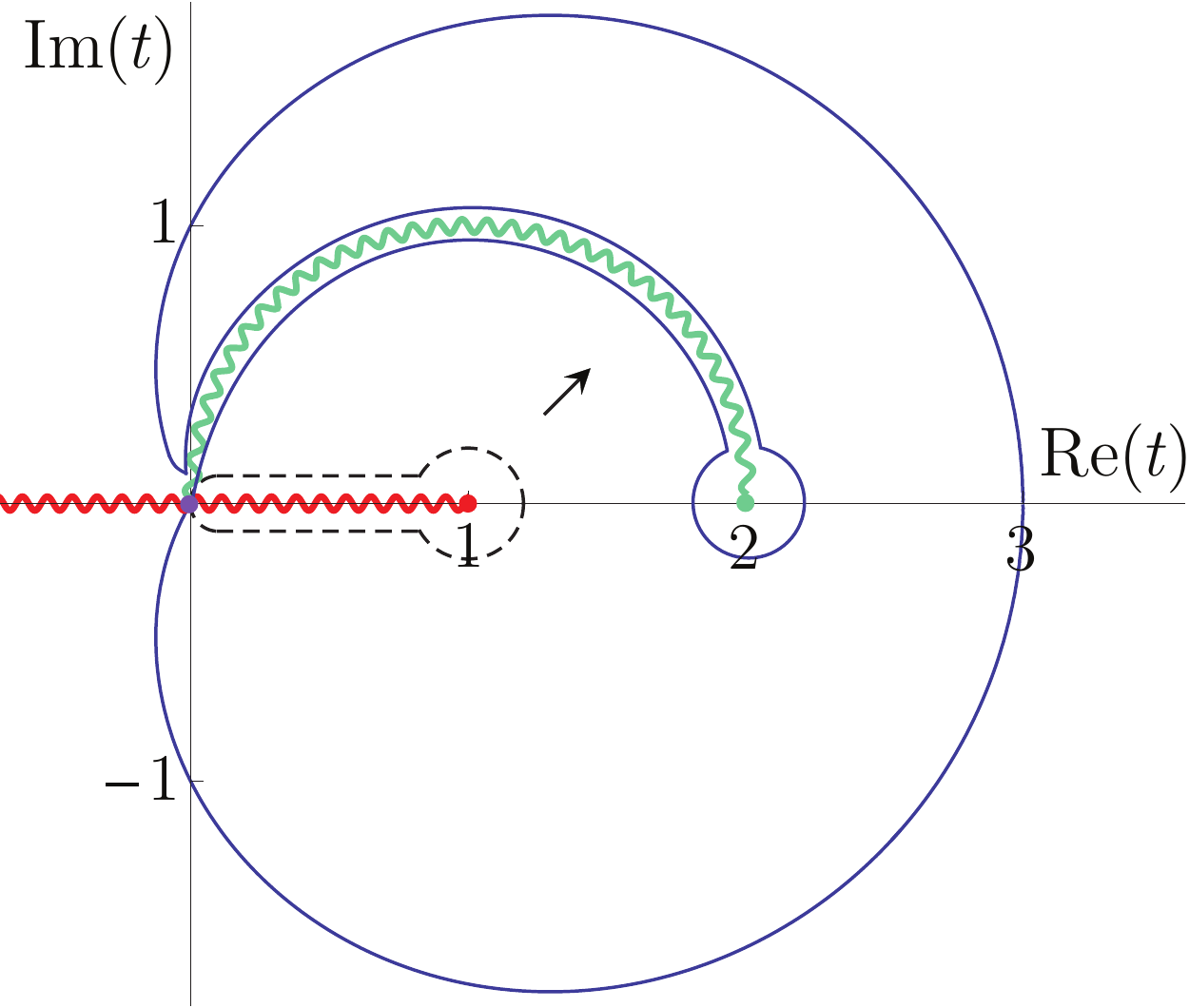}
  \end{minipage} \hspace{0.7cm}
  \begin{minipage}[b]{0.4\textwidth}
    \caption{Integration path in the $ t $--plane when $ b \notin \mathbb{Z} $, $ \ep = 3 $ and $ z = 1/2 $. Curly red: the branch cut from $ - \infty $ to 1. Curly green: the branch cut from $ t_{ \mathrm{c} } $ aligned along the steepest descent curve through $ t_{ \mathrm{c} } $.  Dashed black: the integration loop of \eref{eq:intrepB} and \eref{eq:int:ep>1} by definition. Solid blue: the steepest descent curve through $ t_0 = \ep $, deformed additionally to avoid the branch cut.  
\label{fig:4}}
  \end{minipage} \hfill
\end{figure}

The final case is when $ b $ is a real non-integer, which leads to a branch point instead of a pole at $ t_{ \mathrm{c} } $. In this case, if we were to deform the path as in \fref{fig:3}(b), then we would pass over the branch point at $t_{\mathrm{c}}$ and the branch cut emanating from $t_{\mathrm{c}}$. To  avoid this, we deform the integration path additionally in order to bypass the branch cut (see \fref{fig:4}),
by placing part of it along the branch cut and around the branch point. The branch cut itself is chosen to lie along the path of steepest descent through the point $ t_{ \mathrm{c} } $. The additional contribution to the integral from this additional deformation can then be found by applying Watson's lemma to this augmented part of the path $ C' $  \cite[pp.\ 125-147]{Miller:2006:225}. As shown in \ref{sec:appC}, this contribution proves to be exactly equal to the second term of \eref{eq:AE:ep>1}, so we get the same asymptotic expansion \eref{eq:AE:ep>1}. 

\begin{figure}[hb]
\vskip4pt
\centering
\subfloat[]{\includegraphics[width=0.49\textwidth]{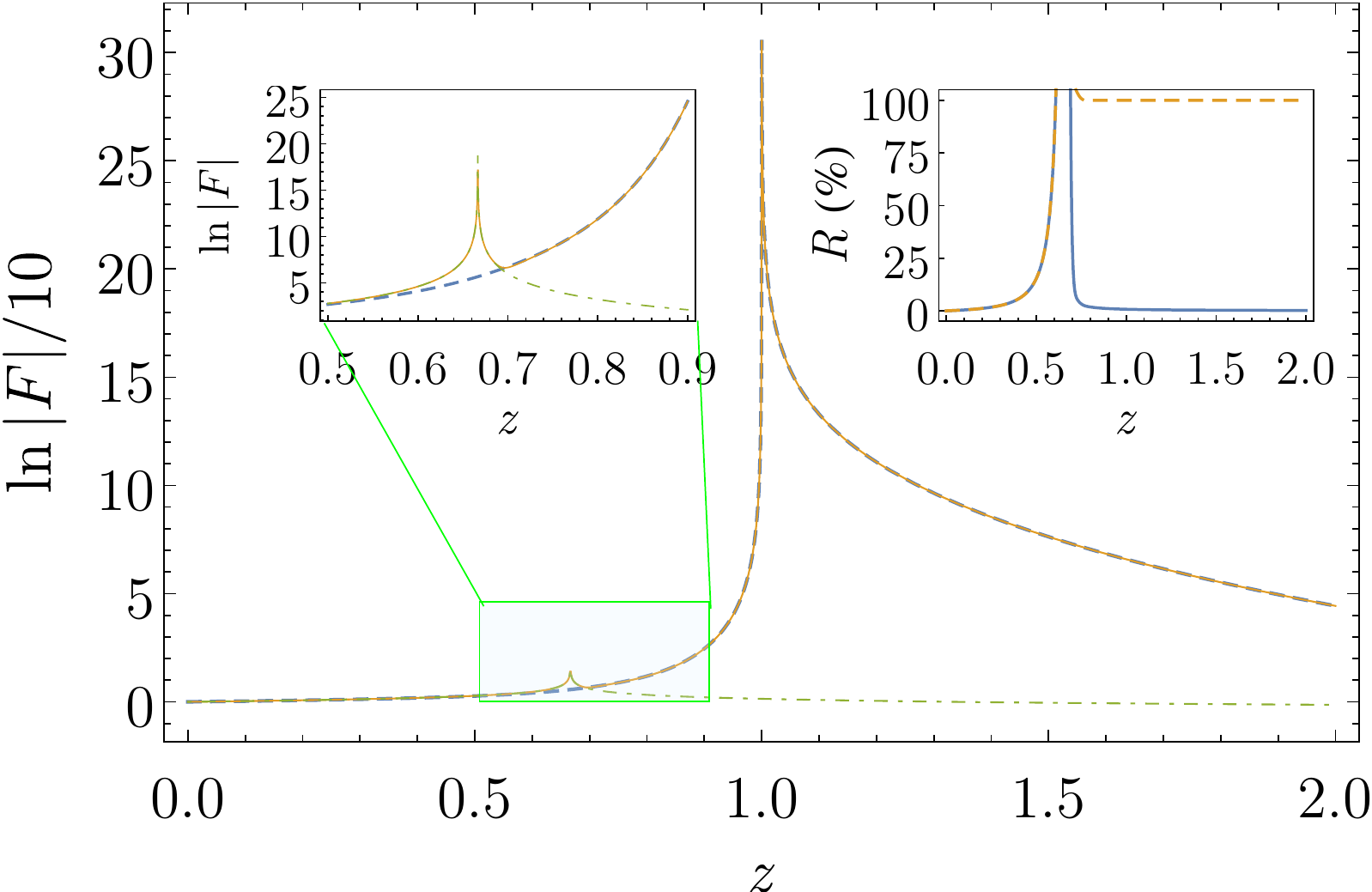}}
\hfill
\subfloat[]{\includegraphics[width=0.49\textwidth]{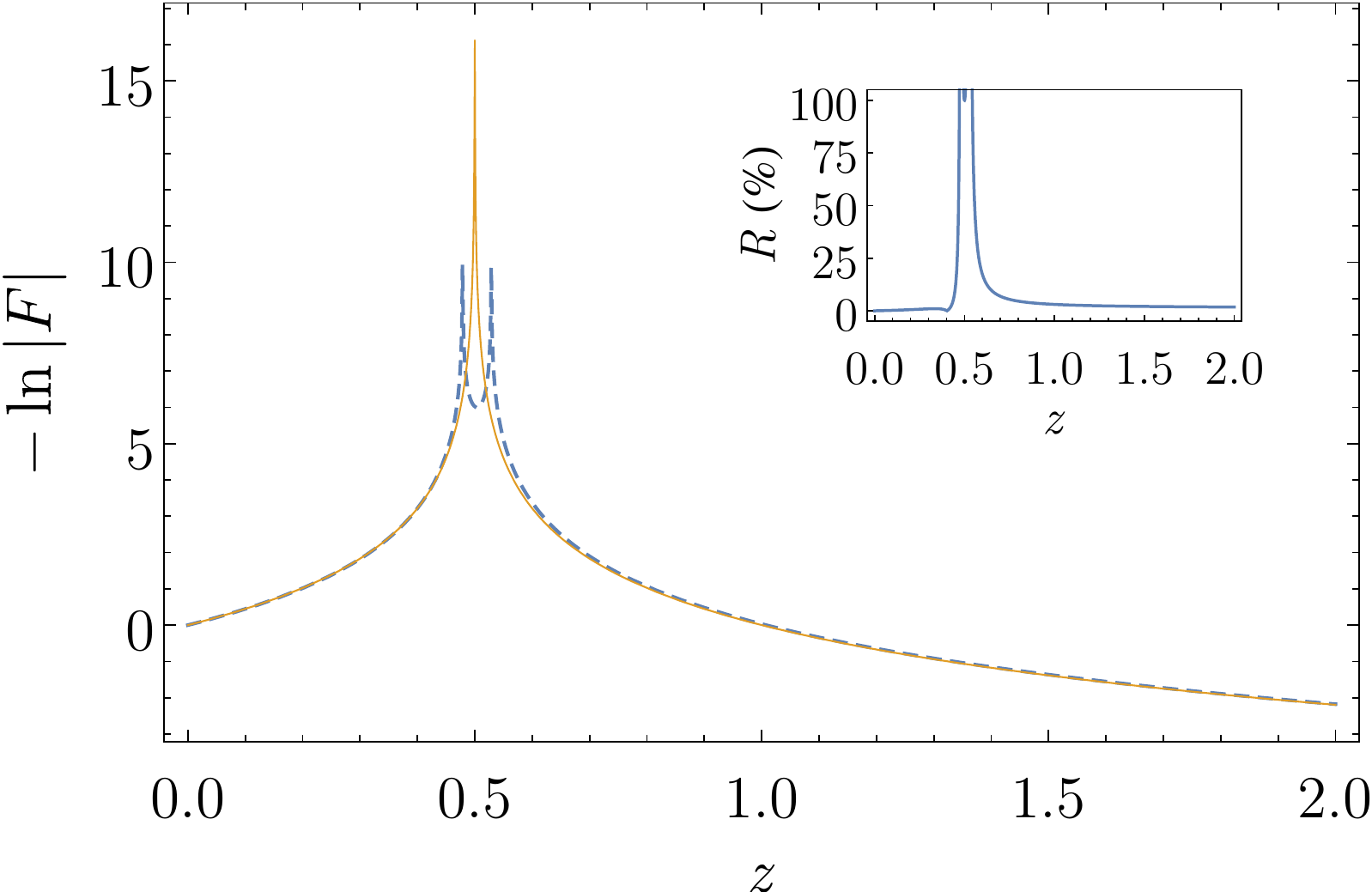}}\\
\subfloat[]{\includegraphics[width=0.49\textwidth]{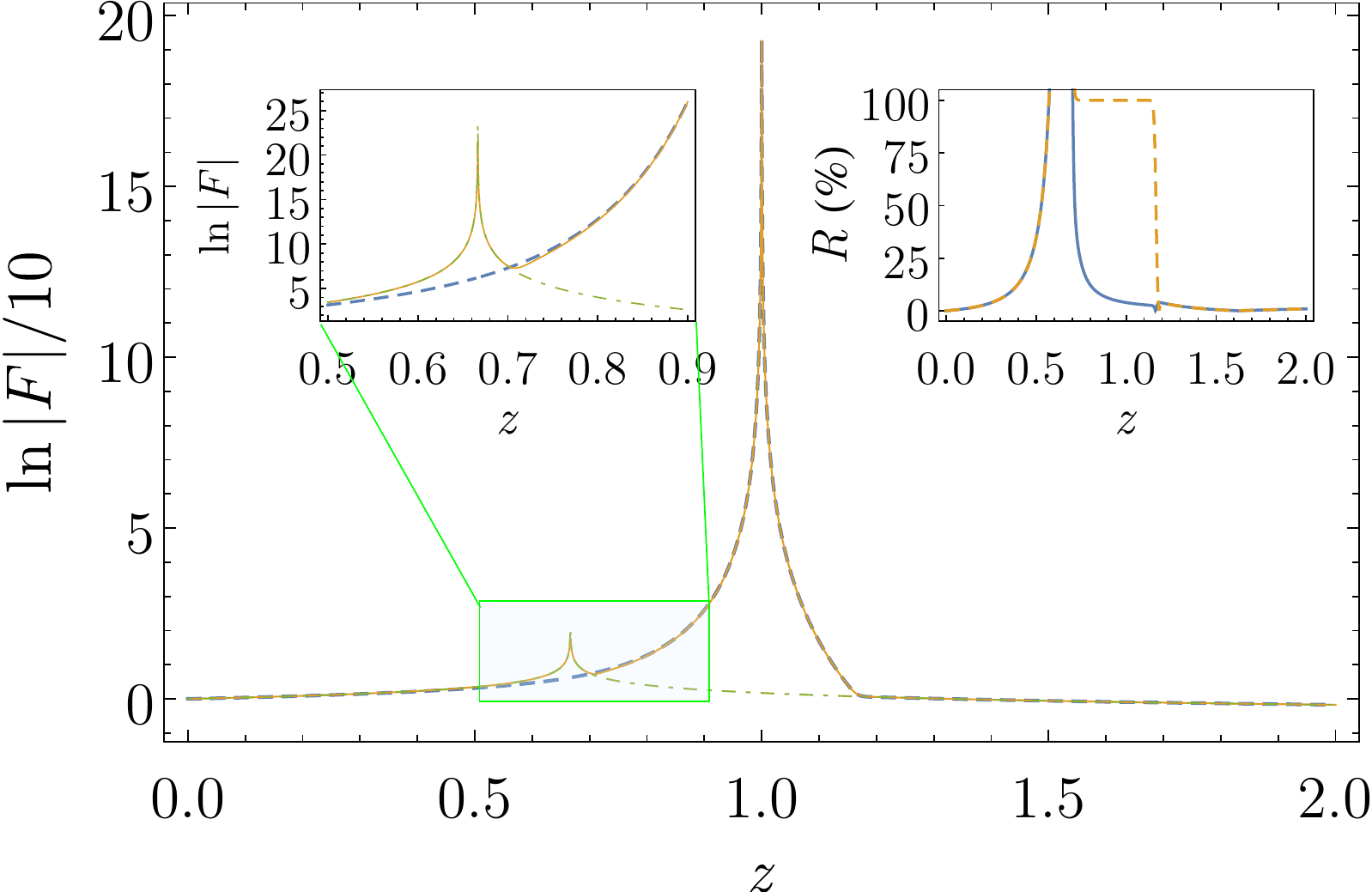}}
\hfill
\subfloat[]{\includegraphics[width=0.49\textwidth]{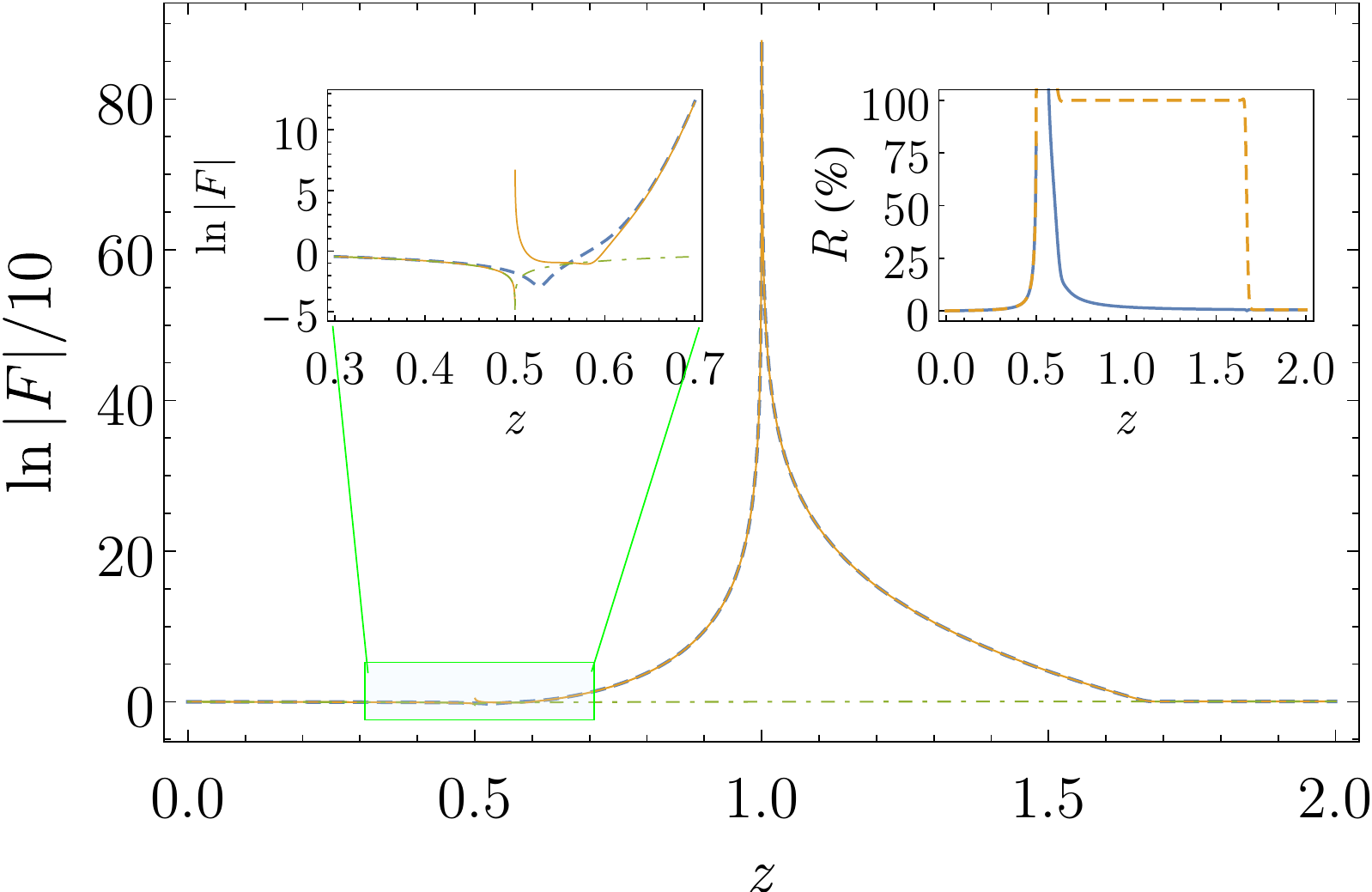}}
\caption{Graphs of the HGF for $ a = 0 $, $ c = 1 $, and different $ b $, $ \ep $ and $ \lambda $: (a) $ ( b, \ep, \lambda ) = ( 2, \frac{ 3 }{ 2 }, 50 (1+3\rmi / 2) ) $, (b) $ ( b, \ep, \lambda ) = ( -2, 2, 100 ) $, (c) $ ( b, \ep, \lambda ) = ( \frac{ 5 }{ 2 }, \frac{ 3 }{ 2 }, 50 ) $, (d) $ ( b, \ep, \lambda ) = ( - \frac{ 1 }{ 2 }, 2, 100  (1+\rmi / 2) ) $. The following description is common to all the plots. Dashed blue: the corresponding HGF. Solid orange: the AE \eref{eq:AE:ep>1} with pole/branch cut contribution included. Dot-dashed green: the AE \eref{eq:AE:ep<1}. Left inset: the vicinity of $ z = 1 / \ep $ enlarged. Right inset: the relative error of limited (eq.\ \ref{eq:AE:ep<1}, dashed orange) and full (eq.\ \ref{eq:AE:ep>1}, solid blue) expansions. All the figures look the same if $ a $ and/or $ c $ and/or $ \lambda $ are non-integers (see \fref{fig:SI1}). Note the logarithm of the absolute values of the HGFs in the plots, which in part (b) results in two spikes at the points where the HGF changes sign. \label{fig:5}}
\end{figure}

Since $ h_{ \ep } (z) > 1 $ for  $z\to 1$ (and for $z<1$ if $z$ is real; compare \ref{sec:appA}), the AE \eref{eq:AE:ep>1} diverges as $|\lambda| \to \infty$, and diverges quicker the closer $|z|$ gets to 1. This is the same asymptotic behaviour as that of the HGF itself, and the AE \eref{eq:AE:ep>1}, in fact, reproduces very well the divergent behaviour of the HGF in the domain $ 1 / \ep < | z | < 1 $. The accuracy of the AE \eref{eq:AE:ep>1} can be seen in the excellent overlap between the HGF and the AE \eref{eq:AE:ep>1} (dashed blue and solid orange curves, respectively) in \fref{fig:5}, 
as well as in  a plot of the relative error between the two curves (right insets in \fref{fig:5}), which remains small even near the point $z = 1$ where the HGF diverges, and only becomes significant in the neighbourhood of $z = 1/\ep$. Therefore, the expansion  \eref{eq:AE:ep>1} is valid when $ b \notin \mathbb{Z}\backslash\mathbb{N} $ and $ \ep > 1 $, for any $ |z| > 1/\ep$, while for $|z| < 1/\ep$ it reduces to the right hand side of \eref{eq:AE:ep<1}.

As in section \ref{ssec:ep<1}, negative values of $\ep$ can be accounted for by the use of \eref{eq:trans1} and \eref{eq:trans2}, with which one can transform the cases $ ( - \ep, 0, 1 ) $ and $ ( \pm \ep, 0, -1 ) $ to $ ( \ep, 0, 1 ) $.

\subsection{Expansions of the HGF for large \texorpdfstring{$a$}{a} and \texorpdfstring{$b$}{b}}\label{sec:ab}

We again assume $\ep > 0$, and use the transformation formulae in \eref{eq:trans1} and \eref{eq:trans2} to handle the case $\ep < 0$. The integral representation suitable for large $a$ and $b$ is  \eref{eq:intrepB}, which in this case reads
\begin{equation}\label{eq:int:ab}
F \! \left( \atp{ a + \ep \lambda, b + \lambda }{ c } ; z \right) 
= \frac{ \Gamma (c) \, \Gamma ( a - c + \ep \lambda + 1 ) }{ 2 \pi \rmi \, \Gamma ( a + \ep \lambda ) } 
\int_0^{(1+)} f(t) \, \e^{ \lambda g(t) } \mathrm{d} t,
\end{equation}
where $ f(t) $ and $ g(t) $ are defined to be
\begin{align} 
f(t) &= \frac{ t^{ a - 1 } ( t - 1 )^{ c - a - 1 } }{ ( 1 - z t )^b } , \label{eq:f:ab} \\
g(t) &= \ln \left( \frac{ t^{\ep} }{ ( t - 1 )^{\ep} ( 1 - z t ) } \right). \label{eq:g:ab}
\end{align}
One branch cut is $(-\infty,1]$. Another branch cut, if $(b+\lambda) \in\mathbb{R}\backslash \mathbb{Z}$, is from $1/z$ to $ \infty $ in a suitable direction. The condition $ g' (t) = 0 $ yields as the saddle points
\begin{equation}\label{eq:saddles}
t_{\pm} = \frac{ 1 - \ep }{ 2 } \pm \frac{ |1 - \ep | }{ 2 } \sigma (z), \quad \sigma (z) = \sqrt{ 1 + \frac{ 4 \ep }{ ( \ep - 1 )^2 z } }. 
\end{equation}

\begin{figure}
\centering
\subfloat[$z= \frac{2}{3}$]{\includegraphics[width=0.45\textwidth]{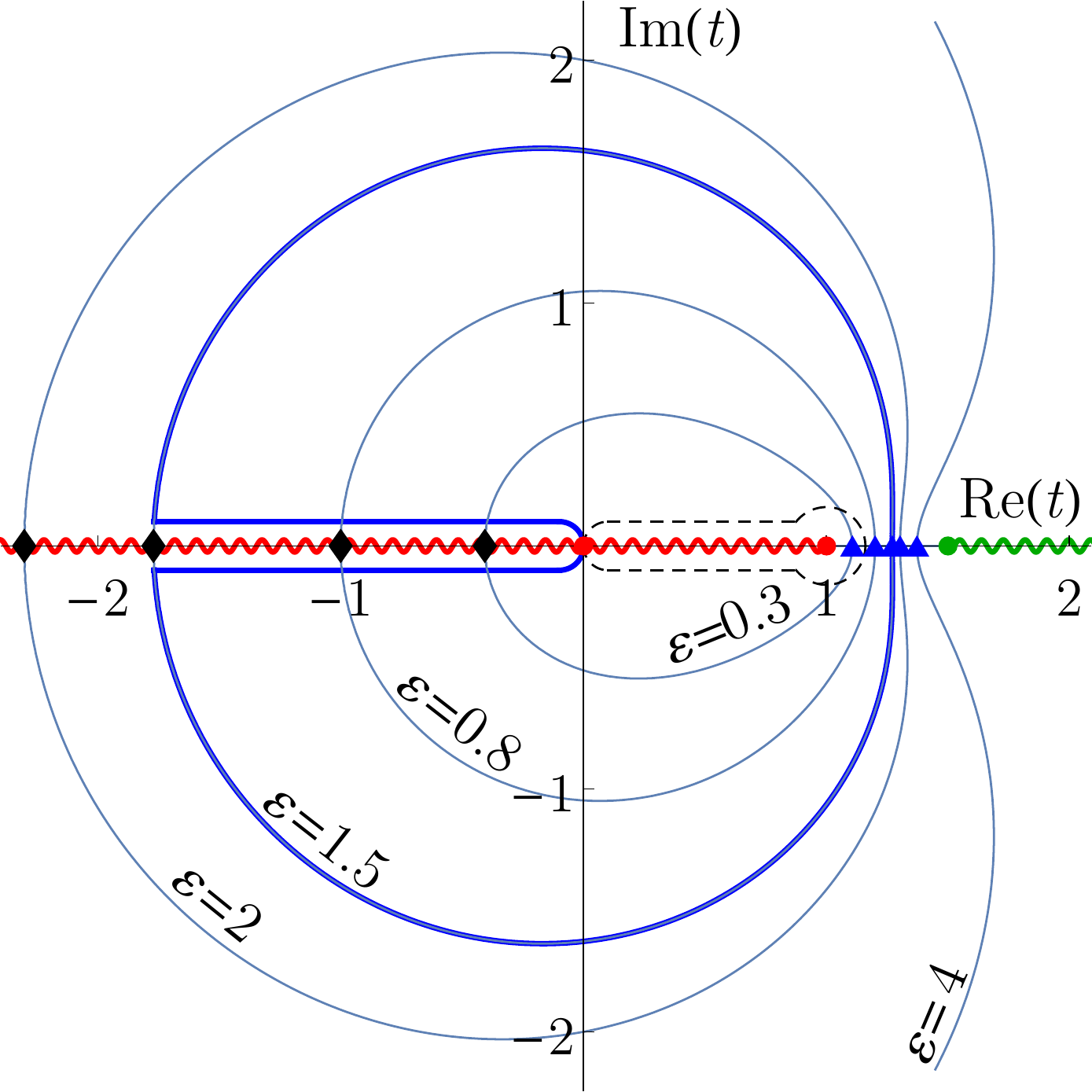}}
\hfill
\subfloat[$z=2$]{\includegraphics[width=0.45\textwidth]{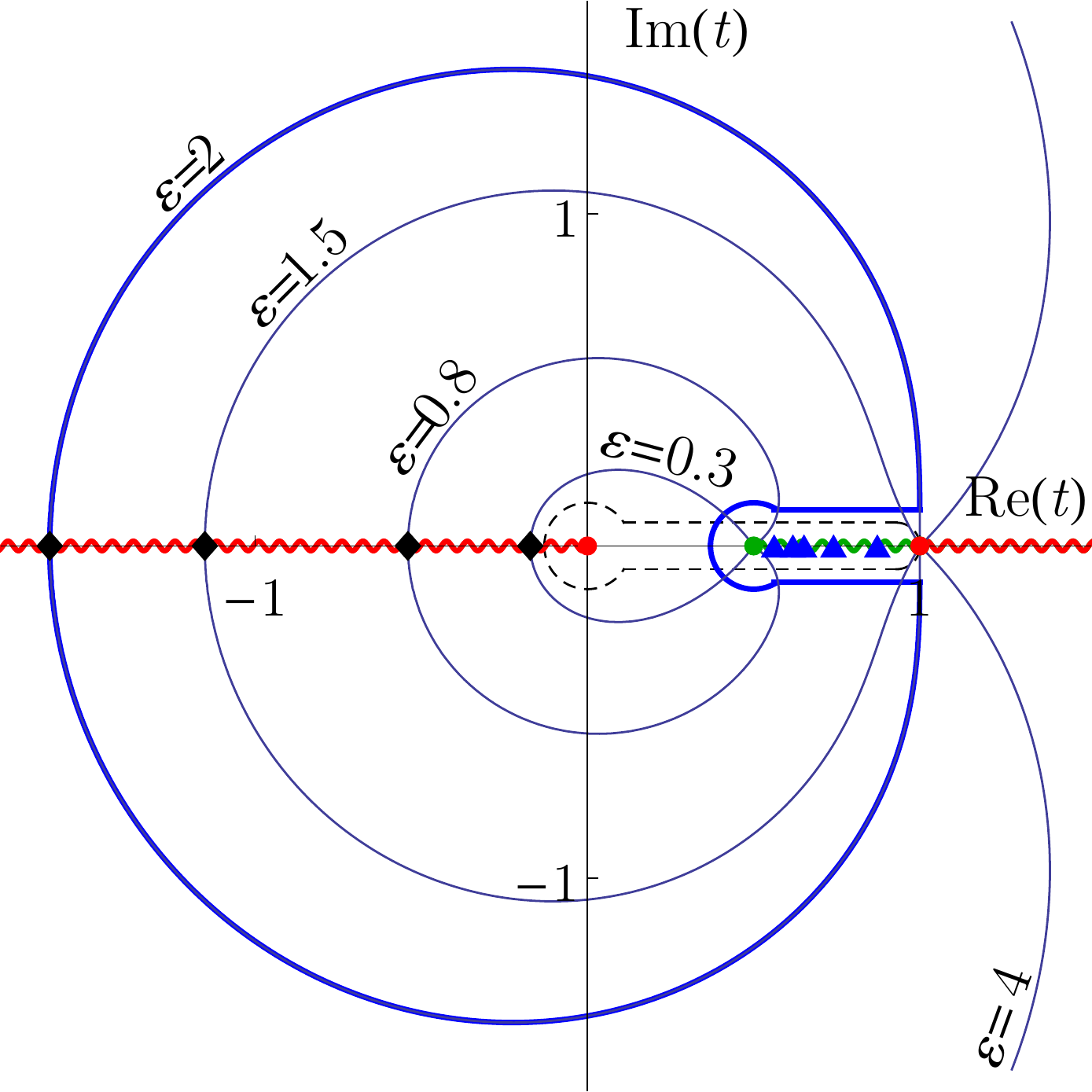}}
\caption{The steepest descent paths through the dominant saddle point for different $ \ep $ and $z$ and real $\lambda$. Thick blue:  the integration path for chosen $\ep$. Curly red: the branch cut $ ( - \infty, 1 ] $. Curly green: the branch cut $ [ 1/z, + \infty ) $. Dashed black: the original integration paths of \eref{eq:intrepB} and \eref{eq:intrepC}. Blue triangles: $ t_+ ( \ep, z ) $. Black diamonds: $ t_- ( \ep, z ) $.  \label{fig:6}}
\end{figure}

If $ \Im (z) \neq 0 $, then $z$ as well as $ t_{\pm} $ lie off the real axis, meaning that one can deform the integration path to coincide with the steepest descent path through $t_{\pm}$ without passing over the branch cut from $ 1/z $ to $ \infty $ (see figures \ref{fig:6}, \ref{fig:SI4}, \ref{fig:SI5} and \ref{fig:SI6}). In this case, therefore, the MSD can be applied directly. Using the fact that $g'(t_{\pm}) = 0$, the second derivative at $t_{\pm}$ reads
\begin{equation}\label{eq:g''}
g'' ( t_ \pm ) = \frac{ \pm \ep | 1 - \ep | \sigma(z) }{ t_\pm^2 ( t_\pm - 1 )^2 } ,
\end{equation}
and the MSD results in the following AE:
\begin{multline}\label{eq:AE:ab}
F \! \left( \atp{ a + \ep \lambda, b + \lambda }{ c } ; z \right)  \sim 
\frac{ \Gamma (c) ( \ep \lambda )^{ \frac{ 1 }{ 2 } - c } }{ \sqrt{ 2 \pi | \ep - 1 | \sigma(z) } } 
\left( \frac{ t_+^{ a + \ep \lambda } ( t_+ - 1 )^{ c - a - \ep \lambda } }{ ( 1 - z t_+)^{ b + \lambda } } + \right.  \\
+\left. \frac{ t_-^{ a + \ep \lambda } ( t_- - 1 )^{ c - a - \ep \lambda } }{ ( 1 - z t_-)^{ b + \lambda } } \right), 
\quad \textrm{as }  |\lambda | \to\infty,
\end{multline}
which is valid for any $z$ and $\lambda $ such that $\Im (z) \neq 0$ and $\Im(\lambda) \neq 0$. 
Here we have used the Stirling approximation to write the prefactor of the integral in \eref{eq:int:ab} as $ \Gamma (c) ( \ep \lambda )^{ 1 - c } / ( 2 \pi \rmi ) $.
For real $\lambda$ and $\Im (z) \neq 0$, only the saddle point at $t_+$ contributes in the MSD, resulting in the AE \eref{eq:AE:z<1}, i.e. \eref{eq:AE:ab} without the second term.

If $z$ is real, then $ t_{\pm} $ are real, with $t_-$ negative and $t_+$ positive for all $z > 0$. (The only possible case for concern here would be if $t_\pm$ coalesced, which only happens for $z = -4\ep / (\ep-1)^2 < 0$ and is ruled out since we assume in this derivation that $\Re(z) > 0$.) Since in this case all the branch cuts lie on the real axis, we have to inspect the applicability of the MSD for real $z$. The comparison of the real parts of $ g(t) $ at $ t_{\pm} $ yields
\begin{equation}\label{eq:recomp:z<1}
 \Re ( g(t_+) ) - \Re ( g(t_-) ) = \ln \! \left( \left| \frac{ t_+ ( t_- - 1 ) }{ t_- ( t_+ - 1 ) } \right|^\ep \left| \frac{ 1 - z t_- }{ 1 - z t_+ } \right| \right) > 0 
\end{equation}
so that the saddle point $t_+$ dominates the contribution to the path integral, while the contribution of $t_-$ can be neglected. 
For $ z < 1 $, $ 1 < t_+ < 1/z$ and $ t_+ $ lies between the branch cuts $ ( - \infty, 1 ] $ and $ [ 1/z, + \infty ) $ (shown as red and green curly curves, respectively, in \fref{fig:6}(a)), meaning that one can again easily deform the integration path to coincide with the steepest descent path through $ t_+ $. 

In contrast, for $ z > 1 $, since $ 1/z < t_+ < 1 $ (see \fref{fig:6}(b)), one cannot enclose the point 1 without passing over the branch cut from $ 1/z $ to $ \infty $. This inability to deform the path to go through $t_+$ when $z$ is real and larger than $1$ means that the integral representation \eref{eq:int:ab} cannot be used in this case, which must therefore be treated differently. We note parenthetically that it is not possible to avoid this difficulty by orienting the branch cut from $1/z$ in some direction other than the positive real axis, as in section \ref{ssec:ep>1}, since the branch point is now logarithmic and the procedure therein is valid for algebraic branch points. It is equally unhelpful to transform the HGF with $ z>1 $ to an HGF with $ z < 1  $ with the help of some transformation formulae (e.g.\ \cite[$\S$15.3.7 or $\S$15.3.9]{Abramowitz:1972:218}), for doing so results in all the three parameters being large, a situation beyond the scope of our work.

Therefore, we now separate the cases of real $z$ to those with $z < 1$ and $z > 1$, and use different integral representations for them.

\subsubsection{Case \texorpdfstring{$z<1$}{z<1}\label{ssec:z<1}}

As for real $z$ from \eref{eq:g''} $ \Im ( g''( t_\pm )) = 0 $, we have $ \alpha = \arg ( g'' ( t_\pm ) ) = 0$ or $ \pi $, depending on the sign of $ g'' ( t_\pm ) $, so that the angles at which the steepest descent paths pass through the saddle points are, from \eref{eq:theta},
$ \vartheta ( t_+ )  \in  \left\{ \pi/2, 3\pi/2 \right\} $ and $ \vartheta ( t_- )  \in  \{ 0,  \pi \}$, while the paths of steepest ascent are perpendicular to those of steepest descent. As the saddles are connected by each path, this means that the steepest descent curve at one saddle becomes the steepest ascent one at the other, and vice versa. Application of the MSD then gives for this case 
\begin{equation}\label{eq:AE:z<1}
F \! \left( \atp{ a + \ep \lambda, b + \lambda }{ c } ; z \right) \sim 
\frac{ \Gamma (c) ( \ep \lambda )^{ \frac{ 1 }{ 2 } - c } }{ \sqrt{ 2 \pi | \ep - 1 | \sigma } } \frac{ t_+^{ a + \ep \lambda } ( t_+ - 1 )^{ c - a - \ep \lambda } }{ ( 1 - z t_+)^{ b + \lambda } } 
\quad \textrm{as }  \lambda\to\infty  .
\end{equation}

\begin{figure}[!ht]
\centering
\subfloat[]{\includegraphics[width=0.49\textwidth]{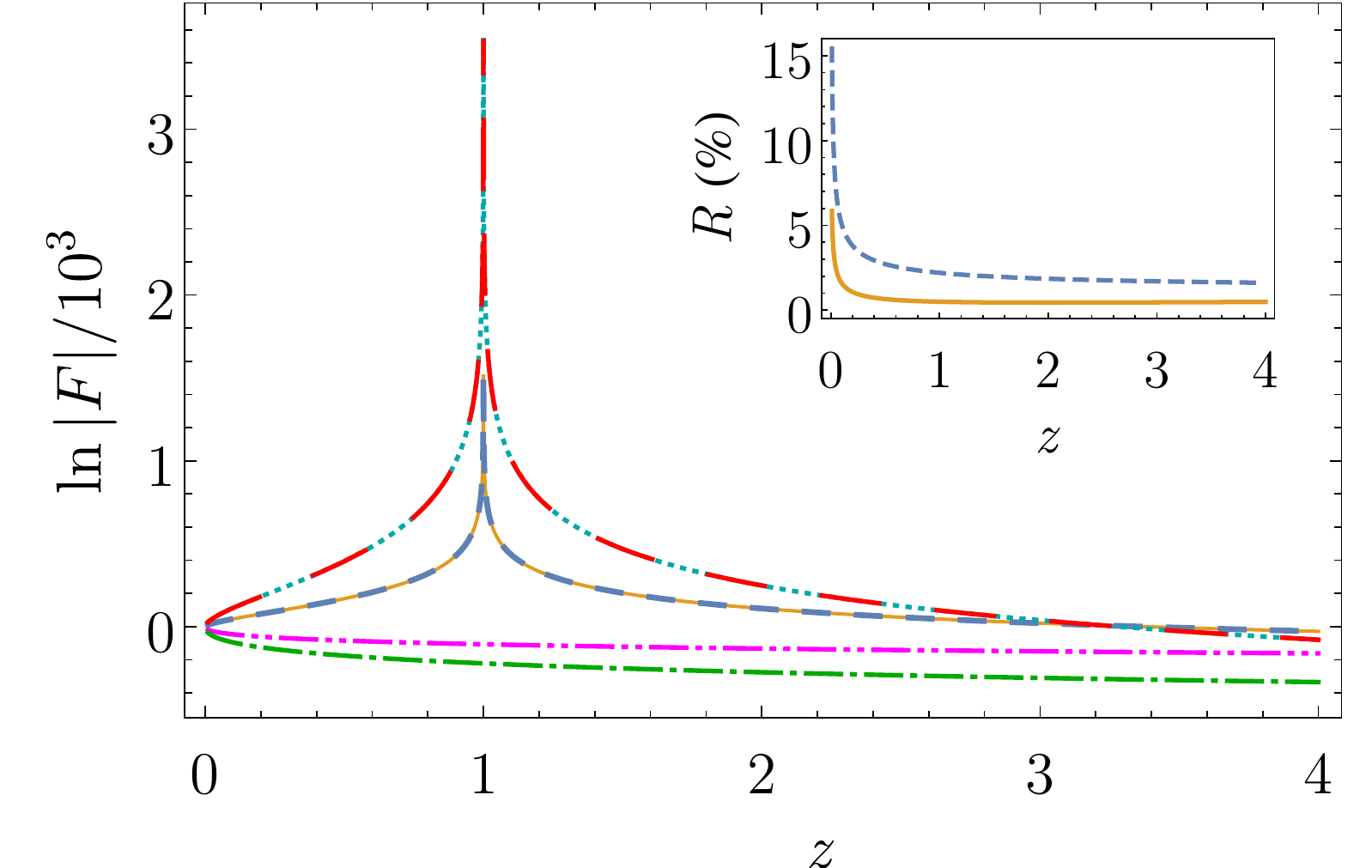}}
\hfill
\subfloat[]{\includegraphics[width=0.49\textwidth]{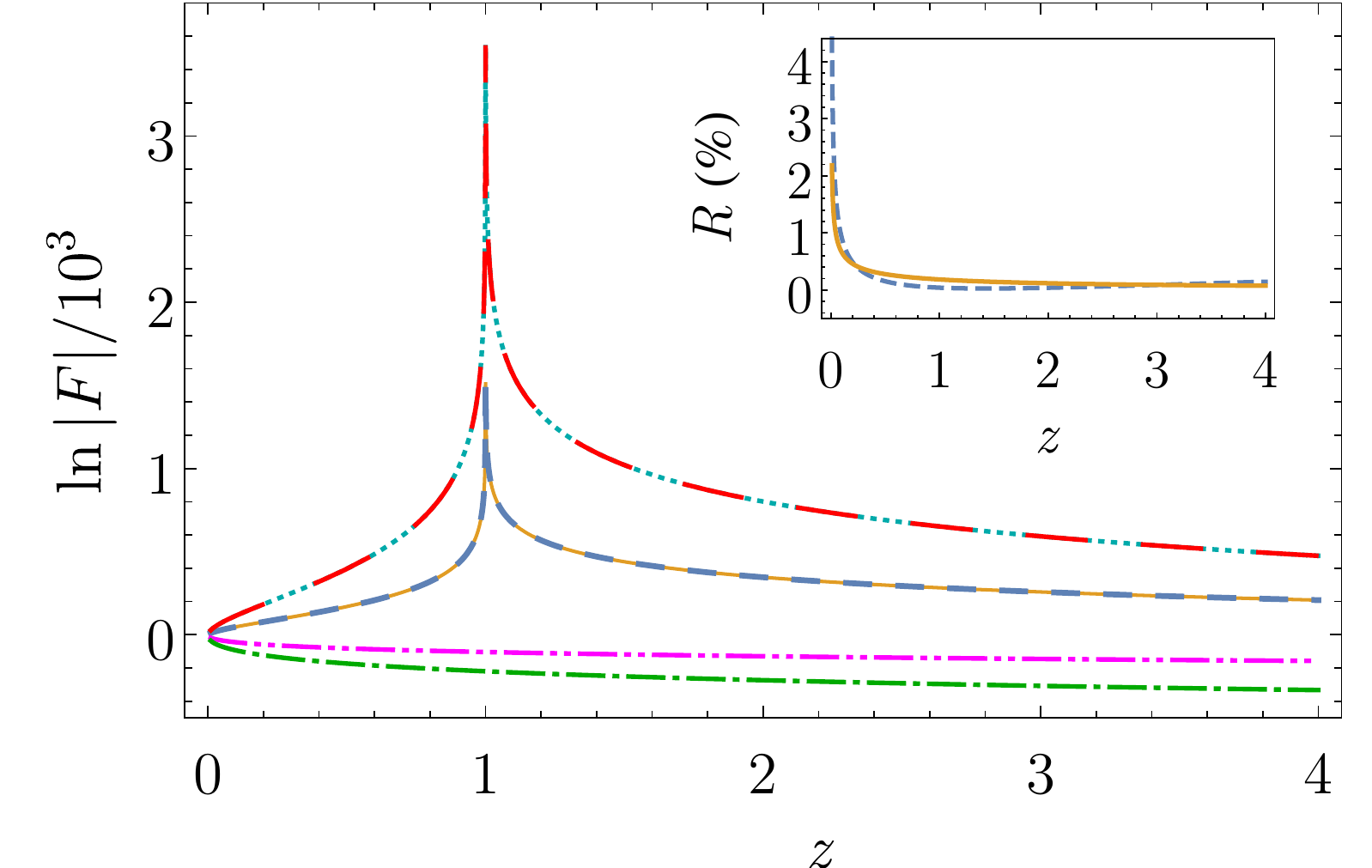}}
\caption{The HGF (dashed blue and long--dashed red) and expression \eref{eq:AE:z<1} (solid orange and dotted cyan) for $ \ep = 1/2 $ and $ \ep = 5/2 $, respectively: (a) $ ( a, b, c, \lambda ) = ( 2, 1, 3, 100 ) $, (b) $ ( a, b, c, \lambda ) = ( \frac{2}{3}, \frac{4}{3}, \frac{7}{3}, 100+50\rmi ) $. The contribution of $ t_- $ (double--dot--dashed magenta and dot--dashed green) is negligible as compared to that of $t_+$. \label{fig:7}}
\end{figure}

As shown in \fref{fig:7}, 
the agreement of \eref{eq:AE:z<1} and the full HGF is excellent, and this is borne out by the relative error curve (inset, \fref{fig:7}), which continually and rapidly decreases to 0, including at the singularity $z = 1$. Unlike for the case $ ( \ep, 0, 1 ) $ in section \ref{sec:ac}, this expansion works even at $ z = 1 / \ep $.

\subsubsection{Case \texorpdfstring{$z>1$}{z>1}}\label{ssec:z>1}

As the integral representation \eref{eq:intrepB} cannot be applied to the case of real $z >1$, and in  \eref{eq:intrepA} and \eref{eq:intrepC} the conditions on the parameters are not satisfied, the best approach is to find the AE of the HGF for $ ( - \ep, -1, 0 ) $, and then transform this to the AE for $ ( \ep, 1, 0) $ using \eref{eq:trans1}.

The integral representation suitable for this case is \eref{eq:intrepC}, in which the HGF reads
\begin{equation}\label{eq:int:z>1}
F \! \left( \atp{ a - \ep \lambda, b - \lambda }{ c } ; z \right) = 
\rme^{ ( \ep \lambda - a ) \pi \rmi } \frac{ \Gamma ( 1 - a + \ep \lambda ) \, \Gamma (c) }{ 2 \pi \rmi \, \Gamma ( c - a + \ep \lambda ) } \!
\int_1^{(0+)} \!  \! f(t) \, \rme^{ \lambda G(t) } \mathrm{d} t,
\end{equation}
with
\begin{align}\label{eq:fg:z>1}
f(t)  &= \frac{ t^{ a - 1 } ( 1 - t )^{ c - a -1 } }{ ( 1 - z t )^b } , \\
G(t) &= \ln \left[ \left( \frac{ 1 }{ t } - 1 \right)^\ep ( 1 - z t ) \right] . 
\end{align}
The branch cuts in the $ t $--plane are $ ( - \infty, 0 ] $ and $ [ 1, + \infty ) $ on the real axis, and from $ 1 / z $ to $ \infty $ in a suitable direction, which we choose to be along the positive real axis. 

Since $ G(t) = \ep \pi \rmi - g(t) $, where $ g(t) $ is as defined in \eref{eq:g:ab}, the saddle points, satisfying $ G'(t) = 0 $, are the same as in \eref{eq:saddles}. But as $ G'' (t_\pm) = - g'' (t_\pm) $, the angles of the steepest descent path at the saddle points are swapped in comparison to the previous section, so that $\vartheta (t_+) \in \{0,\pi\}$ and $\vartheta (t_-) \in \{  \pi/2, 3\pi/2 \}$. It is moreover clear that \eref{eq:recomp:z<1} implies $\Re(G(t_-)) > \Re(G(t_+))$. Therefore, this case can be viewed as a reflection of the previous case about the imaginary axis, i.e., the saddle point $t_-$ dominates and is crossed by the steepest descent path at $ \vartheta = \pi/2 $. 

Now, since $ \Re (G(t)) \to - \infty $ at $ t = t_\mathrm{c} $ and $ t = 1 $, the contribution of these two points in comparison to the saddle point $t_-$ is negligible. Therefore we can let the integration path turn about $ 1 / z $ without incurring any cost; an example of such a path is shown in \fref{fig:6}(b) (thick blue curve). The Stirling approximation gives $ \rme^{ ( \ep \lambda - a ) \pi \rmi } \Gamma (c) ( \ep \lambda )^{ 1 - c } / ( 2 \pi \rmi ) $ for the prefactor of \eref{eq:int:z>1}, and the application of the MSD then yields
\begin{equation}\label{eq:interAE}
F \! \left( \atp{ a - \ep \lambda, b - \lambda }{ c } ; z \right) \sim 
\frac{ \Gamma (c) ( \ep \lambda )^{ \frac{ 1 }{ 2 } - c } }{ \sqrt{ 2 \pi | \ep - 1 | \sigma } } 
\frac{  ( 1 - t_- )^{ c - a + \ep \lambda } }{ ( - t_- )^{  \ep \lambda-a } ( 1 - z t_- )^{ b - \lambda } } 
\quad\!\!\!\!\!  \textrm{as }  \lambda\to\infty.
\end{equation}
Using \eref{eq:trans1}, we can transform from $ ( \ep, 1, 0 ) $ to this case of $ ( - \ep, -1, 0 ) $ so that
\begin{multline} \label{eq:transapp}
 F  \left( \atp{ a + \ep \lambda, b + \lambda }{ c } ; z \right)  
= ( 1 - z )^{ c - a - b - ( \ep + 1 ) \lambda } F \! \left( \atp{ c-a - \ep \lambda, c-b - \lambda }{ c } ; z \right)   \\
 \sim \frac{ \Gamma (c) ( \ep \lambda )^{ \frac{ 1 }{ 2 } - c } }{ \sqrt{ 2 \pi | \ep - 1 | \sigma } }  
\frac{ ( - t_- )^{ c - a - \ep \lambda } ( 1 - t_- )^{ a + \ep \lambda } }{ ( 1 - z t_- )^{ c - b - \lambda } ( 1 - z )^{ a + b - c + ( \ep + 1 ) \lambda } } \quad \textrm{as }  \lambda\to\infty .
\end{multline}
This expression can be simplified with the help of the following relation
\begin{equation}\label{eq:tpm}
\frac{ ( - t_- )^{ c - a - \ep \lambda } ( 1 - t_- )^{ a + \ep \lambda } }{ ( 1 - z t_- )^{ c - b - \lambda } ( 1 - z )^{ a + b - c + ( \ep + 1 ) \lambda } }
= \frac{ t_+^{ a + \ep \lambda } ( t_+ - 1 )^{ c - a - \ep \lambda } }{ ( 1 - z t_+ )^{ b + \lambda } },
\end{equation}
using which \eref{eq:transapp} reduces to \eref{eq:AE:z<1}. We can therefore conclude that \eref{eq:AE:z<1} provides the general asymptotic expansion of the HGF, for the case $(\ep, 1, 0)$ with $\ep > 0$, $a,\ b$ and $c$ real and any $z$. The agreement between the HGF and the AE \eref{eq:AE:z<1} is depicted in \fref{fig:7}(a).

\subsubsection{Note on \texorpdfstring{$z=1$}{z=1} and \texorpdfstring{$z\to\infty$}{z=infinity}\label{ssec:notez1}}

There are some cases where the MSD procedure breaks down. These are (i) $z=1$, and (ii) $z \to \infty$ with $\ep < 1$ (and with $z \in \mathbb R$). In (i), we have $t_+ = 1$ and $t_- = -\ep $, so that  $t_+$ coalesces with the critical point $t_{\mathrm{c}} = 1$ as well as $t_{\mathrm{c}} = 1/z = 1$, at which both the HGF and its integral representation \eref{eq:int:ab} diverge. Similarly, in (ii) the critical point $t_{\mathrm{c}} = 0$ coalesces with the saddle point $t_- =0$ (and with the saddle point $t_+ = 0$ if $\ep > 1$, but that is not a problematic case since the dominant saddle point is $t_-$). In neither of the two cases do the saddle points $t_+$ and $t_-$ coalesce with each other. We exclude both cases (i) and (ii) from the present study, but note that even in these problematic cases the AEs \eref{eq:AE:ab} and \eref{eq:AE:z<1} provide good approximations to the HGF. This is because in the limit $z \to 1$, both the HGF and the AE \eref{eq:AE:z<1} diverge as $\sim 0^{-\lambda}$, while in the limit $z\to\infty$ the limit of the HGF as well as of the AE is $0$.

A rigorous treatment of these cases would require the use of methods proposed by Chester, Friedman, and Ursell \cite{Chester:1957:288} for the coalescence of two saddle points, and by Bleistein \cite{Bleistein:1996:286} for the coalescence of a saddle point and a singularity. In both the methods a change of variables is introduced that simplifies the phase function $g(t)$ in the integral $\int f(t) e ^{\lambda g(t)}\mathrm{d} t$ such that the evaluation is not influenced by the proximity of a saddle point and a singularity. The integrals so attained result in special functions such as Airy and Bessel functions. (For a review, see \cite{Temme:1995:288,Temme:2013:285,FaridKhwaja:2013:286}.) Bleistein's method was used by Olde Daalhuis to derive the general uniform AEs of integrals with $N$ coalescing saddle points \cite{Daalhuis:2000:287,FaridKhwaja:2016:287}. In several recent works this approach was applied to find the uniform AEs of $F ( \atp{ a, b -\lambda }{ c+\lambda } ; -z )  $ \cite{Daalhuis:2003:220}, $F ( \atp{ a+\lambda, b + 2\lambda }{ c } ; -z )  $ \cite{Daalhuis:2003:221}, $F ( \atp{ a\pm\lambda, b \pm \lambda }{ c \pm\lambda} ; -z )  $ \cite{Daalhuis:2010:222, FaridKhwaja:2014:223}, $F ( \atp{ a ,  b}{ b+\lambda } ; -z ) $ \cite{FaridKhwaja:2013:286} and $F ( \atp{ a+\lambda, a-\lambda }{ c } ; \frac{1-z}{2} ) $ \cite{FaridKhwaja:2014:223,FaridKhwaja:2016:287}. All these studies had a saddle point coalescing with either a singularity or another saddle point in the integral representation of the HGFs.

However, in the cases treated in \cite{Daalhuis:2003:220,Daalhuis:2003:221,Daalhuis:2010:222,FaridKhwaja:2013:286,FaridKhwaja:2014:223,FaridKhwaja:2016:287}, $\ep$ was assumed to take some of the values from $\{0,\pm 1, 2\}$. Since these values of $\ep$ are excluded from the present study, treatment of the problematic cases (i) and (ii) listed above would need a generalisation of the expressions from \cite{Daalhuis:2003:220,Daalhuis:2003:221,Daalhuis:2010:222,FaridKhwaja:2013:286,FaridKhwaja:2014:223,FaridKhwaja:2016:287} for $\ep \neq \pm 1$ by the use of Bleistein's method. This is expected to result in expansions which reduce continuously to \eref{eq:AE:z<1} for $z\gtrless 1$ (except for $z \to \infty$), since the AE \eref{eq:AE:z<1} is valid for all values of $z \in \mathbb R$ except the problematic points of $z = 1$ and $z \to \infty$. Such a study is left for future work.

\section{Calculation of the partition function of a 2D lattice gas in a field of traps} 

We may now return to the problem of finding the canonical partition function of a 2D gas on a lattice with traps, from \eref{eq:partfun} and \eref{eq:partitionfunction}, and evaluate its various physically-relevant limits on the basis of the results discussed in the previous sections. The parameters $p$ and $t$ in \eref{eq:partfun}, which refer respectively to the number of particles and of traps on the lattice, can vary from 1 to $N$, where $N$ is the number of nodes on the lattice and is taken to be large (up to $10^{23}$). This means that any of the three parameters of the HGF in \eref{eq:partitionfunction} can be large. 

The first and simplest case is when both $p$ and $t$ are small, and only the third parameter of the HGF in \eref{eq:partitionfunction} is large. From \eref{eq:largec} the partition function in this case trivially reads
\begin{equation}\label{eq:example0}
\frac{Z}{{N-t \choose p}} = F \! \left( \atp{-p,-t}{N-p-t+1} ; z \right) 
\sim 1+\frac{pt}{N-p-t+1} z \approx 1+\frac{pt}{N}z ,
\end{equation}
where $z = 1-\delta_{1,P_{\mathrm{on}}} + P_\mathrm{on}/P_\mathrm{off} $. 

Two other cases are when $p$ is small while $t$ is large, and vice versa. These two cases correspond to the dominant effect in the lattice gas being the trapping of the particles and collisions amongst the particles, respectively. For small $p$ (of the order of 10, say) and large $t$ ($ = N/3$, say), the canonical partition function $Z$ becomes
\begin{multline}\label{eq:example1}
\frac{Z}{{N-t \choose p}} = F \! \left( \atp{-p,-t}{N-p-t+1} ; z \right) 
=  (1-z)^p\; F \! \left( \atp{N-p+1,-p}{N-p-t+1} ; \frac{z}{z-1} \right) \\
\sim  \left( 1+\frac{t z}{N-p-t} \right)^p ,
\end{multline}
where we have first applied the transformation \eref{eq:trans1} on the HGF from sec.\ \ref{ssec:ep>1} with $(a,b,c,\ep,\lambda) = (1,1,1,\frac{N-p}{N-p-t},N-p-t)$, and then used the corresponding AE to find the closed asymptotic form. The function $Z$ for large $p$ and small $t$ is \eref{eq:example1} with $p$ and $t$ swapped, by the symmetry of the first two parameters of the HGF. 

The fourth and final case is when both $p$ and $t$ are large, i.e.\ when both trapping and collisions of the particles play a significant role in the physics of the gas. Here in general the HGF has three large parameters, a situation which is out of the scope of this work, but if we restrict the parameters to $p+t = N$ (or to $N - p - t \simeq 1$), then only the first two parameters of the HGF in \eref{eq:partitionfunction} are large, a case treated in sec.\ \ref{sec:ab}. For instance, for $p=N/3$ and $t=2N/3$, by transformation \eref{eq:trans1} the partition function $Z$ becomes
\begin{equation} \label{eq:example2a}
\frac{Z}{{N-t \choose p}} = F \! \left( \atp{-p,-t}{1} ; z \right) 
=  (1-z)^{p+t+1}\; F \! \left( \atp{p+1,t+1}{1} ; z \right).
\end{equation}
The function on the right is the HGF discussed in sec.\ \ref{sec:ab} for real $z$, with $(a,b,c,\ep,\lambda) = (1,1,1,p/t,t)$. The application of \eref{eq:AE:z<1} by a straightforward manipulation therefore yields
\begin{equation}\label{eq:example2b}
\frac{Z}{{N-t \choose p}} 
\sim \sqrt{ \frac{(1+\sigma) t_+}{4\pi p \sigma} } \left(  \frac{(1-z) t_+}{t_+-1} \right)^p \left( \frac{1-z}{1-z t_+} \right)^{t+1}
\end{equation}
with 
\begin{equation}
\sigma = \sqrt{ 1+ \frac{4pt}{(t-p)^2 z} } \qquad \mathrm{and}\qquad 
t_+ = \frac{t-p}{2t} (1+\sigma).
\end{equation}

The partition function \eref{eq:partitionfunction} as well as its asymptotic expansions \eref{eq:example0}, \eref{eq:example1} and \eref{eq:example2b} are clearly not valid if $p+t > N$ since the HGF is then not defined. In this case, however, we can turn the problem to a physically complementary one, namely the lattice gas of $N-p$ holes in a field of $N-t$ free sites which in this view we imagine as traps for the holes. Since only the binding and unbinding probabilities and the duration of the random walk step rescale, just by a change of the parameters and a proper scaling of the variable the developed partition function and its AEs can be used in this case as well.


\section{Discussion and conclusion}\label{sec:conclusion} 

We have here described the partition function of a gas on a 2D lattice interspersed with traps, which evaluates to the Gauss hypergeometric function (HGF) with one or more large parameters for physically realistic system sizes. The calculation of the partition function is facilitated greatly by asymptotic expansions (AEs) of the HGF in the appropriate large parameter limits, but these expansions are available in the literature only for limited parameter and variable values. 
The main problem in this respect is the presence of poles or branch cuts in the integral representations of the HGF, avoidance of which shrinks the sectors of validity of the AEs. Here we have used the MSD to calculate the AEs of the HGF $F \! \big( \atp{a,b}{c};z \big)$ when any two of the parameters $a$, $b$ and $c$ are large, which are valid for the entire $z$-plane except for a few points, as well as for a much wider range of the parameters than in the existing expressions. We have overcome the problem of poles and branch points by estimating their contribution to the path integral relative to that of the saddle point(s), and when they do contribute to the integral, we have calculated these contributions exactly. 

For large $a$ and $c$, we have shown that if $ | \ep | < 1  $, the pole or branch cut contribution is negligible, while for $ | \ep | > 1$ the contribution is identical for every case in which it exists, and evaluates to the expression \eref{eq:AE:ep>1}. The AE \eref{eq:AE:ep>1} works well in the limit $\ep \to 1$ and exactly at $\ep = 1$, when the limit from both sides amounts to $(1-z)^{-b}$. For large and complex $a$ and $b$ and complex $z$, the MSD yields the AE \eref{eq:AE:ab} for any $\ep >0$, including $\ep = 1$. If either $a$ and $b$ (that is to say $\lambda$) or $z$ is real, the AE reduces to \eref{eq:AE:z<1}.  

\begin{figure}[!ht]
\centering
\includegraphics[width=0.6\textwidth]{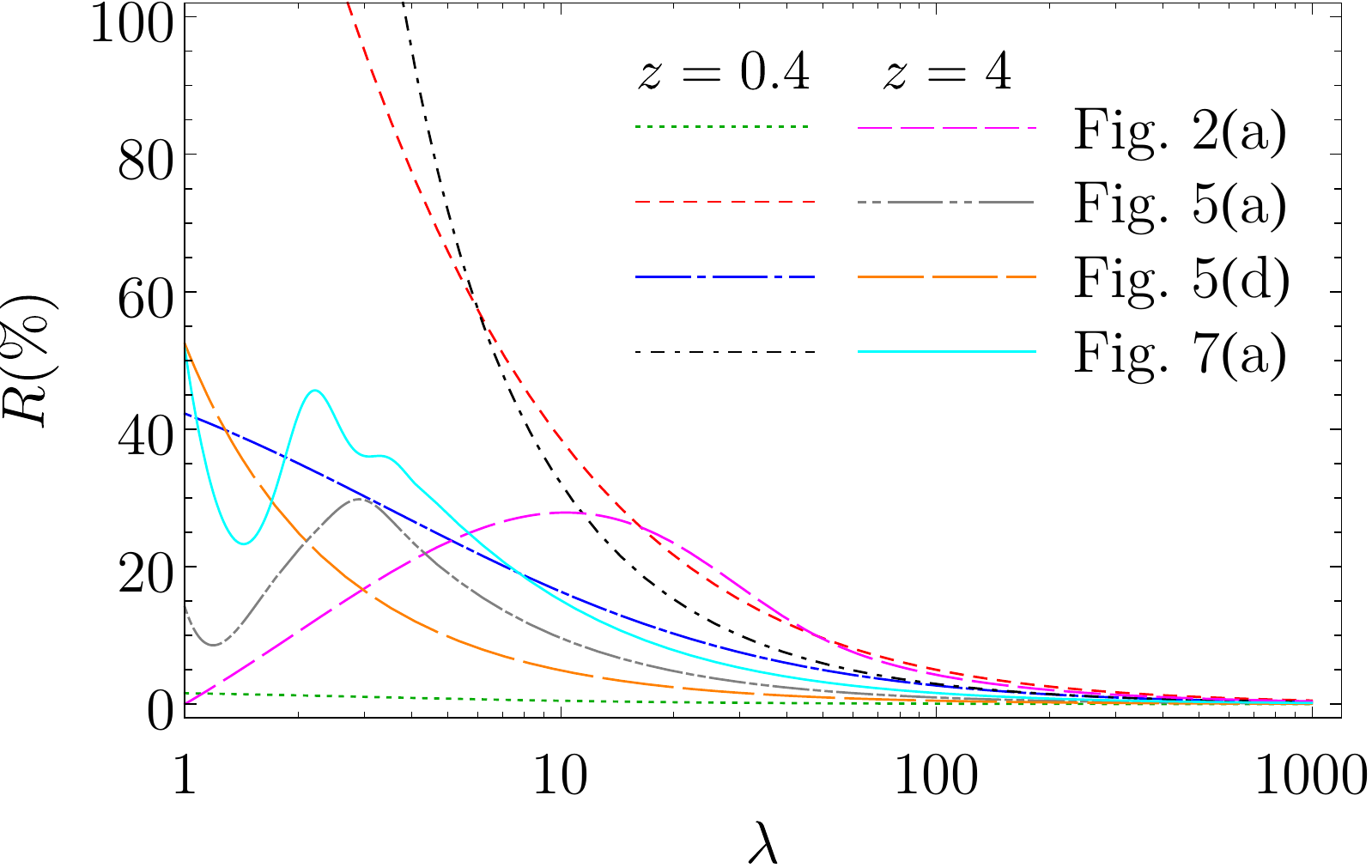}
\caption{The $ \lambda $--dependence of the relative error $R$ for the cases plotted throughout the article for two different $ z $ values. \label{fig:8}}
\end{figure}

The resulting AEs in all the various cases considered have been compared with the corresponding exact HGFs for different values of the parameters and the variable. All the AEs show excellent agreement with the corresponding HGFs. Moreover, the AEs work better as the expansion parameter $ \lambda $ gets larger, as can be seen in \fref{fig:8}, which shows the $\lambda$-dependence of the relative error between the AE and the HGF for the different cases considered in the paper. 

With our study of the AEs of the HGF in all the possible cases of two large parameters, we have succeeded in calculating the partition function of the lattice gas in a field of traps for both low and high concentrations of particles and of traps. Our expressions for the AEs prove equally useful in many other physics problems. To name a recent example, which comes from a different branch of physics but where the mathematical expressions involved turn out to be similar, the conditional probability $p(m|n)$ of emitting $m$ quanta of radiation when $n$ quanta of frequency $\omega$ and azimuthal quantum number $\tilde{m}$ are incident on a Kerr black hole with Hawking temperature $T_\mathrm{H}$, rotational angular frequency $\Omega$ and absorptivity $\Gamma$ evaluates to \cite{Bekenstein:2015:241}
\begin{equation}\label{eq:Bekenstein}
p(m|n) = \frac{(\rme^x-1) \, \rme^{nx} \, \Gamma^{m+n}}{(\rme^x -1+\Gamma)^{m+n+1}} \: F \! \left( \atp{-m,-n}{1} ; (\rme^{\beta}-1)(\rme^{\mu}-1) \right), 
\end{equation}
where $\rme^{-\beta}$ and $\rme^{-\mu}$ are the elementary probabilities to jump down and up, respectively, between two adjacent states, and $x = \hbar (\omega - \tilde{m}\Omega)/T_{\mathrm{H}}$ is the characteristic parameter of transition. Macroscopic black holes correspond to the limit of large $m$ and $n$. Since the HGF in \eref{eq:Bekenstein} is precisely the same as the one in \eref{eq:example2a}, the form \eref{eq:example2b} can be directly applied to get the algebraic asymptotic form of the main result of \cite{Bekenstein:2015:241} for a macroscopic black hole. 

Our expressions also provide an alternate way to find the AEs for some HGFs with non-integer parameters that have previously appeared in physics. Examples of such HGFs are 
$F \big( \atp{ n/2+s_+, n/2+s_-  }{n+1}; x \big) $, with $ s_{\pm} =  \big(b \pm \sqrt{1+(a^2 -1)n^2} \big)/[2(a -1)] $, for real $a$, $b$, $x$ and $n\to \infty$ in \cite{Seifert:1947:302}; 
$ F \big( \atp{  (n+\mathrm{i}s) /2, (n-\mathrm{i}s )/2 }{n+1}; {\frac{c}{n^{2/3}}-\frac{b^2}{a^2}} \big) $, with $s = \sqrt{a^2 n^2-1} / b$  for real $a$, $b$, $c$ and $n\to \infty$ in \cite{Denzler:1997:301}; 
$  F \big( \atp{ n-n, n+1  }{m+1}; \frac{1-z}{2} \big) $ and $  F \big( \atp{(n-m)/2+1, (n-m+1)/2  }{n+3/2}; \frac{1}{z^2} \big) $ for $z$ complex, $n,m\to \infty$ and $m/(n+1/2)$ constant in \cite{Thorne:1597:303};  
$ F \big( \atp{ - a n, -b n  }{1-a n}; -x \big) $ for real $a$, $b$, $x$ and $n\to \infty$ in \cite{Witte:1995304}; and  
$ F \big( \atp{a,b}{c+1}; -\frac{z}{4ab} \big) $ for complex $z$ and $|a|, |b|\to \infty$, in \cite{Thorsley:2001:244,Nagel:2001:247}. 
All of these can be mapped to the cases evaluated here by substitutions for the parameters and/or the variable. 
The various examples discussed here testify to the importance in contemporary physics research of asymptotic forms of the large-parameter hypergeometric function, which we have endeavoured to provide comprehensively in this work.

\section*{Acknowledgements}

A.-S.\ S.\ and M.\ C.\ thank ERC Starting Grant MembranesAct 337283 and DFG RTG 1962 `Dynamic Interactions at Biological Membranes --- from Single Molecules to Tissue' for support. A.-S.\ S.\ and J.\ P.\ thank the Cluster of Excellence: Engineering of Advanced Materials for funding.


\makeatletter
\newcounter{myappcount}
\setcounter{myappcount}{0}
\renewcommand\appendix{\par
\renewcommand \theequation {\@Alph\c@myappcount{.}\@arabic\c@equation}
\stepcounter{myappcount}
  \gdef\thesection{Appendix \@Alph\c@myappcount}}
\makeatother

\appendix

\section{Evaluation of the partition function} \label{sec:appE}
Here we prove that \eref{eq:partfun} evaluates to \eref{eq:partitionfunction}. Let us define $\zeta = 1+ P_{\mathrm{on}}/P_{\mathrm{off}}$ and $m = \min \{p,t \}$. Since $E_{\mathrm{b}} = -k_{\mathrm{B}} T \ln (P_{\mathrm{on}}/P_{\mathrm{off}})$, the last sum in \eref{eq:partfun} becomes
\begin{equation}\label{eq:sum1}
\sum_{k=0}^n {n \choose k} \exp \left( - \frac{k E_{\mathrm{b}}}{k_{\mathrm{B}} T} \right) 
= \sum_{k=0}^n {n \choose k} \left( \frac{P_{\mathrm{on}}}{P_{\mathrm{off}}} \right)^k = \left(1+ \frac{P_{\mathrm{on}}}{P_{\mathrm{off}}}  \right)^n = \zeta^n.
\end{equation}
The partition function is now 
\begin{equation}
Z = \sum_{ n  = 0}^{m} { N - t  \choose { p - n }}
{ t  \choose  n } \zeta^{ n } 
= \sum_{ n  = 0}^{m}
\frac{( N - t )!}{( p - n )!\, (N - t - p + n )!}
\frac{ t !}{ n !\, ( t - n )!} \, \zeta^{ n } ,
\end{equation}
which can be written as
\begin{equation}\label{eq:Zder1}
Z = \frac{(N - t )!}{ p ! (N - t - p )!} \sum_{ n  = 0}^{m}
\frac{(-1)^{ n }  p !}{( p - n )!} 
\frac{(-1)^{ n }  t !}{( t - n )!} 
\frac{(1+N - t - p -1)!}{(1+N - t - p + n -1)!} 
 \cdot \frac{ \zeta^{ n }}{ n ! } .
\end{equation}
We introduce the Pochhammer symbol $ (x)_n = \Gamma (x + n) / \Gamma (x) $. Using its property 
\begin{equation}
(-x)_n = \frac{(-1)^n x!}{(x-n)!},
\end{equation}
we have from \eref{eq:Zder1}
\begin{equation}\label{eq:Z_penultimate_step}
Z  = {N - t  \choose  p } \sum_{ n  = 0}^{m} \frac{(- p )_{ n } (- t )_{ n }} {(N - t - p +1)_{ n }} \frac{ \zeta^{ n }}{ n !}.
\end{equation}
The sum in \eref{eq:Z_penultimate_step} (and in \eref{eq:Zder1}) can be extended to infinity, since the summand in \eref{eq:Zder1} becomes 0 for $n > m$ because then either $( t - n )!$ or $( p - n )!$ is infinite. Comparison with the definition of the hypergeometric function in \eref{eq:hgfdef} then yields
\begin{equation} \label{eq:Zderived}
Z = { N - t  \choose  p } \cdot  F \! \left( \atp{- p ,- t}{N - t - p+1} ; \zeta \right). 
\end{equation}
If $P_{\mathrm{on}} = 1$, then all the particles that enter the traps bind to the traps, i.e.\ there is no sum \eref{eq:sum1} in \eref{eq:partfun}, but only $\exp \left[-n E_{\mathrm{b}}/(k_{\mathrm{B}} T) \right] = (\zeta-1)^n$ instead. The possibility of this case is covered by replacing $\zeta$ by $\zeta -\delta_{1,P_{\mathrm{on}}}$ in \eref{eq:Zderived}. This is precisely \eref{eq:partitionfunction}.

\stepcounter{myappcount}
\section{Analysis of \texorpdfstring{$\Im(g(t_{\mathrm{c}}))$}{Im(g(t_c))} in \texorpdfstring{\eref{eq:rediff1}}{(2.17)}} \label{sec:appD}

Let  $g(t)$ be the function defined in \eref{eq:fg:ep<1}. The critical point is $t_\mathrm{c} = 1/z$, and the parameter $\ep$ satisfies $\ep \in \mathbb{R}$, $0< \ep < 1$. We write $z$ as $z = r \, \rme^{\rmi \vartheta}$ with $r  \in  \mathbb{R}$ and $\vartheta \in (-\pi/2,\pi/2)$ and define $f_{\ep}(r,\vartheta) =  \Im (g(t_{\mathrm{c}})) $, i.e.\ 
\begin{equation}\label{eq:A0:fdef}
f_{\ep}(r,\vartheta) 
=  \Im \left[\ln \left( \frac{(z-1)^{1-\ep}}{z}\right)  \right]
=  (1-\ep) \arctan\left( \frac{r \sin\vartheta}{r\cos\vartheta-1} \right) - \vartheta .
\end{equation}
Here we show that for $r > 1/\ep$, $f_\ep(r,\vartheta) > 0$ if $\vartheta \in (-\pi/2, 0)$, and $f_\ep(r, \vartheta) < 0$ if $\vartheta \in (0, \pi/2)$. These two conditions on $\vartheta$ correspond to $t_\mathrm{c}$ that lies on the \textit{deformed} integration path of \eref{eq:int:ep<1}  for $\arg(\lambda) >0$ and $\arg(\lambda) <0$, respectively. 

The derivative of $f_\ep(r,\vartheta)$ with respect to $r$
\begin{equation}
\frac{\rmd }{\rmd r} f_{\ep} (r,\vartheta) = \frac{-(1 - \ep) \sin\vartheta}{(r - 1)^2 + 2 r (1 - \cos\vartheta)},
\end{equation}
shows immediately that 
\begin{equation}\label{eq:ddrg}
\frac{\rmd }{\rmd r} f_{\ep} (r,\vartheta) > 0 \textrm{ for } \vartheta \in (-\pi/2, 0) \textrm{ and } \frac{\rmd }{\rmd r} f_{\ep} (r,\vartheta) < 0 \textrm{ for } \vartheta \in (0, \pi/2).
\end{equation}
If we look at the function on the boundary, i.e.\ at $r = 1/\ep$, we see that
\begin{equation}
\frac{\rmd }{\rmd \vartheta} f_{\ep}(1/\ep,\vartheta) = \frac{- (\ep + \ep^2) (1 - \cos\vartheta)}{(1 - \ep)^2 + 2 \ep (1 - \cos\vartheta)} < 0 \textrm{ for } \vartheta \in (-\pi/2, \pi/2).
\end{equation}
Since $f_{\ep}(1/\ep, 0) = 0$, this immediately shows that 
\begin{equation}
f_{\ep}(1/\ep,\vartheta) > 0 \textrm{ if } \vartheta \in (-\pi/2, 0) \textrm{, and } f_{\ep}(1/\ep, \vartheta) < 0 \textrm{ if } \vartheta \in (0, \pi/2).
\end{equation}
Combined with \eref{eq:ddrg}, this completes the proof.

\stepcounter{myappcount}
\section{Analysis of \texorpdfstring{$h_{\ep}(x)$}{h(ε,x)} in \texorpdfstring{\eref{eq:recomp:ep<1}}{(2.18)} } \label{sec:appA}
In this section, we explore the behaviour of $h_{\ep}(x)$, defined as 
\begin{equation}\label{eq:hep1}
h_{\ep}(x) = \frac{ 1 }{ \ep^{\ep} |x| } \left| \frac{ x - 1 }{ 1 - \ep } \right|^{ 1 - \ep },
\end{equation}
for different $\ep$ and $x$, in particular the ranges when it is larger and smaller than $1$. One solution to the equation $h_{\ep}(x) = 1$ is $x_{\mathrm{c}} = 1/\ep$, and it is clearly also a solution to
\begin{equation}\label{eq:hep2}
h_{\ep}'(x) = h_{\ep}(x) \frac{1-\ep x}{x(x-1)} = 0.
\end{equation}
As $h_{\ep}'' (x_{\mathrm{c}})$ evaluates to $\ep^3/(\ep-1)$, $x_{\mathrm{c}} = 1/\ep$, where $h_{\ep}(x_{\mathrm{c}}) = 1$, is a maximum of $h_{\ep}(x)$ for $\ep < 1$ and a minimum for $\ep >1$. For $\ep <1$, we see from \eref{eq:hep2} the following: 
(i) for $ 0 < x < 1 $, $h_{\ep}'(x) < 0 $ and $\infty > h_{\ep} > 0$;  
(ii) for $ 1 <x < 1/\ep\  = x_{\mathrm{c}} $, $h_{\ep}'(x) > 0 $ and $ 0 < h_{\ep}(x) < 1 $; and 
(iii) for $ 1/\ep < x < \infty $, $h_{\ep}'(x) < 0 $ and $ 1 > h_{\ep}(x) > 0 $. 
For $\ep >1$, we have analogously: 
(i) for $ 0 < x < 1/\ep $, $h_{\ep}'(x) < 0 $ and $ \infty > h_{\ep}(x) > 1$; 
(ii) for $ 1/\ep <x < 1 $, $h_{\ep}'(x) > 0 $ and $ 1 < h_{\ep}(x) < \infty $; and 
(iii) for $ 1 < x <  \infty $, $h_{\ep}'(x) < 0 $ and $ \infty > h_{\ep}(x) > 0$, with $h_{\ep}(x) $ passing through another root of the equation $h_{\ep}(x) = 1$. 

In conclusion, 
\setlength{\arraycolsep}{2pt}
\begin{equation}\label{eq:hep3}
\Biggl\{ \begin{array}{ l@{\quad}cl } 
h_{\ep}(x) \leq 1                                         & \forall \: \ep < 1 \textrm{ and } x > 1; \\
h_{\ep}(x) \geq 1                                         & \forall \: \ep > 1 \textrm{ and } x < 1. \\
\end{array}
\end{equation}
The above analysis is graphically represented in \fref{fig:SI3}.

 \stepcounter{myappcount}
\section{Proof of \texorpdfstring{\eref{eq:res}}{(2.22)}} \label{sec:appB}

We want to calculate the residue of the function
\begin{equation}
h(t) = f(t) \rme^{ \lambda g(t) } = \frac{ t^{ a - 1 + \ep \lambda } ( t - 1 )^{ c - a - ( \ep - 1 ) \lambda -1 } }{ ( 1 - z t )^b }.
\end{equation}
We will abbreviate $ ( \ep - 1 ) \lambda$ as $\bar \lambda $. As the pole at $ 1 / z $ is of order $ b $, the residue formula reads
\begin{eqnarray}
\res_{ 1/z } h(t) 
&=& \frac{ 1  }{ ( b - 1 ) ! } \lim_{ t \to 1/z } \frac{ \rmd^{ b - 1 } }{ \rmd t^{ b - 1 } } \left[ \left( t - \frac{ 1 }{ z } \right)^b h(t) \right] \\
&= & \frac{ ( -z )^{-b} }{ \Gamma (b) } \lim_{ t \to 1/z } \frac{ \rmd^{ b - 1 } }{ \rmd t^{ b - 1 } } \left[ t^{ a - 1 + \ep \lambda } ( t - 1 )^{ c - a - \bar \lambda - 1 } \right].
\end{eqnarray}
From straightforward iterations of the derivative we have
\begin{equation}\label{eq:aA:dn}
\frac{ \rmd^n }{ \rmd x^n } \left( x^\alpha ( x - 1 )^\beta \right) 
= \sum_{k=0}^n \binom{n}{k}  \prod_{i=1}^{k} ( \alpha - i + 1 ) \prod_{i=1}^{ n - k } ( \beta - i + 1 ) \: x^{ \alpha - k } ( x - 1 )^{ \beta + k - n }.
\end{equation}
Using the identities
\begin{equation}\label{eq:aA:pochprop}
\prod_{i=1}^n ( x - i + 1 ) = \frac{ \Gamma ( x + 1 ) }{ \Gamma ( x - n + 1 ) } = (-1)^n ( -x )_n,
\end{equation}
along with \eref{eq:aA:dn}, we get
\begin{eqnarray}
 \res_{ 1/z } h(t)
  &=&  \frac{ (-z)^{-b} }{ \Gamma (b) } \lim_{ t \to 1/z }  \sum_{k=0}^{ b - 1 }  \binom{ b - 1 }{ k } t^{ a - 1 + \ep \lambda - k } ( t - 1 )^{ c - a - b - \bar \lambda + k } \cdot \nonumber \\
  & & \hspace{6cm} \cdot (-1)^{ b - 1 } ( 1 - a - \ep \lambda )_{k} ( a - c + \bar \lambda + 1 )_{ b - k - 1 } \nonumber \\
  &=& \frac{ - ( 1 - z )^{ c - a - b - \bar \lambda } }{ \Gamma (b) \: z^{ c + \lambda - 1 } } \sum_{k=0}^{ b - 1 } \binom{ b - 1 }{ k }   ( 1 - a - \ep \lambda )_{k} ( a - c + \bar \lambda + 1 )_{ b - k - 1 } ( 1 - z )^k \nonumber \\
  &=& \frac{ (-1)^{b} }{ \Gamma (b) } \frac{  ( c - a - b - \bar \lambda + 1 )_{ b - 1 }  }{ z^{  c + \lambda -1 } ( 1 - z )^{ a + b - c + \bar \lambda } } 
  \sum_{k=0}^{ b - 1 } \frac{ (-1)^k ( b - 1 ) ! }{ ( b - 1 - k ) ! } \frac{ ( 1 - a - \ep \lambda )_k ( 1 - z )^k }{ ( c - a - b - \bar \lambda + 1 )_k   k ! } .
\label{eq:aA:ressum}
\end{eqnarray}
There we have used another property of the Pochhammer symbol, namely
\begin{equation}
( 1 - x )_{ m + n } = \frac{ \Gamma (x) }{ \Gamma ( x - m ) } \frac{ (-1)^{ m + n } }{ ( x - m )_{ - n } } = (-1)^{ m + n } \frac{ ( x - m )_m }{ ( x - m )_{ - n } } 
\end{equation}
to evaluate the factor $ ( a - c + \bar \lambda + 1 )_{ b - k - 1 } $. The first factor under the sum in \eref{eq:aA:ressum} allows us to replace $b - 1$ in the upper limit of the sum by $ \infty $; from \eref{eq:aA:pochprop}, that factor can be recognised as $ ( 1 - b )_k $. The sum in \eref{eq:aA:ressum} then is another HGF:
\begin{equation}\label{eq:aA:sumHGF}
\sum_{k=0}^{\infty} \frac{ ( 1 - a - \ep \lambda )_k ( 1 - b )_k }{ ( c - a - b - \bar \lambda + 1 )_k } \frac{ ( 1 - z )^k }{ k ! } 
= F \! \left( \atp{ 1 - a - \ep \lambda, 1 - b }{  c - a - b - \bar \lambda + 1 } ; 1 - z \right).
\end{equation}
Using the following transformation formula \cite[\S 15.8.6]{Olver:2010:219},
\begin{equation}
F \! \left( \atp{ - m , b }{ c } ; z \right) = \frac{ (b)_m }{ (c)_m } (-z)^m F \! \left( \atp{ - m, 1 - c - m }{ 1 - b - m } ; \frac{ 1 }{ z } \right),
\end{equation}
we get
\begin{multline}\label{eq:aA:trans}
 F \! \left( \atp{ 1 - a - \ep \lambda, 1 - b }{  c - a - b - \bar \lambda + 1 } ; 1 - z \right) 
= \frac{ (-1)^{ b - 1 } ( 1 - a -   \ep \lambda )_{ b - 1 } }{ ( c - a - b - \bar \lambda + 1 )_{ b - 1 } }  ( 1 - z )^{ b - 1 } \cdot \\
 \cdot  F \! \left( \atp{ a - c + \bar \lambda + 1, 1 - b }{ a - b + \ep \lambda + 1 } ; \frac{ 1 }{ 1 - z } \right). 
\end{multline}
We introduce $\lambda' = \ep \lambda$ and $\ep' = 1 - 1 / \ep $, so that $\bar \lambda = \ep' \lambda'$. We also introduce $ a' = a - c + 1$, $ b' = 1 - b$, $ c' = a - b + 1$ and $ z' = 1/(1 - z)$. With these changes, the HGF from \eref{eq:aA:trans} becomes  $ F \big( \atp{ a' + \ep' \lambda', b' }{ c' + \lambda' } ; z' \big) $, which is just the parameter set that is discussed in sec.\ \ref{sec:ac}. Here, since $ 0 < \ep'  = 1 - 1 / \ep < 1 $ (as $ \ep > 1 $), and $ 1 / z' $ is neither a pole nor a branch point of the HGF (as $ b' \in \mathbb{Z} \backslash \mathbb{N} $), we can use the result of section \ref{ssec:ep<1} (for $\ep < 1$) to write the AE of $ F \big( \atp{ a' + \ep' \lambda', b' }{ c' + \lambda' } ; z' \big) $ as $ ( 1 - \ep' z' )^{ - b' } $. In terms of the original variables, this gives
\begin{equation}\label{eq:aA:interAE}
F \! \left( \atp{ a - c + \bar \lambda + 1, 1 - b }{ a - b + \ep \lambda + 1 } ; \frac{ 1 }{ 1 - z } \right) 
\approx \left[ 1 - \left( 1 - \frac{ 1 }{ \ep } \right) \frac{ 1 }{ 1 - z } \right]^{ b - 1 } = 
\left( \frac{ \frac{ 1 }{ \ep } - z }{ 1 - z } \right)^{ b - 1 }.
\end{equation}
The factor $ ( 1 - a - \ep \lambda )_{ b - 1 } $ on the right side in \eref{eq:aA:trans} can be simplified using the Stirling approximation. In general, if $ \arg (w) < \pi $ and $ a < 0 $, this approximation gives
\begin{equation}
( b - a w )_n 
       =  \frac{ (-1)^n \Gamma ( a w - b + 1 ) }{ \Gamma ( a w - b - n + 1 ) } 
        \sim \frac{ (-1)^n \sqrt{ 2 \pi } \rme^{ - a w } ( a w )^{ a w - b + \frac{ 1 }{ 2 } } }{ \sqrt{ 2 \pi } \rme^{ - a w } ( a w )^{ a w - b - n + \frac{ 1 }{ 2 } } } 
        = (- a w )^n 
\end{equation}
as  $w\to \infty$. This means $ ( 1 - a - \ep \lambda )_{ b - 1 } \sim (-1)^{ b - 1 } ( \ep \lambda )^{ b - 1 } $ as $ \lambda \to \infty $. Combining this with the results from \eref{eq:aA:ressum} to \eref{eq:aA:interAE}, we obtain
\begin{equation}
\res_{ 1/z } h(t) \sim  - \frac{ \lambda^{ b - 1 } z^{ 1 - c - \lambda } }{ \Gamma (b) ( 1 - z )^{ a + b - c + \bar \lambda } } ( \ep z - 1 )^{ b - 1 }  
 \quad \textrm{as }  \lambda \to\infty,
\end{equation} 
which is what we needed to prove.

\stepcounter{myappcount}
\section{The branch cut contribution in section \texorpdfstring{\ref{ssec:ep>1}}{2.2.2}} \label{sec:appC}

Assume that the function $ f(t) $ has an algebraic branch point of the type $ f(t) = ( t - t_{ \mathrm{b} } )^{\nu} \bar{f} (t)$, where $ \nu\in \mathbb{C} \backslash \mathbb{Z} $ and $ \bar f (t) $ is an analytic function of $ t $ near $ t = t_{ \mathrm{b} } $ for which $ \bar f ( t_{ \mathrm{b} } ) \neq 0 $. Also assume that $ g(t) $ is an analytic function of $ t $ near $ t = t_{ \mathrm{b} } $ for which $ g' ( t_{ \mathrm{b} } ) \neq 0 $. Then the AE of the integral $ F_{ \mathrm{b} } ( \lambda ) = \int_{ C'' } f(t) \rme^{ \lambda g(t) } \: \rmd t$ along $ C'' $, which is the path along the two sides of the branch cut emanating from $ t_{ \mathrm{b} } $, reads \cite[p.\ 131]{Miller:2006:225}
\begin{equation}\label{eq:aB:Fbcut}
F_{ \mathrm{b} }( \lambda ) \sim
s ( 1 - \rme^{ - 2 \pi \rmi \nu } ) \rme^{ \lambda g ( t_{ \mathrm{b} } ) }  
\sum_{n=0}^{\infty}  \frac{ \Gamma ( n + \nu + 1 ) \beta^{ (n) } (0)}{ n ! \lambda^{ n + \nu + 1 } }
\quad \textrm{as }  \lambda\to\infty.
\end{equation}
Here the change of variables $ g(t) - g( t_{ \mathrm{b} } ) = - \tau $ was introduced and 
\begin{equation}
\beta( \tau ) = \bar f ( t ( \tau ) ) \left[ \frac{ t (\tau) - t_{ \mathrm{b} } }{ \tau } \right]^{\nu} t' (\tau)
\end{equation}
was defined; also $s$ was introduced such that $ s = - 1 $ if $ C'' $ lies to the right, and $ s = 1 $ if $ C'' $ lies to the left of the original integration path, when situated on the path and looking in the direction of integration. Comparing to \eref{eq:f:ep>1} and \eref{eq:g:ep>1}, we see that $ s = - 1 $ (see \fref{fig:4}), $ \nu = - b $, and 
\begin{eqnarray}
\bar f (t)                                &=& (-z)^{-b} t^{ a - 1 } ( t - 1 )^{ c - a - 1 },                                                                                                                                                                                                                          \\
g'(t)                                        &=& \frac{ \ep }{ t } - \frac{ \ep - 1 }{ t - 1 },                                                                                                                                                                                                                         \\
\beta (0)                               &=& \bar f ( t_{ \mathrm{b} } ) \left[ \frac{ -1 }{ g'( t_{ \mathrm{b} } ) } \right]^{ \nu + 1 } = (-1)^{-b}\: \frac{ z^{ 1 - c } ( \ep z - 1 )^{ b - 1 } }{ ( 1 - z )^{ a + b - c } }. 
\end{eqnarray}
Here we have recognised that $ t (0) = t_\mathrm{b} =1/z$ and $ t' (0) = - 1 / g' ( t_{ \mathrm{b} } ) $. Combining the above expressions with \eref{eq:aB:Fbcut}, the first-order contribution to $ F_{ \mathrm{b} } ( \lambda ) $ reads
\begin{equation}
F_{ \mathrm{b} }( \lambda ) \sim 
(-1)^{-b}\: \frac{ \rme^{ 2 \pi \rmi b } - 1 }{ z^{ \lambda } ( 1 - z )^{ ( \ep - 1 ) \lambda } } \frac{ \Gamma ( 1 - b ) }{ \lambda^{ 1 - b } } \frac{ z^{ 1 - c } ( \ep z - 1 )^{ b - 1 } }{ ( 1 - z )^{ a + b - c } } 
\quad \textrm{as }  \lambda\to\infty .
\end{equation}
As $ (-1)^{-b} = \rme^{ - \rmi \pi b } $, i.e. 
\begin{equation}
(-1)^{-b} \left( \rme^{ 2 \pi \rmi b } - 1 \right) = 2 \rmi \sin ( \pi b ) = \frac{ 2 \pi \rmi }{ \Gamma (b) \Gamma ( 1 - b ) }
\end{equation}
by the reflection formula of the $ \Gamma $--function \cite[p.\ 256]{Abramowitz:1972:218}, we have 
\begin{equation}\label{eq:aB:Fblim}
F_{ \mathrm{b} } ( \lambda ) \sim 
\frac{ 2 \pi \rmi }{ \Gamma (b) \lambda^{ 1 - b } } \frac{ z^{ 1 - c - \lambda } ( \ep z - 1 )^{ b - 1 } }{ ( 1 - z )^{ a + b - c + ( \ep - 1 ) \lambda } } 
\quad \textrm{as }  \lambda\to\infty.
\end{equation}
Finally, we multiply \eref{eq:aB:Fblim} by \eref{eq:pf} to find the contribution of the branch cut to the HGF expansion:
\begin{equation} 
F_{ \mathrm{b} } ( \lambda ) \cdot ( \ref{eq:pf} ) \sim 
\frac{ \sqrt{ 2 \pi } }{ \Gamma (b) } \frac{ ( \ep - 1 )^{ a - c + \frac{ 1 }{ 2 } } }{ \ep^{ a - \frac{ 1 }{ 2 } } } 
\frac{ z^{ 1 - c } ( \ep z - 1 )^{ b - 1 } }{ ( 1 - z )^{ a + b - c } } \: \lambda^{ b - \frac{ 1 }{ 2 } } \left[ \frac{ 1 }{ \ep^{\ep} z } \left( \frac{ \ep - 1 }{ 1 - z } \right)^{ \ep - 1 } \right]^{ \lambda }  
\end{equation}
as $ \lambda\to\infty$. This is exactly the same as the second term of \eref{eq:AE:ep>1}. Therefore, the contributions due to the enclosed branch point and the enclosed pole when deforming the integration path in \eref{eq:int:ep>1} have identical analytic expressions. 

\bibliographystyle{myunsrtnat}

\bibliography{HypergeometricFunctionsExpansions}


\newpage




\addcontentsline{toc}{section}{Supplementary material}

\begin{center}
{\hspace{1cm} }\\{\hspace{1cm} }\\{\hspace{1cm} }\\Supplementary material for\\
{\Large{{\bf{\textsf{Asymptotic expansions of the hypergeometric \\ \vspace{0.01cm} function with two large parameters --- application to \\ \vspace{0.01cm} the partition function of a lattice gas in a field of \\ \vspace{0.01cm} traps \\ \vspace{0.01cm}}}}}}
\end{center}
\begin{center}
\vskip6pt
Mislav Cvitkovi\'c$^{1,2,\dagger}$, Ana-Sun\v{c}ana\ Smith$^{1,2}$, Jayant Pande$^{1}$

\vskip12pt

$^1${\footnotesize{\emph{PULS Group, Department of Physics \& Cluster of Excellence: EAM, Friedrich-Alexander University Erlangen-N\"urnberg, N\"agelsbachstr.\ 49b, Erlangen, Germany}}}\\
$^2${\footnotesize{\emph{Division of Physical Chemistry, Ruđer Bošković Institute, Bijeni\v{c}ka c.\ 54, Zagreb, Croatia}}}\\
\vskip4pt
$^\dagger${\footnotesize{Author for correspondence. E--mail:\ \href{mailto:mislav.cvitkovic@fau.de}{mislav.cvitkovic@fau.de}. }}
\vskip2pt
\end{center}

\setcounter{figure}{0}
\setcounter{page}{1}
\renewcommand\thefigure{SI\arabic{figure}}

\begin{figure}[h!]
\centering
\subfloat[]{\includegraphics[width=0.49\textwidth]{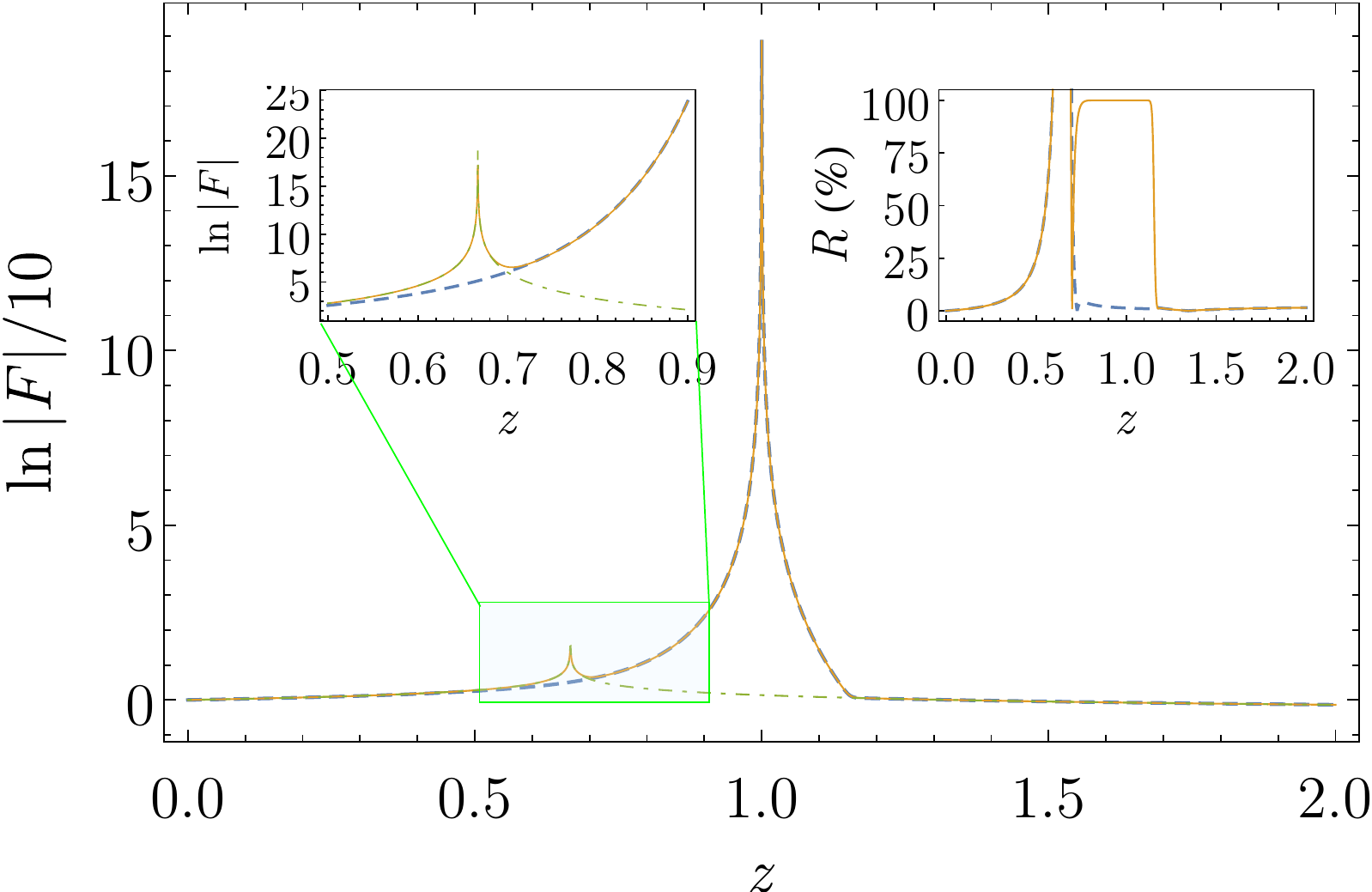}}
\hfill
\subfloat[]{\includegraphics[width=0.49\textwidth]{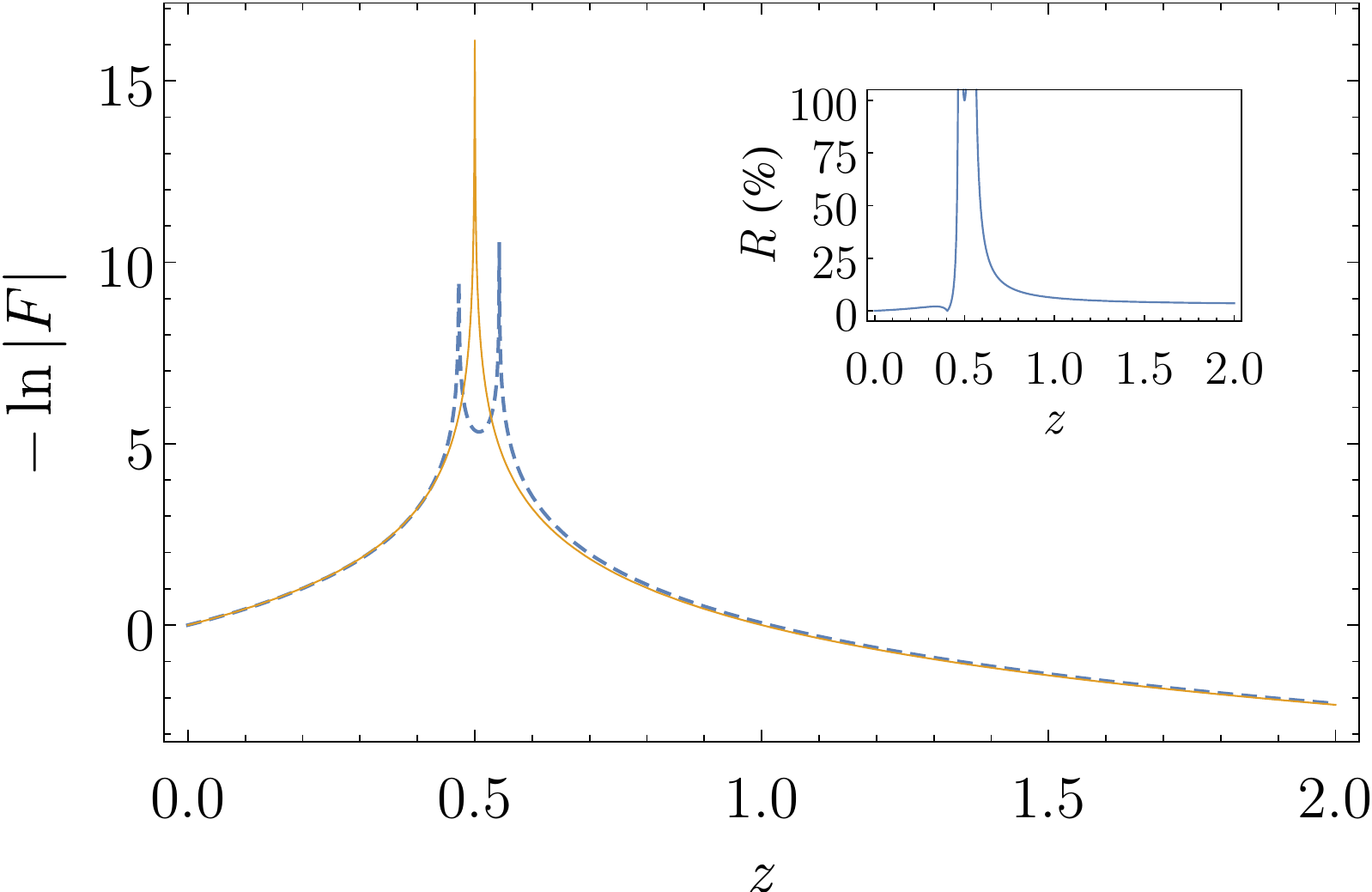}}\\
\subfloat[]{\includegraphics[width=0.49\textwidth]{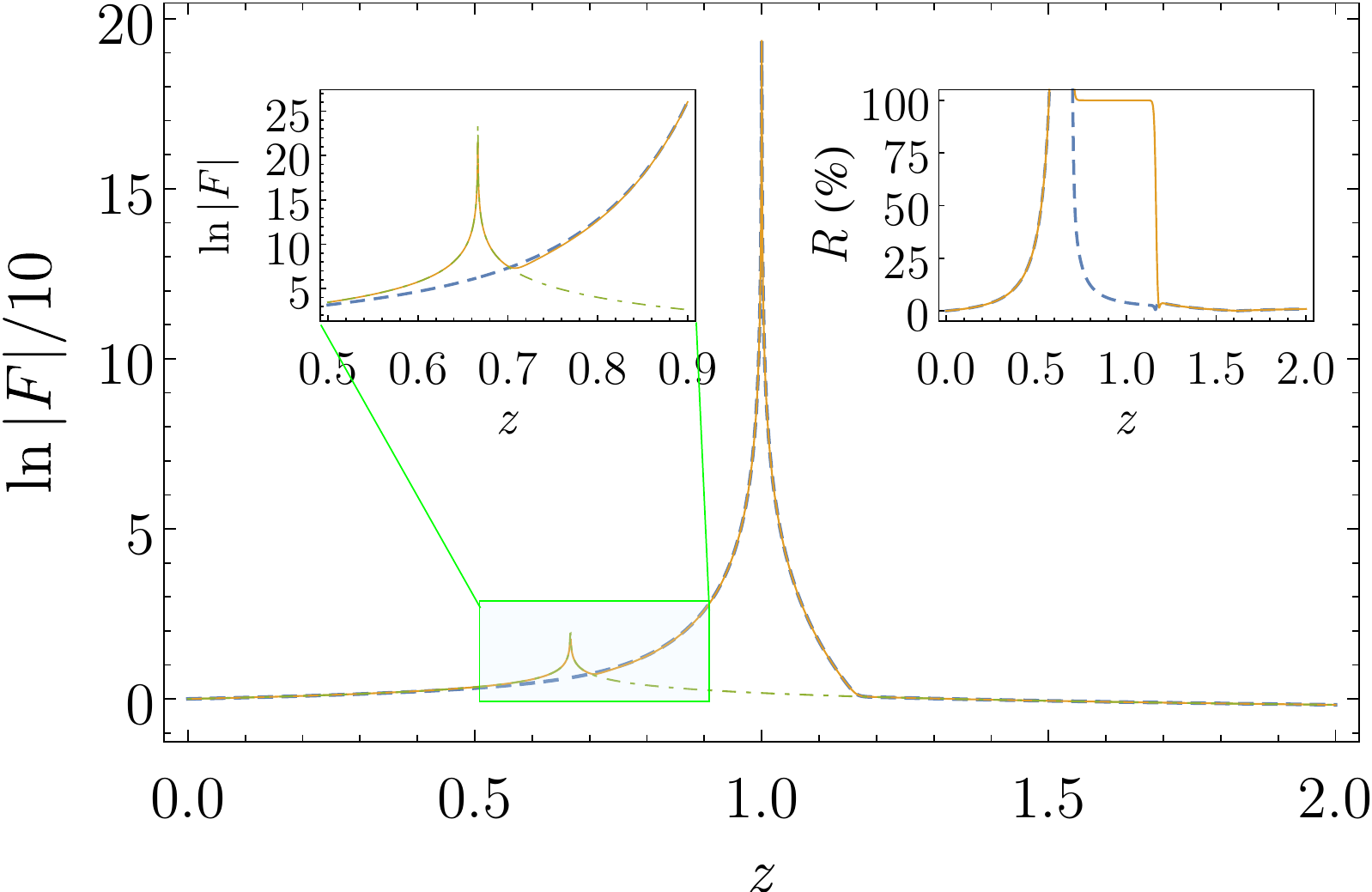}}
\hfill
\subfloat[]{\includegraphics[width=0.49\textwidth]{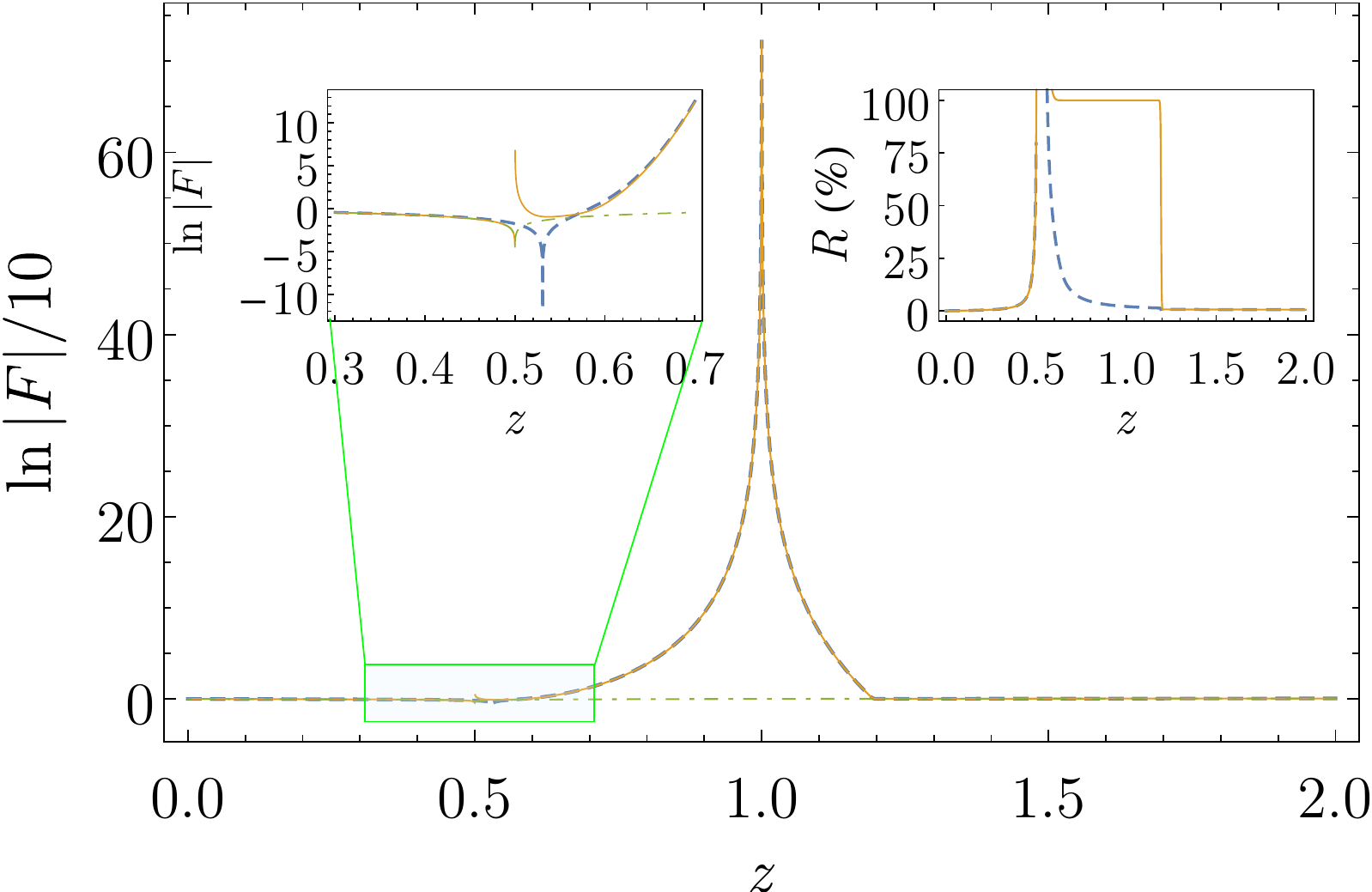}}
\caption{Graphs of the HGF for $ a = 0 $, $ c = 1 $, and different $ b $, $ \varepsilon $ and $ \lambda \neq \mathbb{N} $: (a) $ ( b, \varepsilon, \lambda ) = ( 2, \frac{ 3 }{ 2 }, 50.5 ) $, (b) $ ( b, \varepsilon, \lambda ) = ( -2, 2, 100.25 ) $, (c) $ ( b, \varepsilon, \lambda ) = ( \frac{ 5 }{ 2 }, \frac{ 3 }{ 2 }, 50.25 ) $, (d) $ ( b, \varepsilon, \lambda ) = ( - \frac{ 1 }{ 2 }, 2, 100.25 ) $. The following description is common to all the plots. Dashed blue: the HGF. Solid orange: the AE  \eref{eq:AE:ep>1} with pole/branch cut contribution included. Dot-dashed green: the AE \eref{eq:AE:ep<1}. Left inset: the vicinity of $ z = 1 / \varepsilon $ enlarged. Right inset: the relative error of limited (eq.\ \ref{eq:AE:ep<1}, solid orange) and full (eq.\ \ref{eq:AE:ep>1}, dashed blue) expansions.  Note the logarithm of absolute values on the graphs. \label{fig:SI1}}
\end{figure}

\newpage

\begin{figure}[h!]
\centering
\subfloat[]{\includegraphics[width=0.49\textwidth]{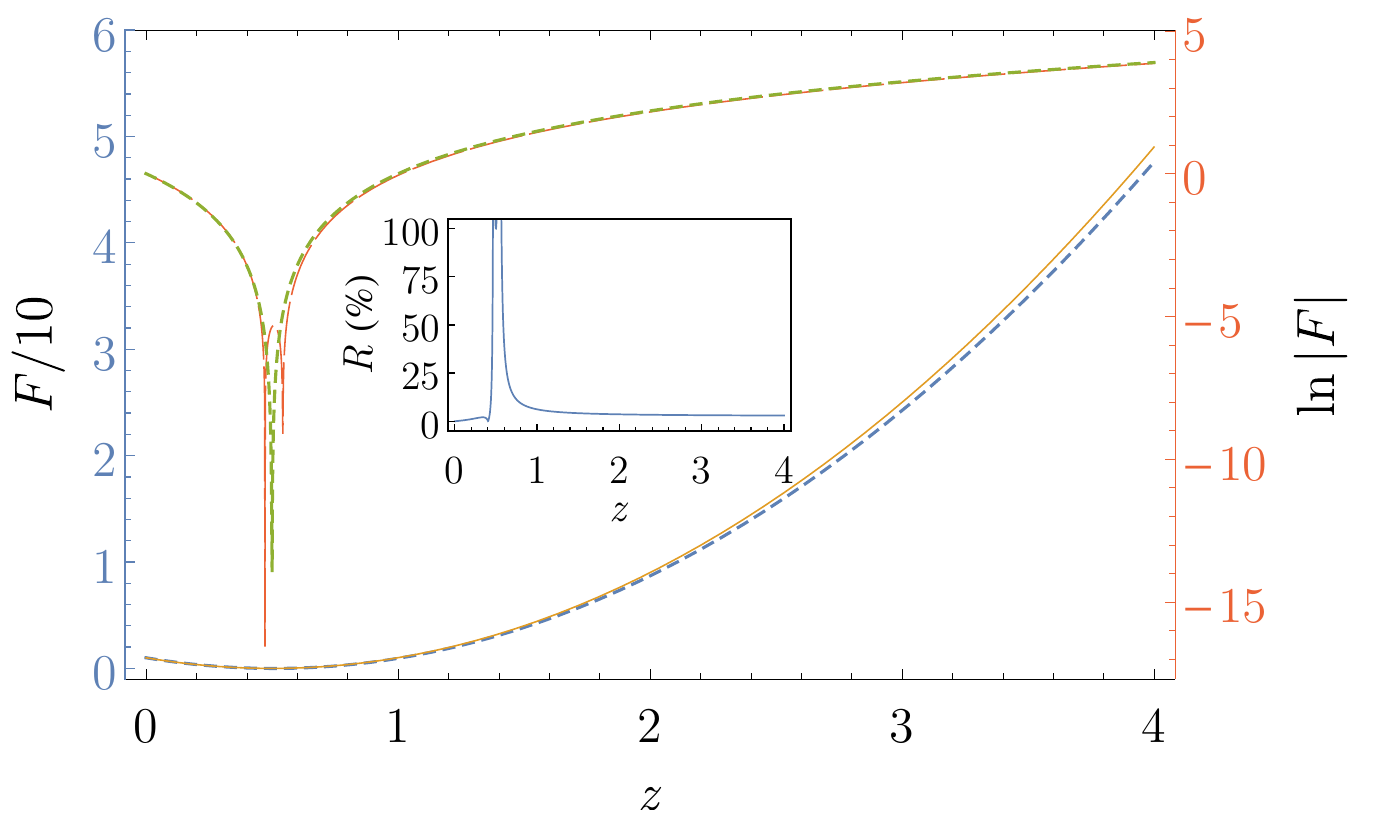}}
\hfill
\subfloat[]{\includegraphics[width=0.49\textwidth]{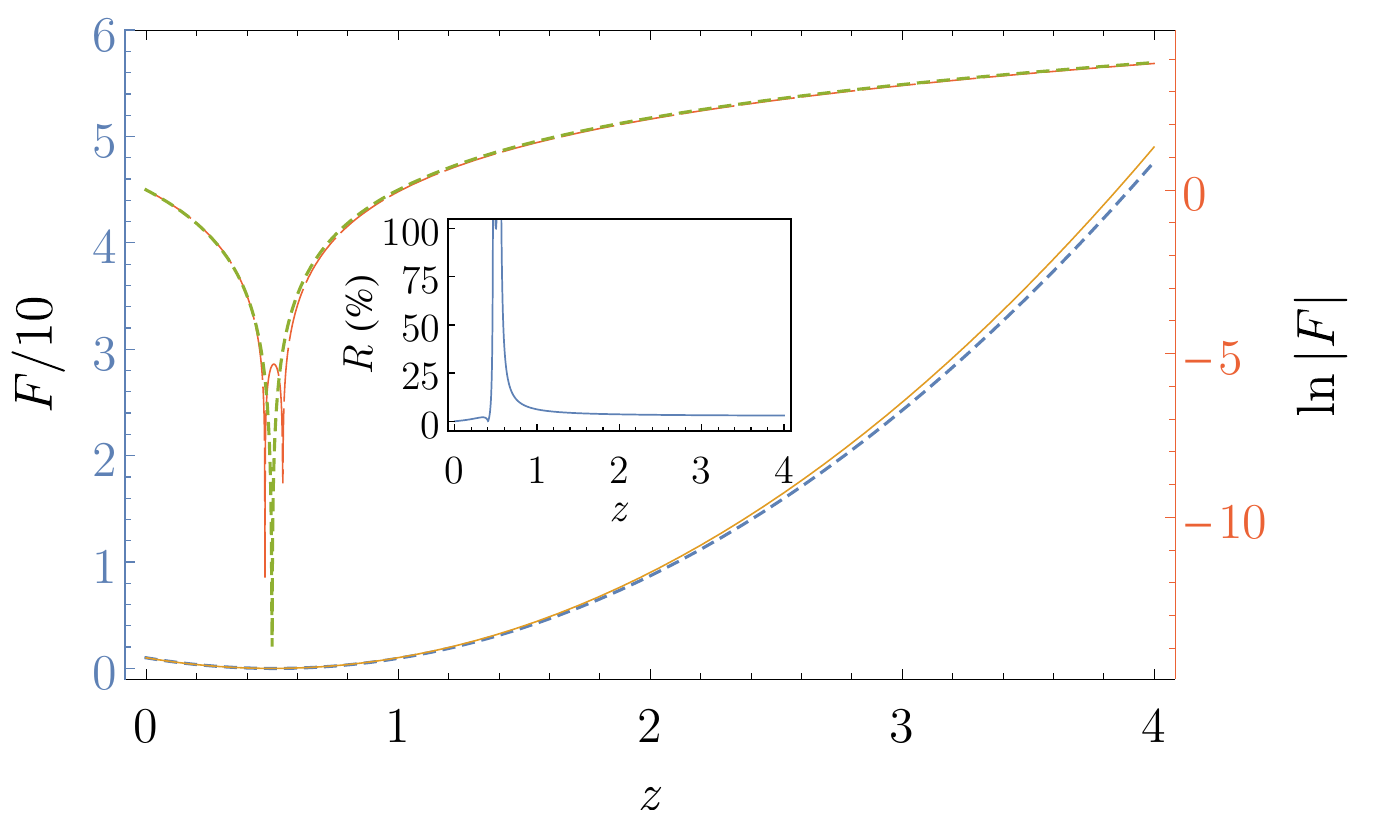}}
\caption{Figures \ref{fig:5}(b) and \ref{fig:SI1}(b) with both linear (left, blue) and logarithmic (right, red) axes. Dashed blue and long--dashed red: the HGF. Solid orange and dotted green: expansions. \label{fig:SI2}}
\end{figure}

\vskip20pt

\begin{figure}[h!]
\centering
\subfloat[]{\includegraphics[width=0.49\textwidth]{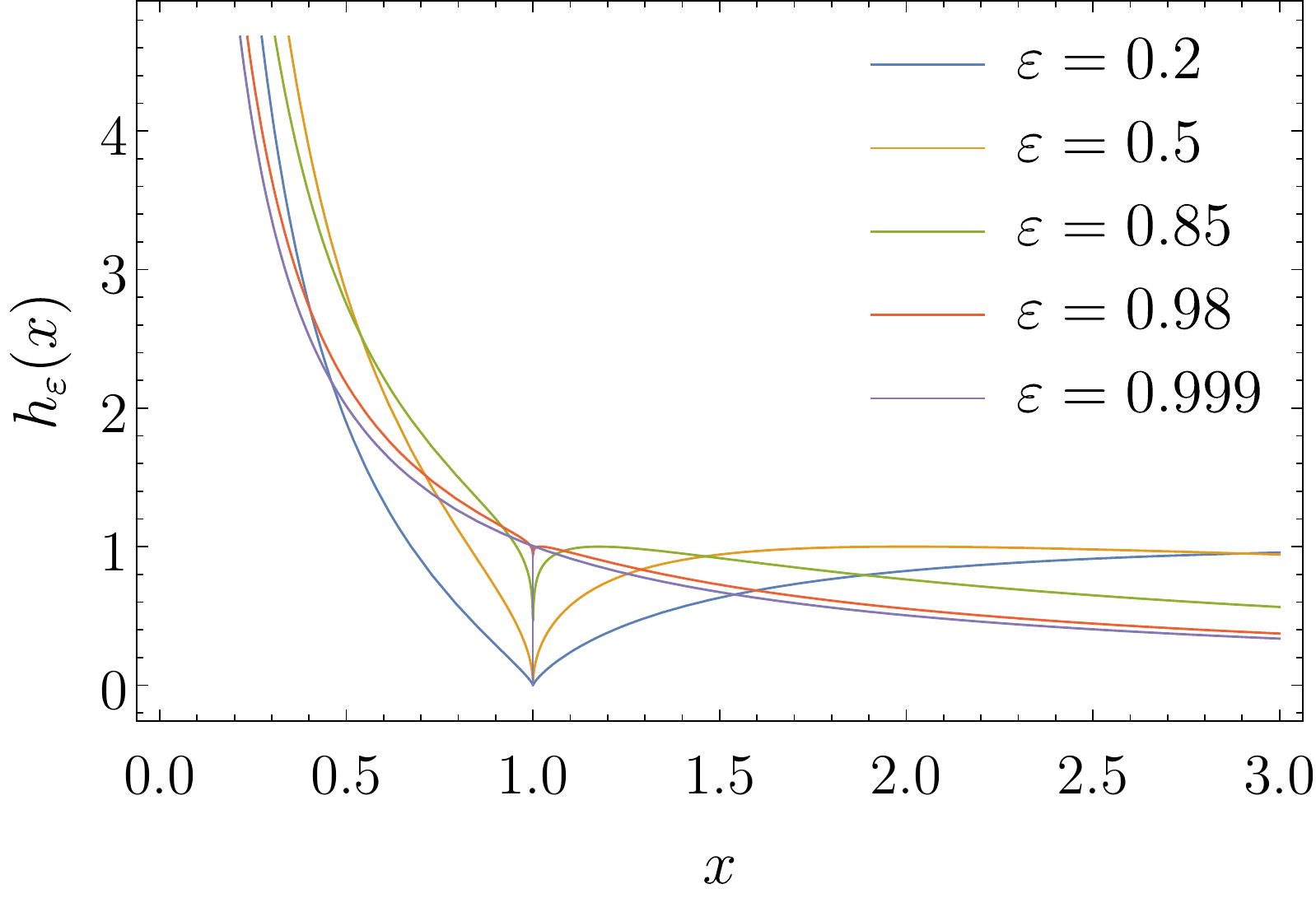}}
\hfill
\subfloat[]{\includegraphics[width=0.49\textwidth]{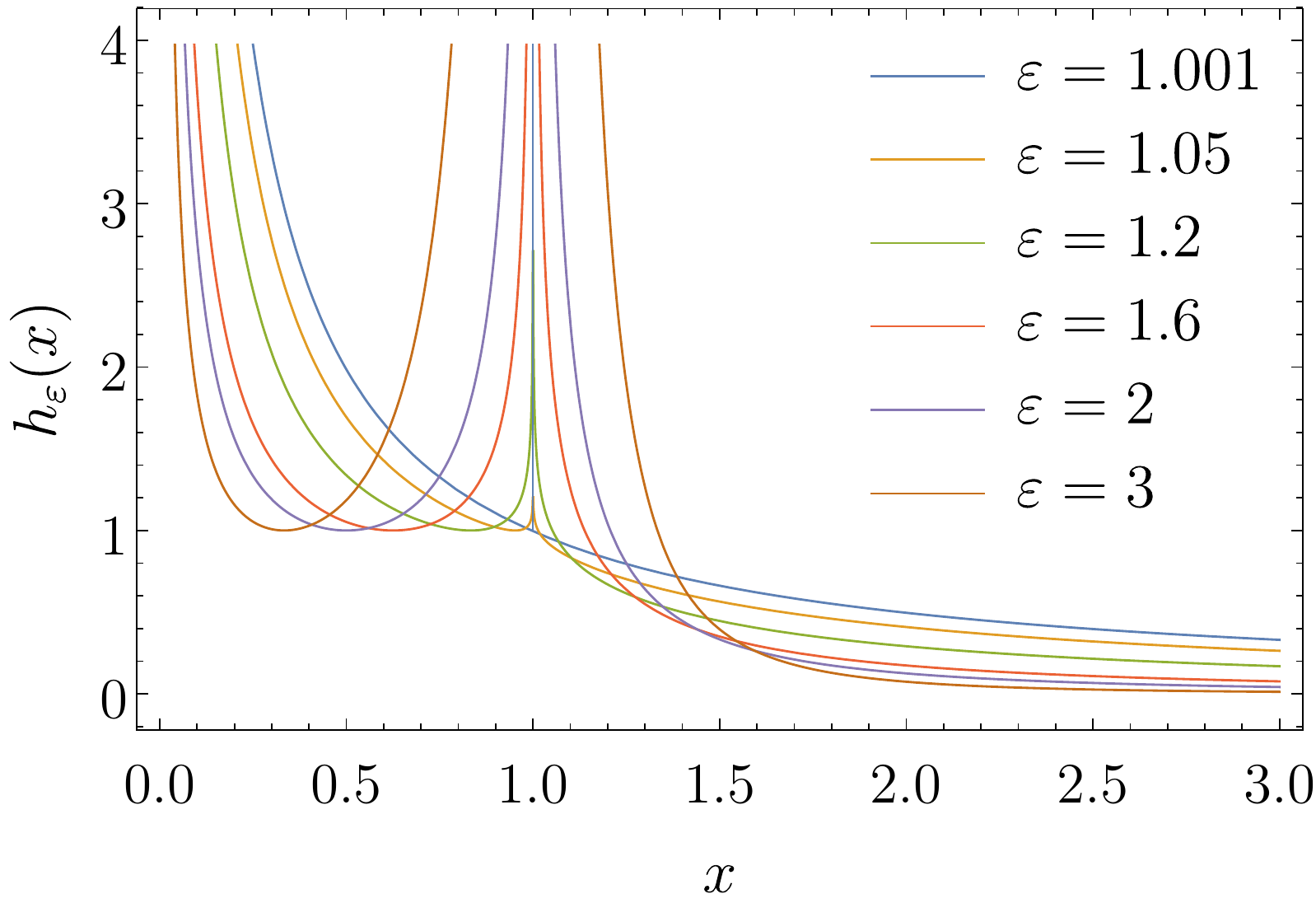}}
\caption{The plots of $h_{\varepsilon} (x)$ from \ref{sec:appB} for various $\varepsilon$. \label{fig:SI3}}
\end{figure}

\newpage

\begin{figure}[h!]
\centering
\subfloat[$\varepsilon=\frac{1}{2}$, $z=\frac{2}{3}+\rmi$]{\includegraphics[width=0.45\textwidth]{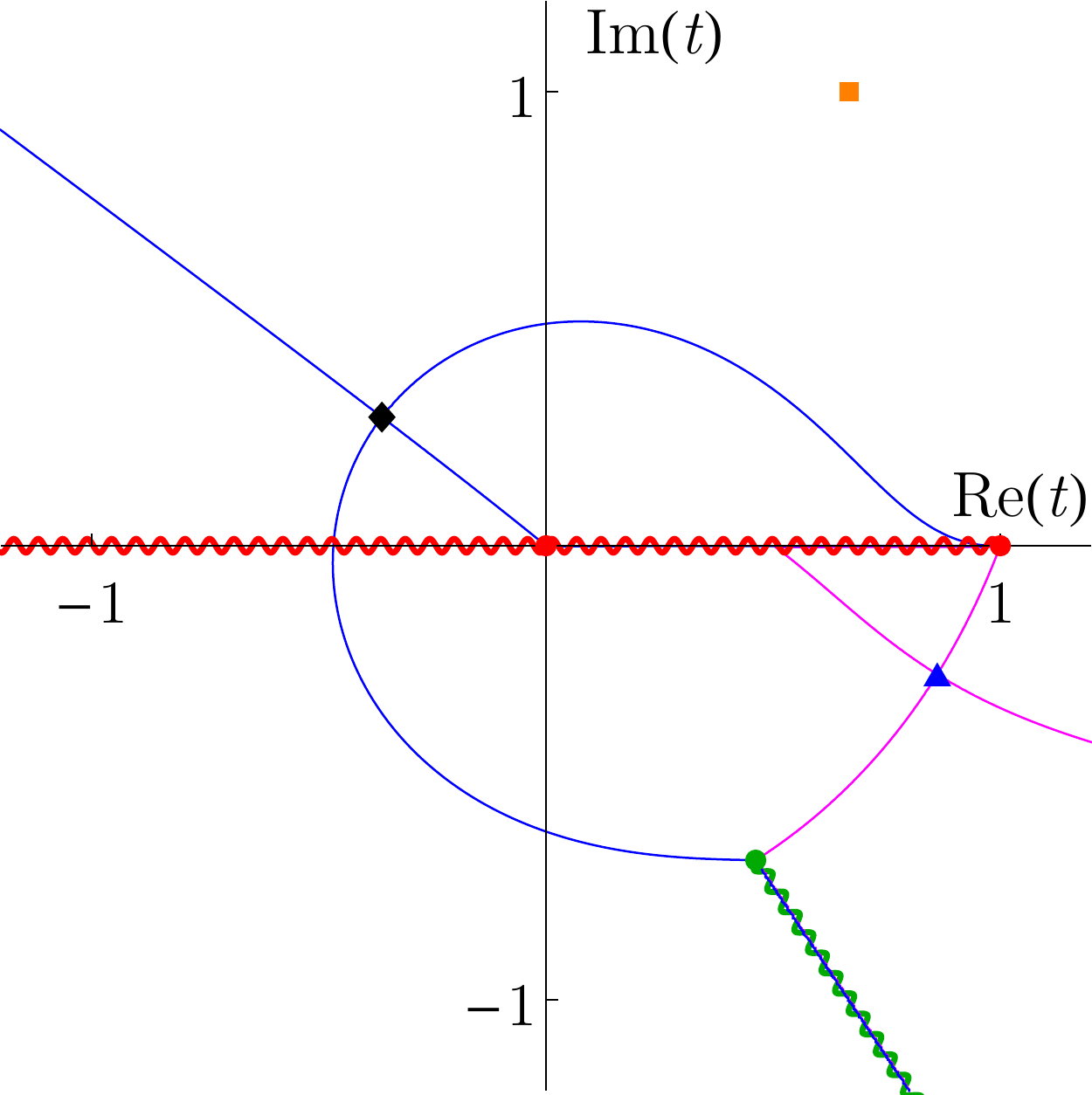}}
\hfill
\subfloat[$\varepsilon=\frac{1}{2}$, $z=\frac{4}{3}+\rmi$]{\includegraphics[width=0.45\textwidth]{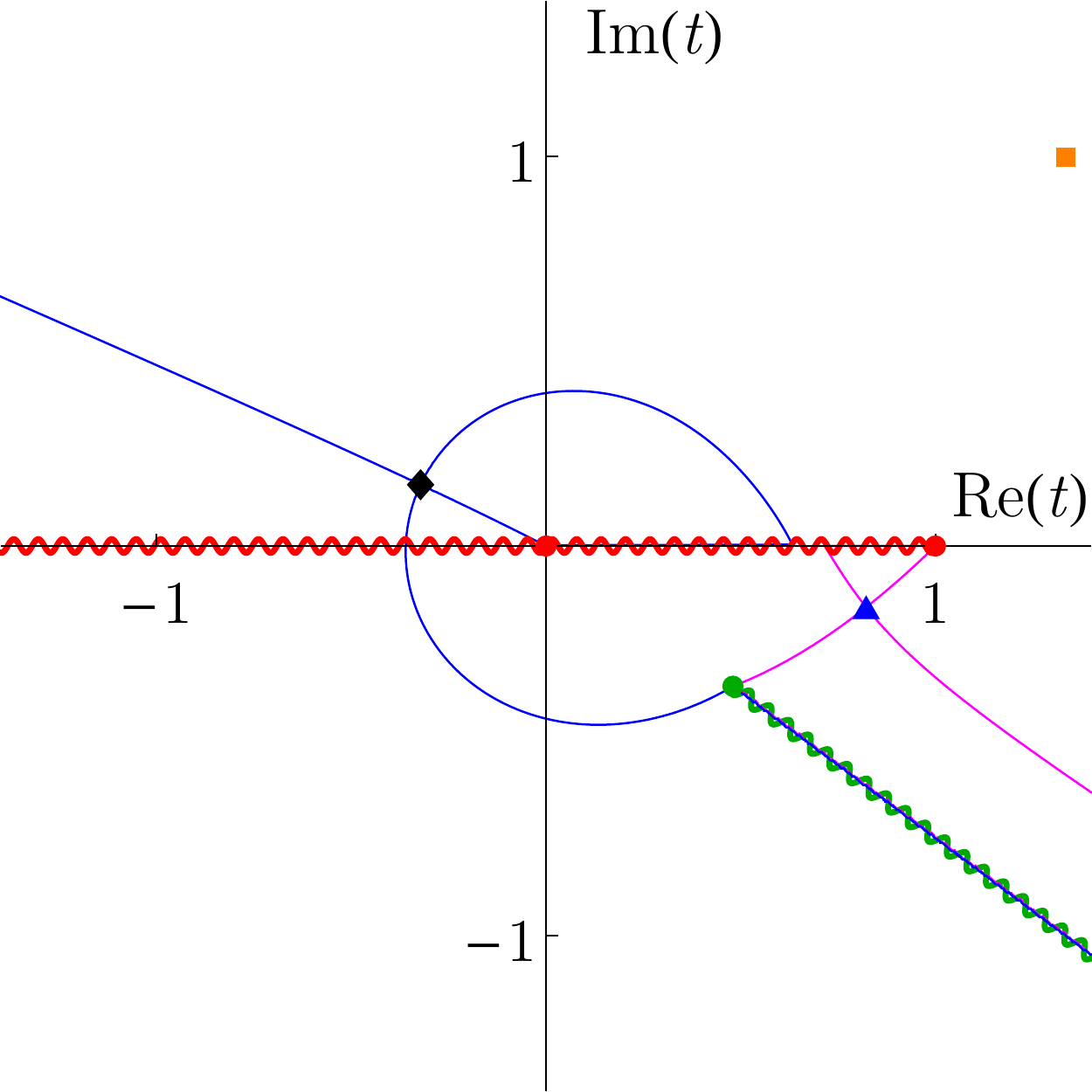}}\\
\subfloat[$\varepsilon=\frac{3}{2}$, $z=\frac{2}{3}+\frac{2}{3}\rmi$]{\includegraphics[width=0.45\textwidth]{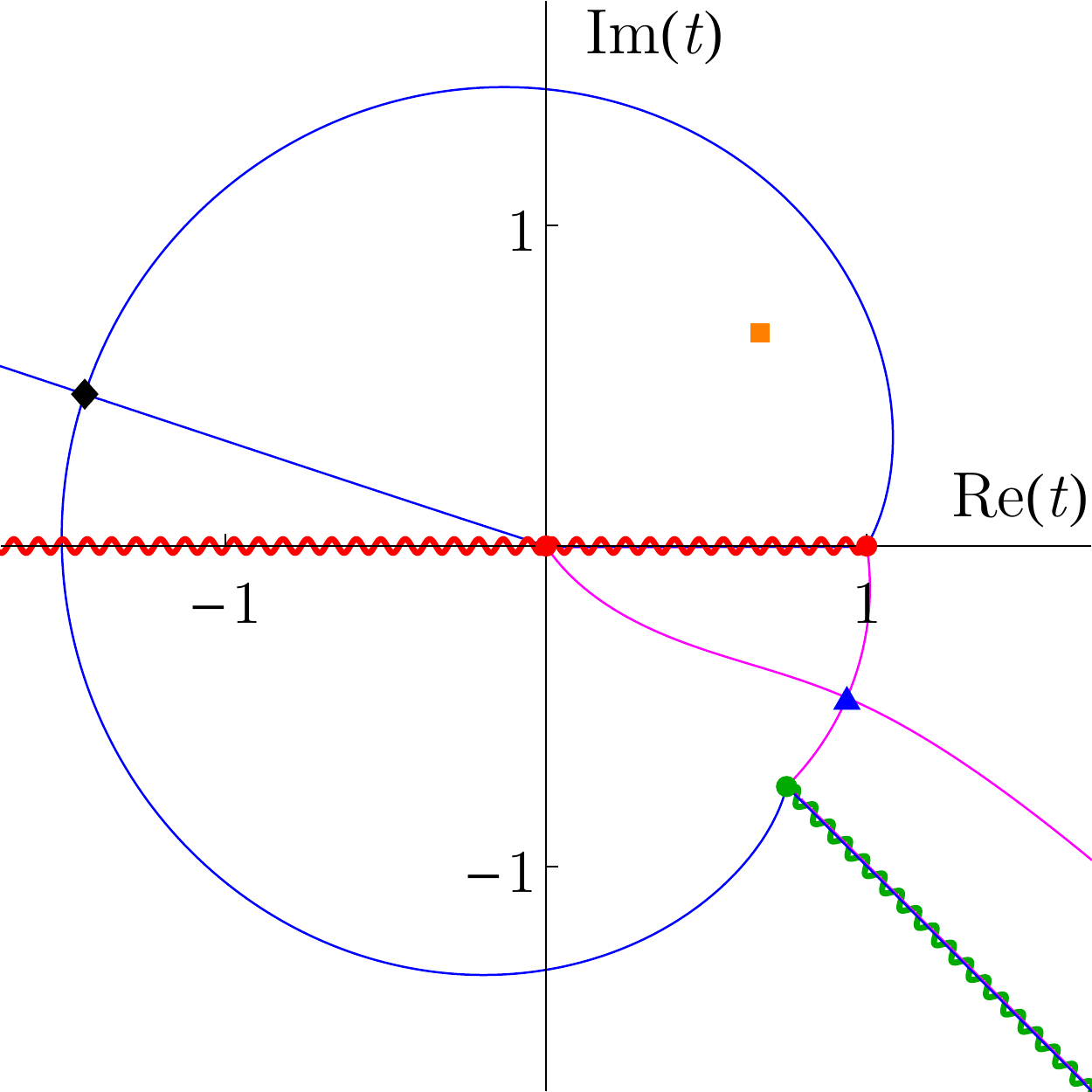}}
\hfill
\subfloat[$\varepsilon=\frac{3}{2}$, $z=\frac{4}{3}+\rmi$]{\includegraphics[width=0.45\textwidth]{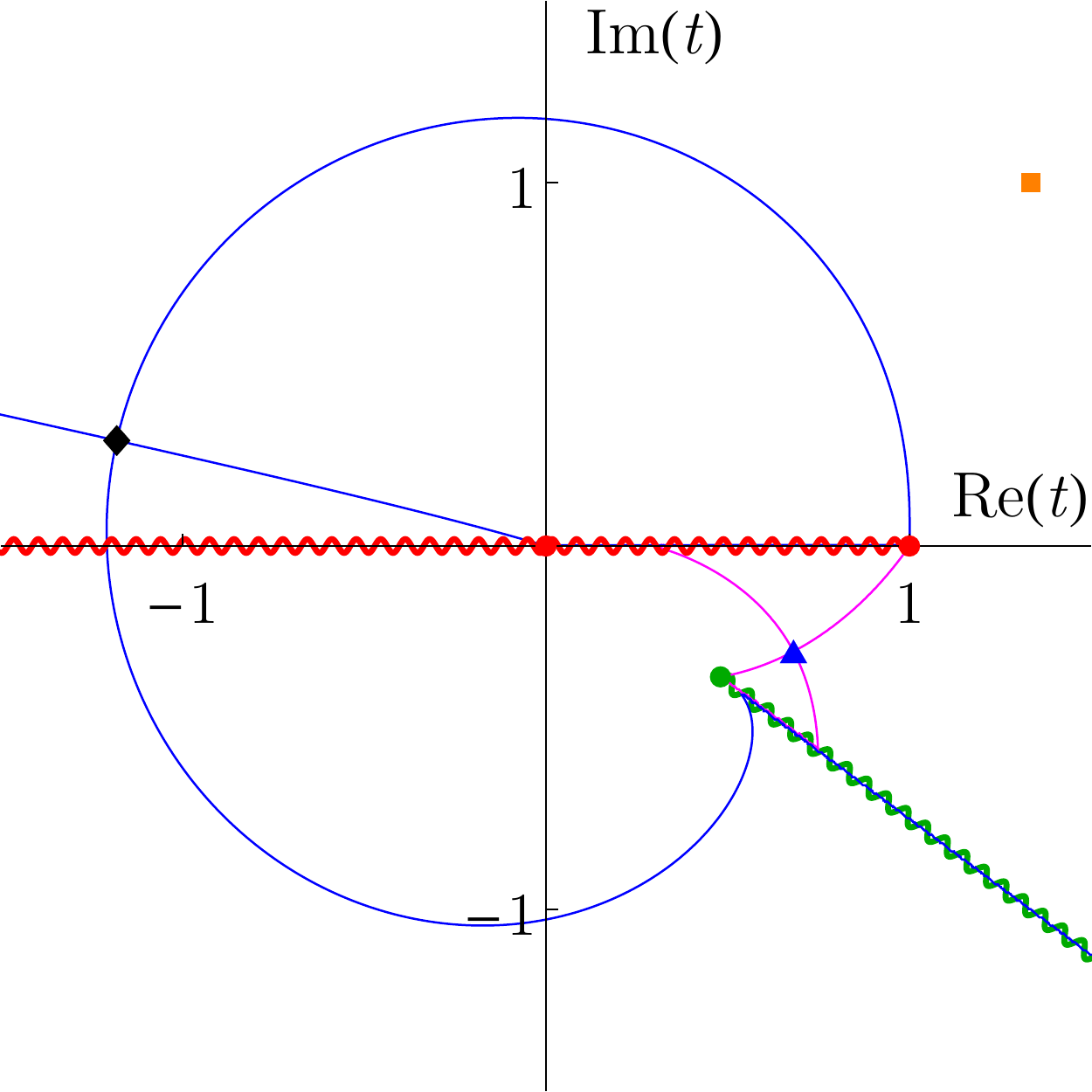}}
\caption{The steepest descent and ascent paths through $t_-$ (black diamond, blue line) and $t_+$ (blue triangle, magenta line). $\lambda$ is real. Orange square: the point $z$. Curly red: the branch cut $ ( - \infty, 1 ] $. Curly green: the branch cut $ [ 1/z, + \infty ) $. \label{fig:SI4}}
\end{figure}

\newpage

\begin{figure}[h!]
\centering
\subfloat[$\Im (z) = 0.003$]{\includegraphics[width=0.4\textwidth]{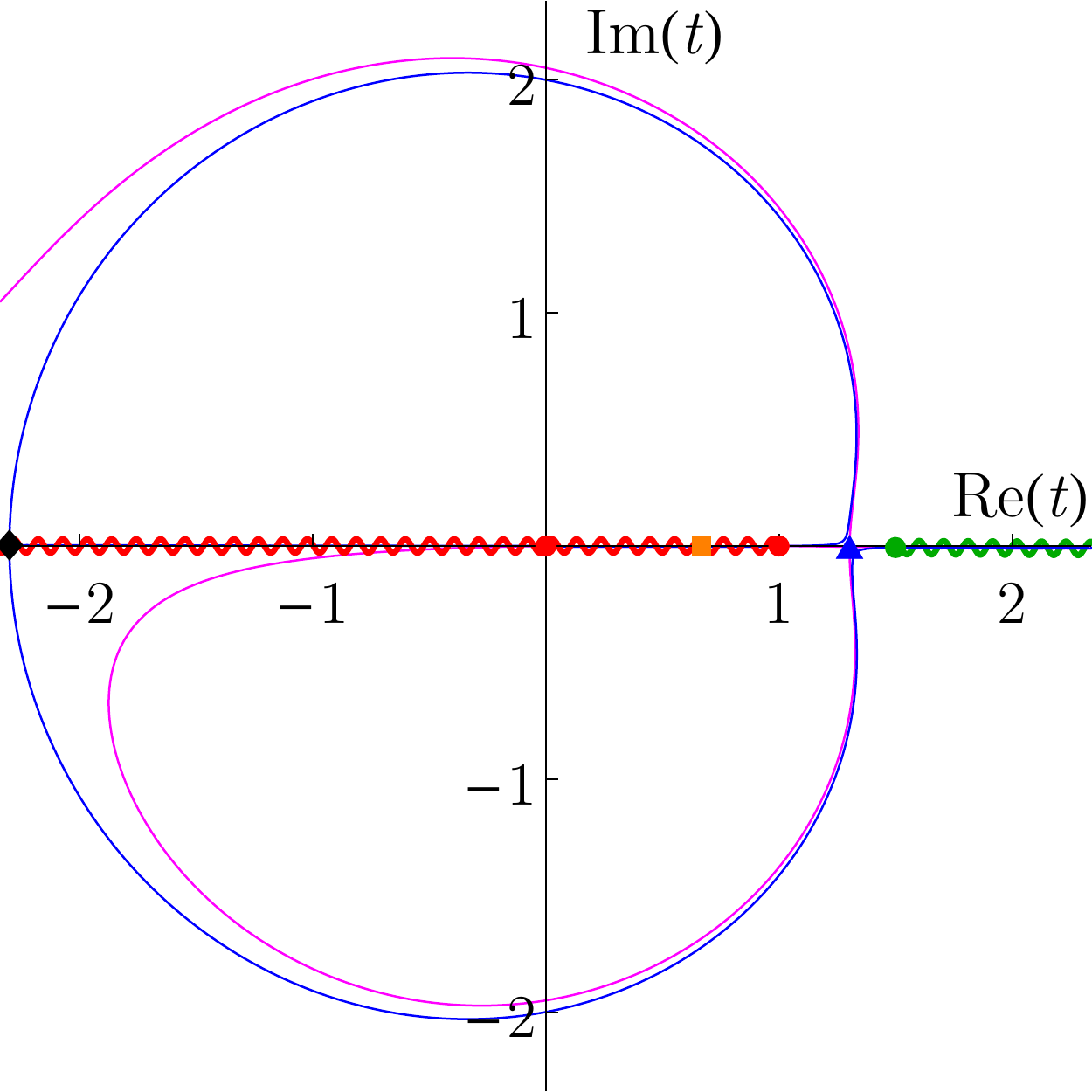}}
\hspace{1cm}
\subfloat[$\Im (z) = 0.03$]{\includegraphics[width=0.4\textwidth]{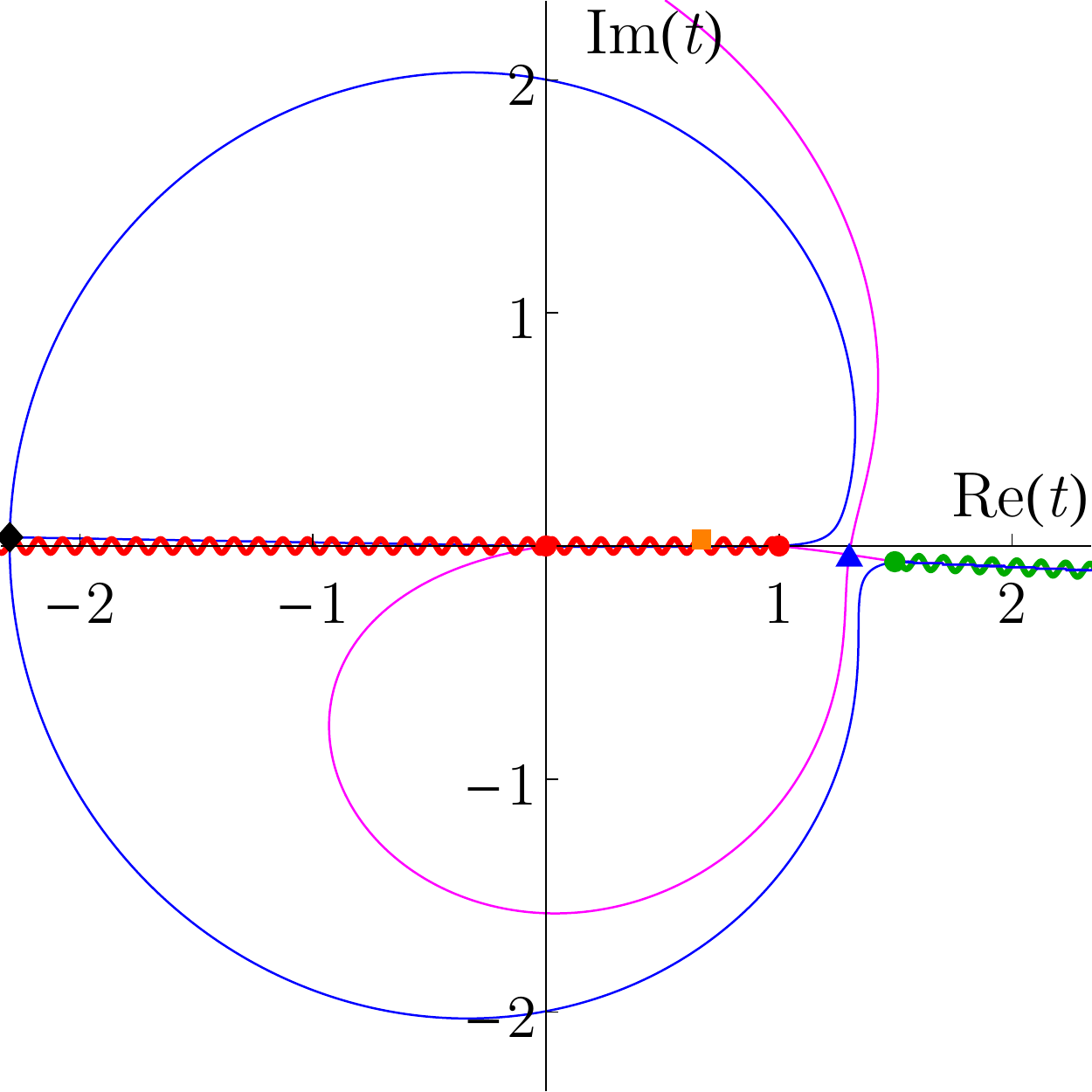}}\\
\subfloat[$\Im (z) = 0.3$]{\includegraphics[width=0.4\textwidth]{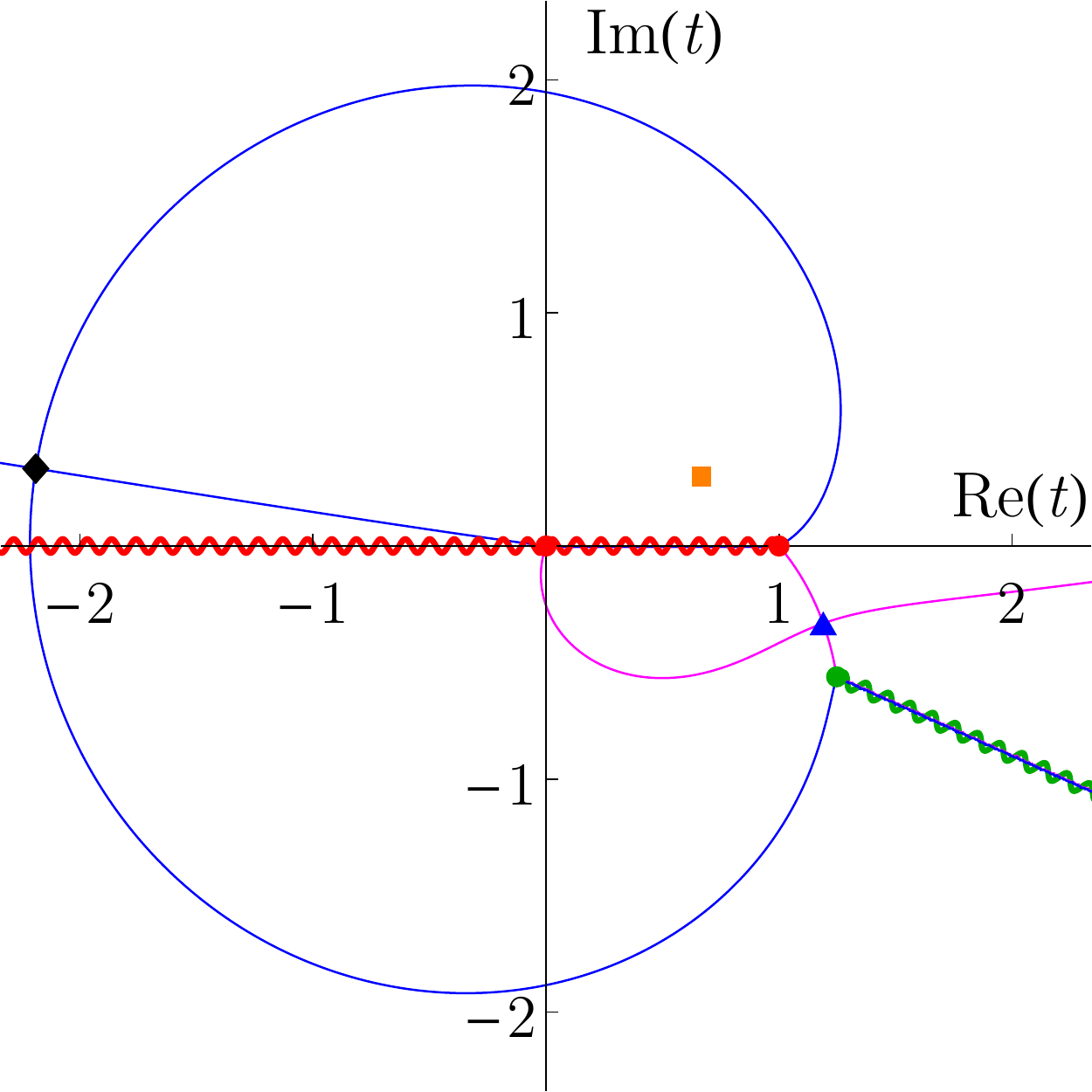}}
\hspace{1cm}
\subfloat[$\Im (z) = 0.5$]{\includegraphics[width=0.4\textwidth]{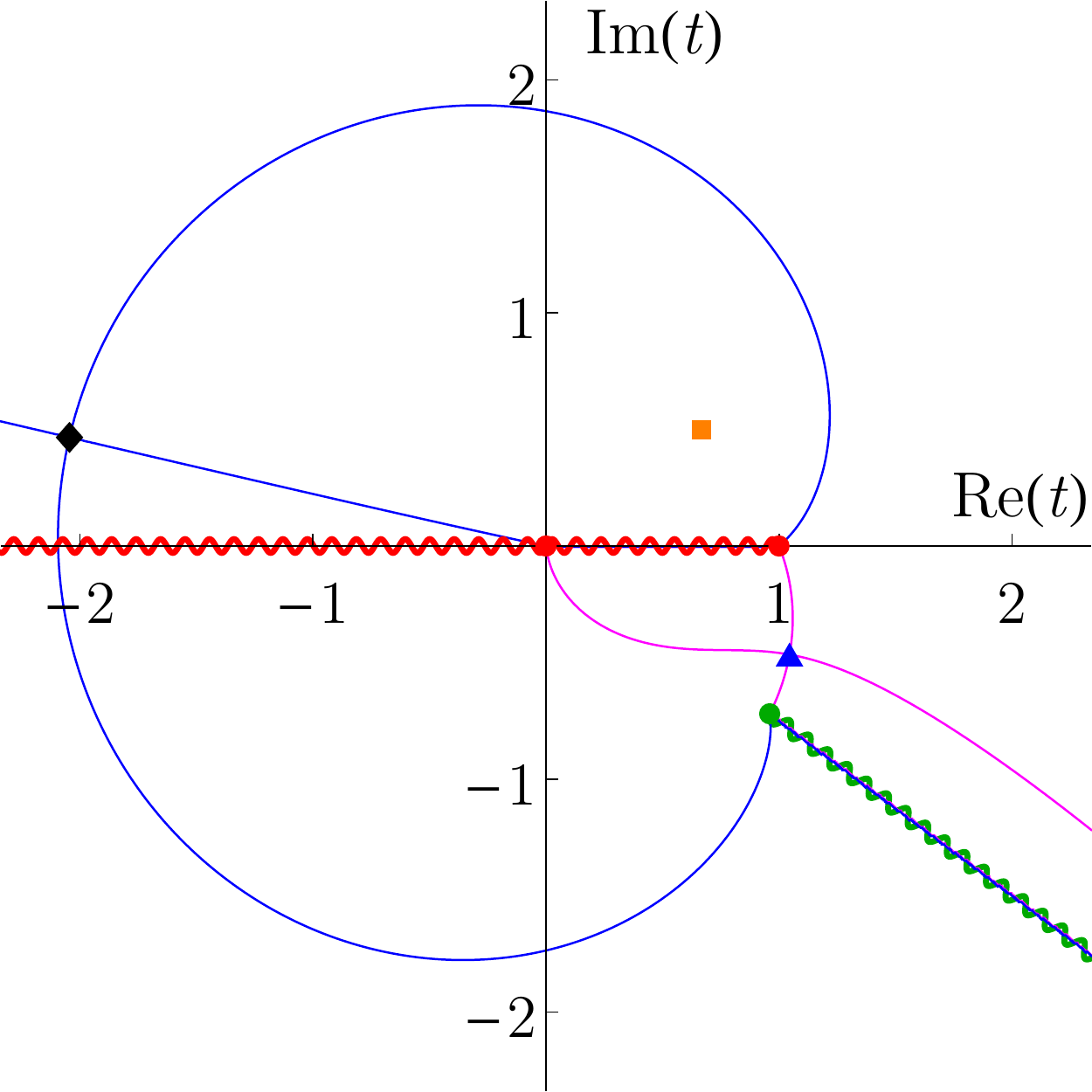}}\\
\subfloat[$\Im (z) = 1$]{\includegraphics[width=0.4\textwidth]{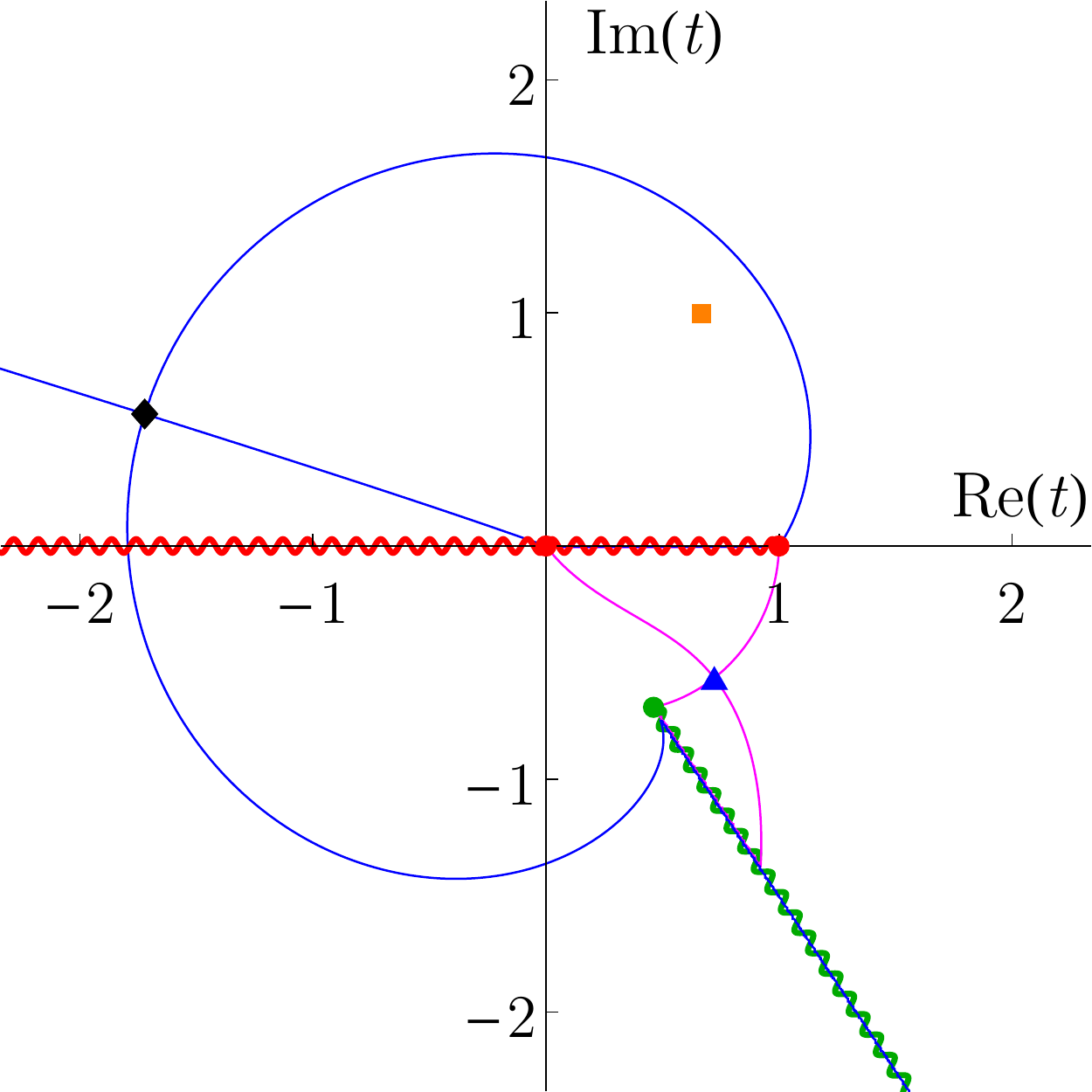}}
\hspace{1cm}
\subfloat[$\Im (z) =3$]{\includegraphics[width=0.4\textwidth]{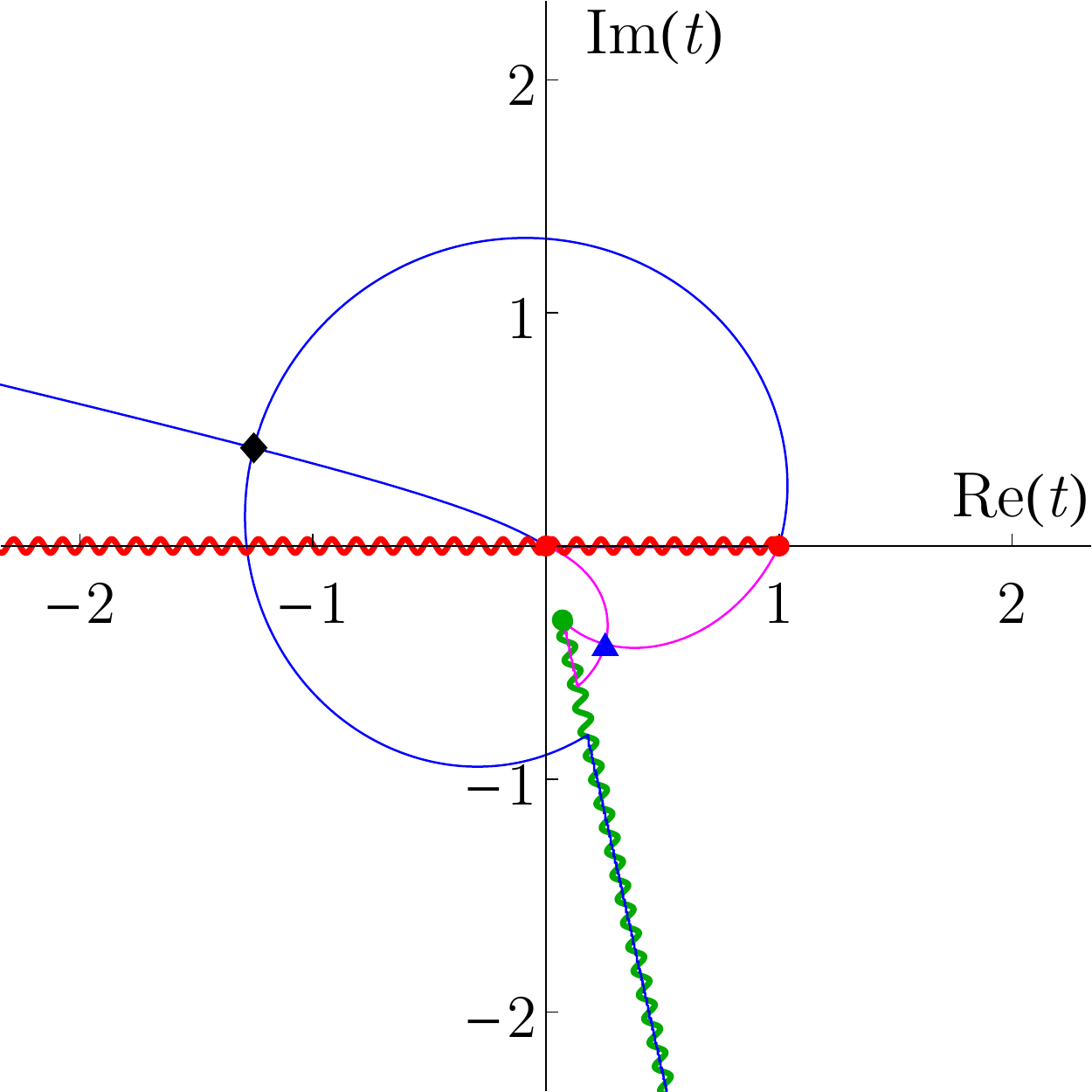}}
\caption{The steepest descent and ascent paths through $t_-$ and $t_+$ for real $\lambda$, $\varepsilon=2$ and $\Re(z)=2/3$, while varying $\Im(z)$. Legend: same as on \fref{fig:SI4}. \label{fig:SI5}}
\end{figure}

\newpage

\begin{figure}[h!]
\centering
\subfloat[$\Im (\lambda) = 0.003$]{\includegraphics[width=0.4\textwidth]{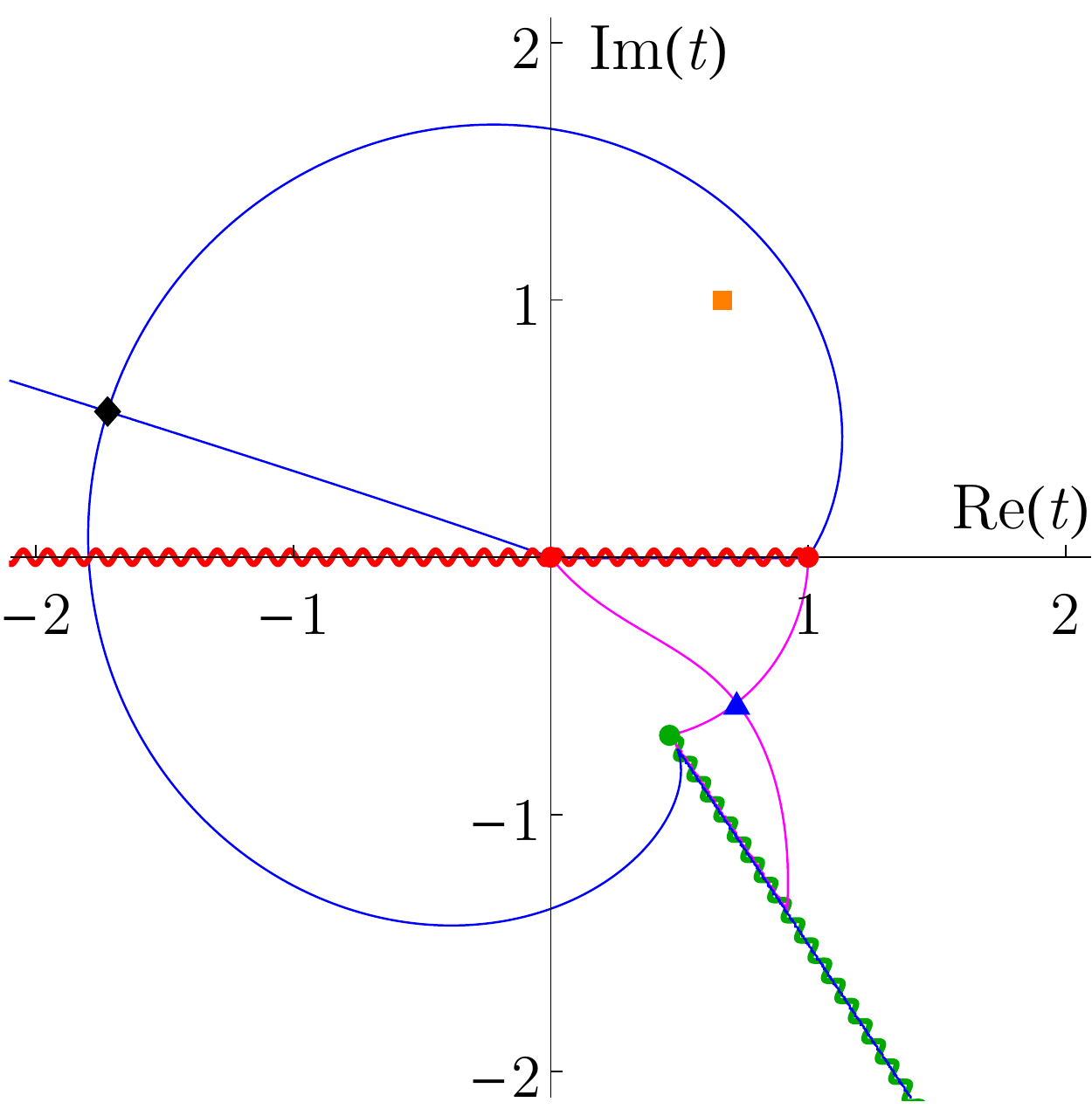}}
\hspace{1cm}
\subfloat[$\Im (\lambda) = 0.03$]{\includegraphics[width=0.4\textwidth]{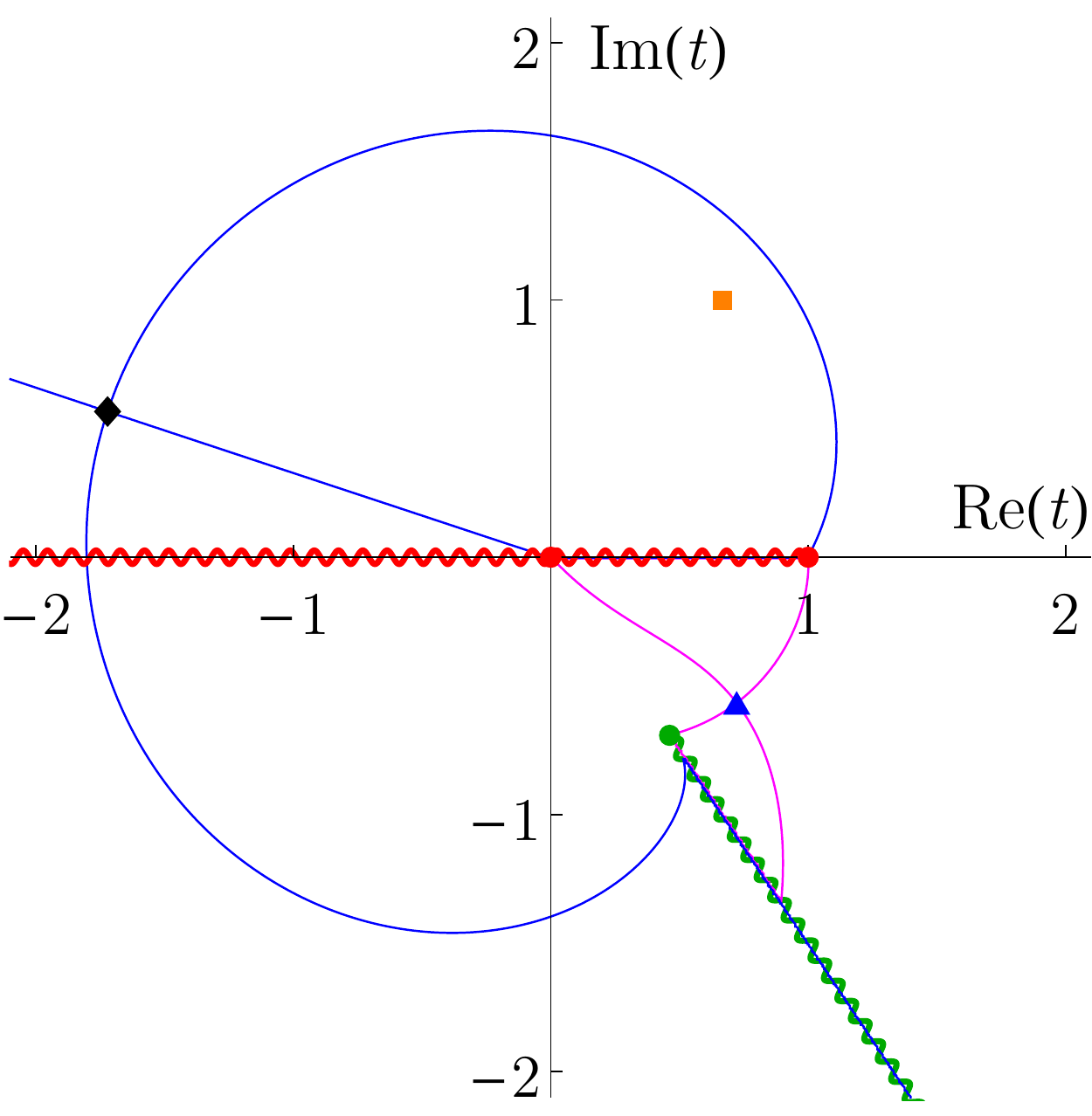}}\\
\subfloat[$\Im (\lambda) = 0.3$]{\includegraphics[width=0.4\textwidth]{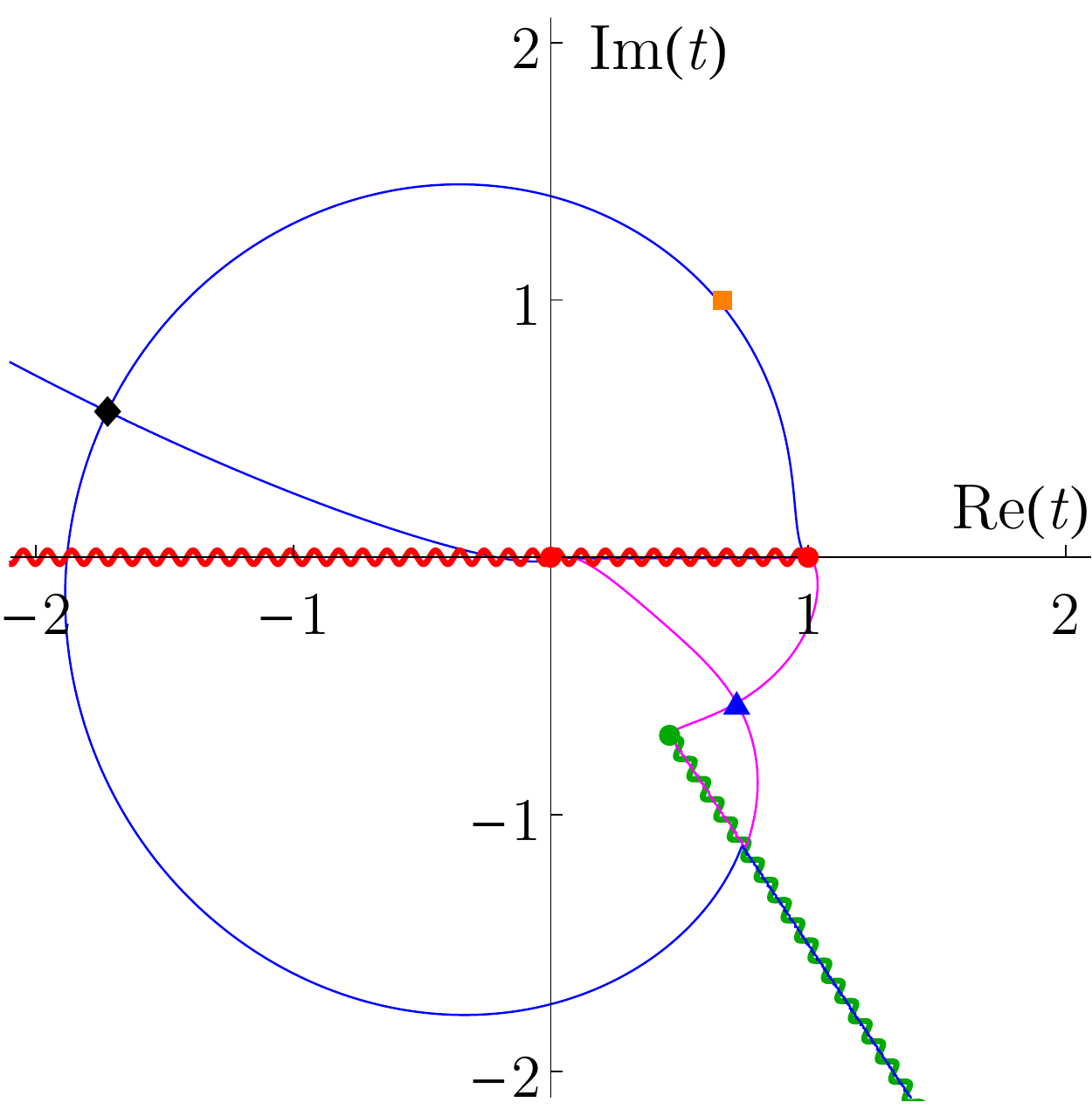}}
\hspace{1cm}
\subfloat[$\Im (\lambda) = 0.5$]{\includegraphics[width=0.4\textwidth]{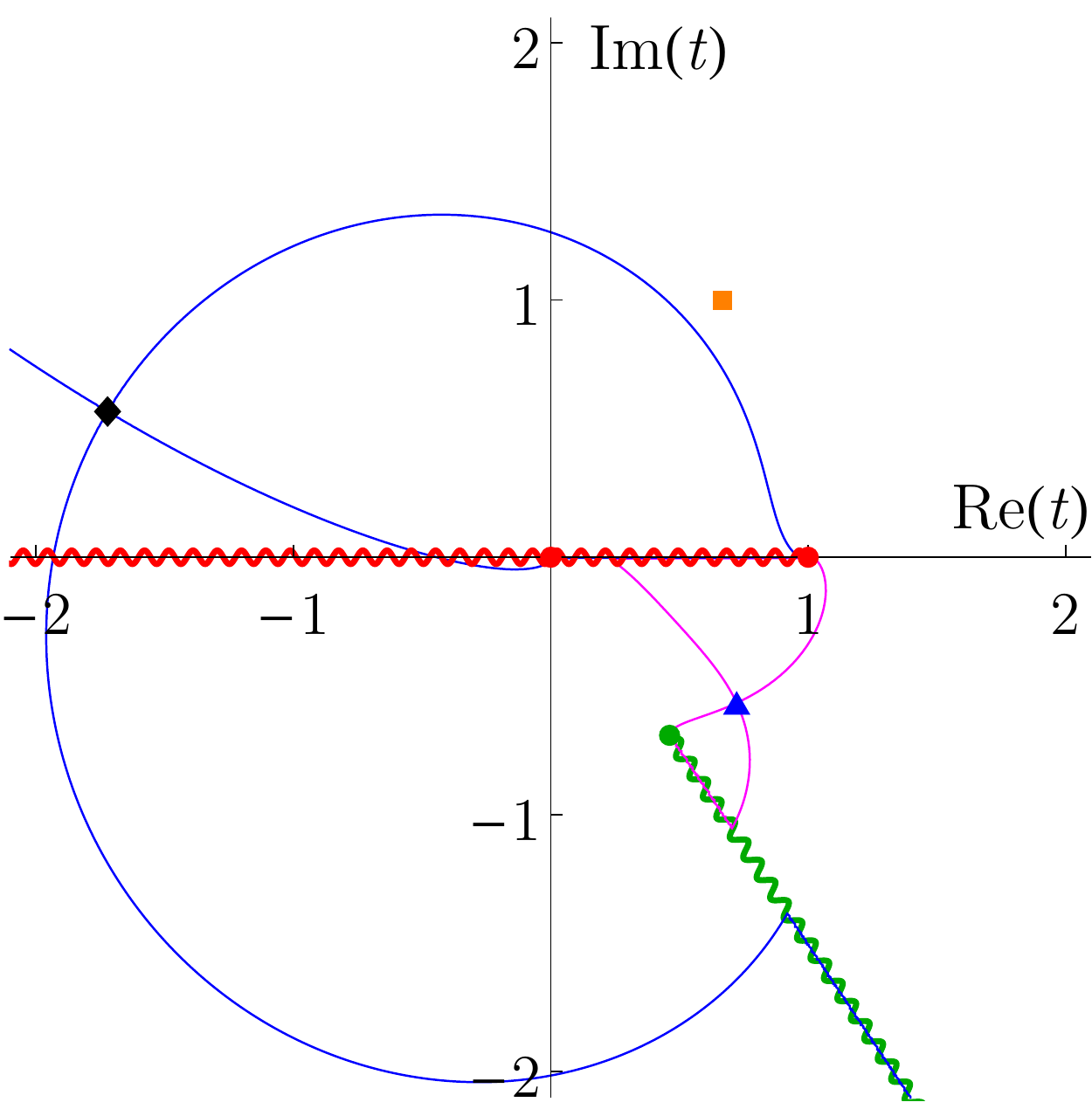}}\\
\subfloat[$\Im (\lambda) = 1$]{\includegraphics[width=0.4\textwidth]{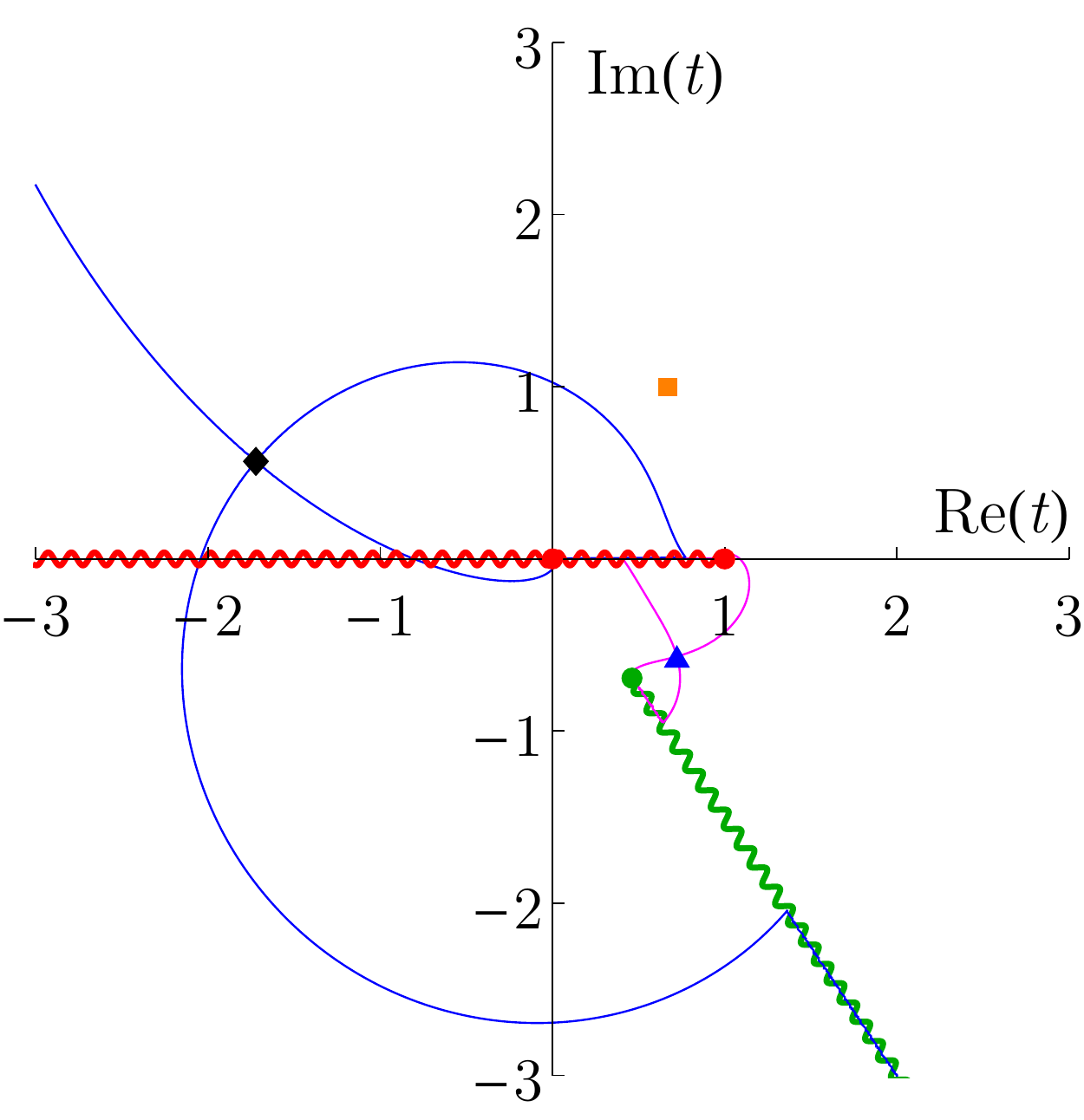}}
\hspace{1cm}
\subfloat[$\Im (\lambda) =3$]{\includegraphics[width=0.4\textwidth]{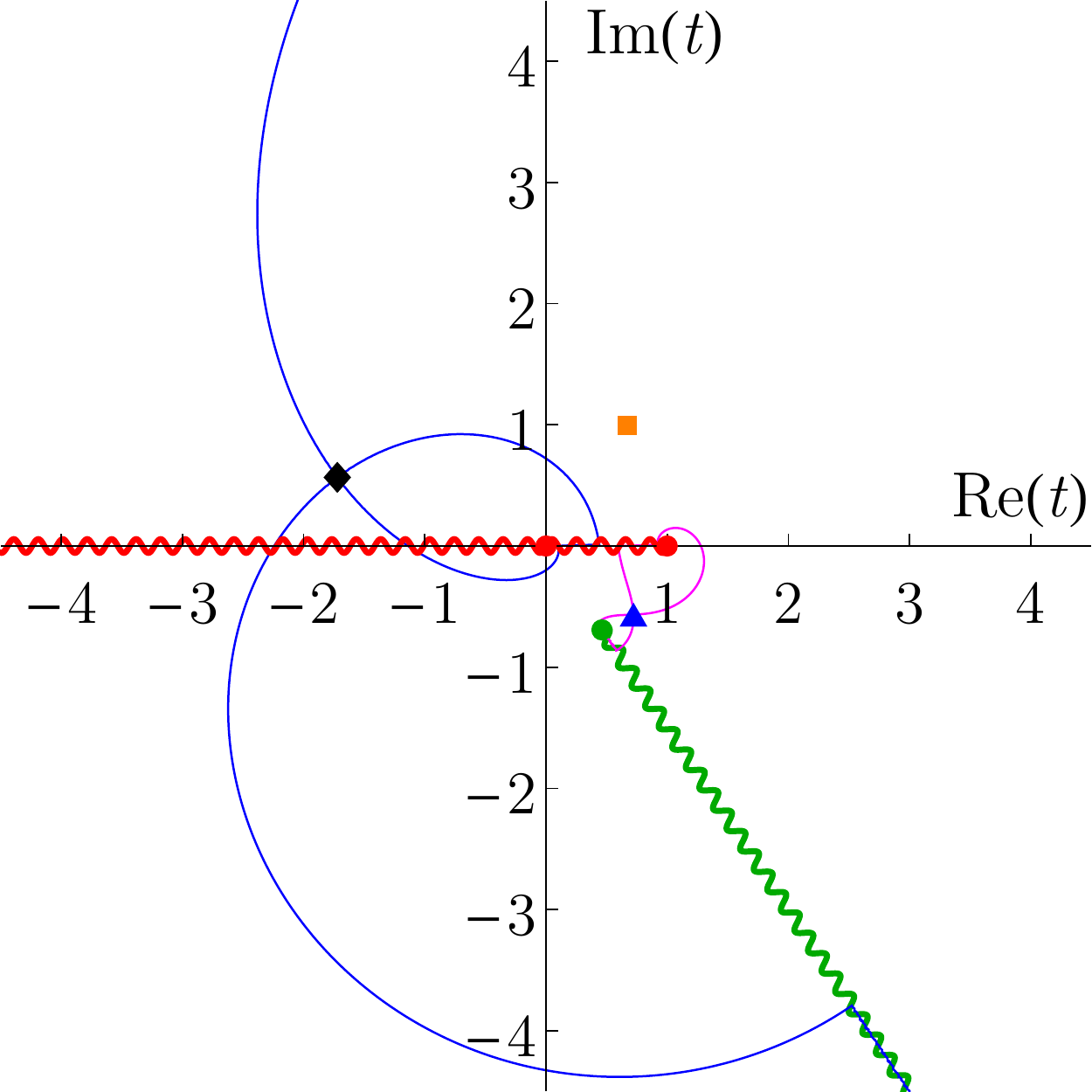}}
\caption{The steepest descent and ascent paths through $t_-$ and $t_+$ for $\varepsilon=2$, $z=2/3+\rmi $, and $\Re(\lambda) =1 $, while varying $\Im(\lambda)$. Legend: same as on \fref{fig:SI4}. \label{fig:SI6}}
\end{figure}

\end{document}